\documentclass[12pt]{article}
\usepackage{cite}
\usepackage{amsmath, amsthm, amssymb,slashed,graphicx}
\parskip=\baselineskip
\def\Bbb{\mathbb}
\def\Tr{{\rm Tr}}
\def\16{{\bf 16}}
\def\1{{\bf 1}}
\def\2{{\bf 2}}
\def\3{{\bf 3}}
\def\4{{\bf 4}}
\def\bar{\overline}
\def\tilde{\widetilde}
\def\R{{\Bbb{R}}}\def\Z{{\Bbb{Z}}}
\def\N{{\mathcal N}}
\def\hat{\widehat}
\font\teneurm=eurm10 \font\seveneurm=eurm7 \font\fiveeurm=eurm5
\newfam\eurmfam
\textfont\eurmfam=\teneurm \scriptfont\eurmfam=\seveneurm
\scriptscriptfont\eurmfam=\fiveeurm

 \font\teneusm=eusm10 \font\seveneusm=eusm7 \font\fiveeusm=eusm5
\newfam\eusmfam
\textfont\eusmfam=\teneusm \scriptfont\eusmfam=\seveneusm
\scriptscriptfont\eusmfam=\fiveeusm
\def\eusm#1{{\fam\eusmfam\relax#1}}
\font\tencmmib=cmmib10 \skewchar\tencmmib='177
\font\sevencmmib=cmmib7 \skewchar\sevencmmib='177
\font\fivecmmib=cmmib5 \skewchar\fivecmmib='177
\newfam\cmmibfam
\textfont\cmmibfam=\tencmmib \scriptfont\cmmibfam=\sevencmmib
\scriptscriptfont\cmmibfam=\fivecmmib

\numberwithin{equation}{section}
\def\neg{\negthinspace}
\begin{document}

\begin{titlepage}
\begin{flushright}
hep-th/yymm.nnnn
\end{flushright}
\vskip 1.5in
\begin{center}
{\bf\Large{Janus Configurations, Chern-Simons Couplings, }\vskip0cm
\bf\Large{And The $\theta$-Angle  in ${\cal N}=4$ Super Yang-Mills
Theory} }\vskip 0.5cm {Davide Gaiotto and Edward Witten} \vskip
0.05in {\small{ \textit{School of Natural Sciences, Institute for
Advanced Study}\vskip 0cm {\textit{Einstein Drive, Princeton, NJ
08540 USA}}}}

\end{center}
\vskip 0.5in

\baselineskip 16pt
\date{April, 2008}

\begin{abstract}
We generalize the half-BPS Janus configuration of four-dimensional
${\cal N}=4$ super Yang-Mills theory to allow the theta-angle, as
well as the gauge coupling, to vary with position.  We show that
the existence of this generalization is closely related to the
existence of novel three-dimensional Chern-Simons theories with
${\cal N}=4$ supersymmetry.  Another closely related problem,
which we also elucidate, is the D3-NS5 system in the presence of a
four-dimensional theta-angle.
\end{abstract}
\end{titlepage}
\vfill\eject

\tableofcontents

\section{Introduction}

\def\N{{\cal N}}
\def\frak{\mathfrak}

In this paper, we will consider, in four-dimensional $\N=4$ super
Yang-Mills theory, several questions  that involve
$\theta$-dependence, different choices of unbroken supersymmetry
algebra, and relations to three-dimensional supersymmetric theories
with Chern-Simons interactions.

We start by reconsidering the Janus solution \cite{BGH,CFKS}. In
$\N=4$ super Yang-Mills theory, the Janus solution corresponds to a
situation in which the coupling constant depends non-trivially on
one of the spatial coordinates, which we will call $y$.  The
original Janus solution  preserved the full $R$-symmetry group of
the theory and violated all supersymmetry. Later, however, variants
were found that preserve some supersymmetry and only part of the
$R$-symmetry \cite{CK}, and in fact it is possible to preserve
one-half of the full supersymmetry \cite{DEG2}.
  The unbroken supersymmetry algebra is then $OSp(4|4)$, which is a
``half-BPS'' subalgebra of the full $PSU(4|4)$ symmetry algebra of
$\N=4$ super Yang-Mills theory. This is the case that we focus on.

Originally,  Janus solutions were found in supergravity, but
 counterparts, which we will call Janus
configurations, also exist \cite{CFKS,DEG} in weakly coupled field
theory. However, the known field theory constructions are not as
general as what has been found in supergravity. The known field
theory constructions are limited to the case that only the gauge
coupling $g$, and not the theta-angle $\theta$, depends
non-trivially on $y$.

One goal of this paper is to generalize the Janus solution to the
case that both $\theta$ and $g$ depend on $y$.  This is accomplished
in section \ref{computation}.  As we explain there, a key input is
the fact that the relevant unbroken supersymmetry algebra has
inequivalent embeddings in $PSU(4|4)$. To make $\theta$ become
$y$-dependent, one must use a $y$-dependent embedding of the
superalgebra.

The problem of making $\theta$ to be $y$-dependent is related to the
problem of Chern-Simons couplings in three-dimensional gauge theory.
Let us consider a four-dimensional gauge theory with $\theta$ a
function of $y$, which is one of the four coordinates. The relevant
part of the action is
\begin{equation}\label{zrypt}I_\theta=-\frac{1}{32\pi^2}
\int \mathrm{d}^4x\, \theta(y)\,
\epsilon^{\mu\nu\alpha\beta}\,\Tr\,F_{\mu\nu}F_{\alpha\beta}.\end{equation}
We write $\mathrm{d}^4x=\mathrm{d}^3x\, \mathrm{d}y$. After
integrating by parts and dropping any surface terms, $I_\theta$ is
equivalent to
\begin{equation}\label{zypt}I_\theta=\frac{1}{8\pi^2}\int \mathrm{d}^3x
\,dy\,
\frac{\mathrm{d}\theta}{\mathrm{d}y}\epsilon^{\mu\nu\lambda}\Tr\,\left(A_\mu\partial_\nu
A_\lambda+\frac{2}{3}A_\mu A_\nu A_\lambda\right).  \end{equation}
This interaction is similar to a three-dimensional Chern-Simons
interaction, so supersymmetrizing a four-dimensional theory with a
$y$-dependent $\theta$ angle is somewhat similar to
supersymmetrizing a three-dimensional theory with a Chern-Simons
interaction.

We therefore re-examine the problem of supersymmetrizing the
three-dimensional Chern-Simons coupling.  Quite a few results are
already known.  $\N=3$ supersymmetry (in the three-dimensional
sense) allows one to add a Chern-Simons coupling to a general
three-dimensional gauge theory that also has the conventional
$F^2$ kinetic energy \cite{Kao:1992ig,KLL,KS,Gaiotto:2007qi}.  It
has been argued \cite{Schwarz} that there are additional
possibilities if one omits the usual kinetic energy, and recently
a Chern-Simons theory with $\N=8$ supersymmetry (and no $F^2$
term) was constructed \cite{BL}.  This construction was very
special: the gauge group is $SO(4)$, and the matter representation
is uniquely determined.

In our problem, the unbroken supersymmetry corresponds to $\N=4$ in
the three-dimensional sense, so in section \ref{chernsimons}, we
consider three-dimensional Chern-Simons theories with this amount of
supersymmetry, and no $F^2$ coupling.  Our approach is to assume
$\N=1$ supersymmetry, which admits a convenient superspace
description, and then restrict the couplings so that a global
$SO(4)$ symmetry appears, promoting $\N=1$ to $\N=4$. Moreover, we
take the superpotential to be quartic so that, just as in
\cite{Schwarz,BL}, the theories we construct are conformally
invariant at the classical level, and presumably also quantum
mechanically.

We are able to completely classify theories of this kind, in terms
of supergroups.  The gauge group is the bosonic part of a
supergroup, and the matter representation is determined by the
fermionic part of that supergroup. Leaving aside theories with
abelian gauge symmetry  or associated with certain exceptional
supergroups, the main examples correspond to the supergroups
$U(N|M)$ (or their cousins $SU(N|M)$ and $PSU(N|N)$) and
$OSp(N|M)$. The gauge groups are $U(N)\times U(M)$ and $O(N)\times
Sp(M)$, and the matter fields are in the bifundamental
representations.

The same groups and representations arise in the theory of
D3-branes interacting with NS5-branes.  This fact suggests that it
would be fruitful to combine the following three problems: Janus
configurations, D3-branes ending on fivebranes, and
three-dimensional Chern-Simons couplings.  This is our goal in the
rest of the paper.  In section \ref{janagain}, we repeat the
analysis of section \ref{computation} using three-dimensional
$\N=1$ superfields. In contrast to section \ref{computation}, in
which we start with the full $R$-symmetry and constrain the
couplings to get supersymmetry, here we start with $\N=1$
supersymmetry (in the three-dimensional sense)  and constrain the
couplings to get the full $R$-symmetry.  The two approaches lead
to the same Lagrangians with the same supersymmetry.

In section \ref{boundary}, we apply this method to the D3-NS5
system.  We find a close parallel with the purely
three-dimensional results of section \ref{chernsimons}, and this
enables us to resolve a riddle.  This system is usually considered
at $\theta=0$, and its appropriate description for $\theta\not=0$
does not seem to be known. The answer is given by a special case
of our construction. Equivalently, we can use our method to
describe at $\theta=0$ a system consisting of a D3-brane and a
$(1,q)$ five-brane (a combination of an NS-fivebrane and $q$
D-fivebranes).  The low energy description of this system has also
not been understood in the literature.  Closing the circle, we
show that the Janus configuration can be recovered from a
knowledge of the D3-NS5 system with general couplings.

\section{Janus Configuration With Spatially Varying Theta Angle }\label{computation}

\subsection{Preliminaries}\label{preliminaries}

${\mathcal N}=4$ super Yang-Mills theory is conveniently obtained
by dimensional reduction from ten dimensions \cite{Brink:1976bc}.
We begin in $\Bbb{R}^{1,9}$, with metric $g_{IJ}$, $I,J=0,\dots,9$
of signature $-++\dots +$. Gamma matrices $\Gamma_I$ obey
$\{\Gamma_I,\Gamma_J\}=2g_{IJ}$, and the supersymmetry generator
is a Majorana-Weyl spinor $\varepsilon$, obeying
$\bar\Gamma\varepsilon=\varepsilon$, where
$\bar\Gamma=\Gamma_0\Gamma_1\cdots\Gamma_9$.  The fields are a
gauge field $A_I$ and Majorana-Weyl fermion $\Psi$, also obeying
$\bar\Gamma\Psi=\Psi$. Thus, $\varepsilon$ and $\Psi$ both
transform in the $\16$ of $SO(1,9)$.  The supersymmetric action is
\begin{equation}\label{tsupac}I=\frac{1}{e^2}\int \mathrm{d}^{10}x
\,\Tr\,\left(\frac{1}{2}F_{IJ}F^{IJ}-i\bar\Psi
\Gamma^ID_I\Psi\right).\end{equation} The conserved supercurrent is
\begin{equation}\label{tupac}J^I =
\frac{1}{2}\Tr\,\Gamma^{JK}F_{JK}\Gamma^I\Psi,\end{equation} and the
supersymmetry transformations are
\begin{align}\label{tzupac}\delta A_I  &= i\bar\varepsilon\Gamma_I\Psi\\
\delta\Psi &=\frac{1}{2} \Gamma^{IJ}F_{IJ}\varepsilon.
\end{align}

We reduce to four dimensions by simply declaring that the fields are
allowed to depend only on the first four coordinates
$x^0,\dots,x^3$. This breaks the ten-dimensional Lorentz group
$SO(1,9)$ to $SO(1,3)\times SO(6)_R$, where $SO(1,3)$ is the
four-dimensional Lorentz group and $SO(6)_R$ is a group of
$R$-symmetries. Actually, the fermions transform as spinors of
$SO(6)_R$, and the $R$-symmetry group of the full theory is really
$Spin(6)_R$, which is the same as  $SU(4)_R$.  The ten-dimensional
gauge field splits as a four-dimensional gauge field $A_\mu$,
$\mu=0,\dots,3$, and six scalars fields $A_{3+i}$, $i=1,\dots,6$
that we rename as $\Phi_i$.  They transform in the fundamental
representation of $SO(6)_R$ . The supersymmetries $\varepsilon$ and
fermions $\Psi$ transform under $SO(1,3)\times SO(6)_R$ as
$(\2,\1,\4)\oplus (\1,\2,\bar\4)$, where $(\2,\1)$ and $(\1,\2)$ are
the two complex conjugate spinor representations of $SO(1,3)$ and
$\4,\,\bar\4$ are the two complex conjugate spinor representations
of $SO(6)_R$.

In a Janus configuration, the coupling parameters depend
non-trivially on one of the four spacetime coordinates, which we
take to be $y=x^3$. To preserve half of the supersymmetry, it is
necessary to break the $R$-symmetry from $SO(6)$ to $SO(3)\times
SO(3)$.  A special case of a Janus configuration is one in which the
couplings jump discontinuously at, say $y=0$.  Such a configuration
is invariant under those conformal transformations that preserve the
plane $y=0$.  The group of such conformal transformations is the
three-dimensional conformal supergroup $SO(2,3)$, whose double cover
is (the split real form) $Sp(4,\R)$. The corresponding supergroup is
$OSp(4|4)$, whose bosonic part is $SO(4)\times Sp(4)$.  The second
factor is the conformal supergroup and the first factor is the
$R$-symmetry group ($SO(4)$ is a double cover of $SO(3)\times
SO(3)$).  The spatial variation of couplings in a conformally
invariant Janus configuration reduces $PSU(4|2,2)$ to $OSp(4|4)$.  A
more general Janus configuration reduces $PSU(4|2,2)$ to the
subalgebra of $OSp(4|4)$ consisting of symmetries that preserve a
metric in spacetime. This subalgebra is usually called the
three-dimensional global supersymmetry algebra with $\N=4$
supersymmetry (8 supercharges) and $R$-symmetry group $SO(4)$.

\subsubsection{Outer Automorphism}\label{outer}

A key feature of this problem is that there is a one-parameter
family of inequivalent embeddings of $OSp(4|4)$ in $PSU(4|2,2)$. The
reason for this is that $PSU(4|2,2)$ has a one-parameter group of
outer automorphisms.  Represent an element $M$ of  $PSU(4|2,2)$ by a
supermatrix
\begin{equation}\label{gorf}M=\begin{pmatrix} S&T\\
U&V\end{pmatrix}\,\end{equation} where $S$ and $V$ are bosonic
$4\times 4$ blocks and $U$ and $T$ are fermionic ones (in
$PSU(4|4)$, $M$ has superdeterminant 1 and is identified with
$\lambda M$ for any scalar $\lambda$). Then $PSU(4|4)$ has a group
$F\cong U(1)$ of outer automorphisms, acting by $M\to VMV^{-1}$ with
\begin{equation}V=\begin{pmatrix}e^{i\beta}& 0 \\ 0 &
1\end{pmatrix},~{\beta\in\R}.\end{equation}  Conjugation by $U(1)$
generates a one-parameter family of embeddings of $OSp(4|4)$ in
$PSU(4|4)$.

\def\4{{\bf 4}}
Concretely, the fermionic generators of $PSU(4|2,2)$ transform under
the bosonic subgroup $SU(4)\times SU(2,2)$ as $\4\otimes
\bar\4'\oplus \bar\4\otimes \4'$, where $\4$ and $\4'$ are the
four-dimensional representations of $SU(4)$ and $SU(2,2)$,
respectively.  Once we reduce $SU(4)$ and $SU(2,2)$ to $SO(4)$ and
$Sp(4,\R)$, the representations $\4$ and $\4'$ become real. So, as a
representation of $SO(4)\times Sp(4,\R)$, the fermionic generators
of $PSU(4|2,2)$ consist of two copies of the real representation
$\4\otimes \4'$.  These two copies are rotated by the outer
automorphism group $F\cong SO(2)$.  If we pick any linear
combination of the two copies of $\4\otimes \4'$, then this,
together with the Lie algebra of $SO(4)\times Sp(4,\R)$, gives an
$OSp(4|4)$ subalgebra of $PSU(4|4)$.

Though we have described this in the conformally-invariant case,
conformal invariance is not essential.  All statements remain valid
if we replace $Sp(4,\R)$ by its subgroup that preserves a metric;
this is simply the three-dimensional Poincar\'e group. In its action
on the fermions, the Poincar\'e group reduces to the
three-dimensional Lorentz group $SO(1,2)$.

 Conformally-invariant Janus
configurations are particularly interesting, but they are not
generic.  We will not impose conformal invariance in the following
analysis.

\subsubsection{Notation}\label{notation}

It is convenient to  split the scalars $\Phi_i,\,i=1,\dots,6$ into
two groups acted on respectively by the two factors of
$SO(3)\times SO(3)\subset SO(6)_R$.   We take these two groups to
consist of the first three and last three\footnote{In
ten-dimensional notation, $\vec X$ is related to $x^4,x^5,x^6$ and
$\vec Y$ to $x^7,x^8,x^9$.} $\Phi$'s; we rename
$(\Phi_1,\Phi_2,\Phi_3)$ as $\vec X=(X_1,X_2,X_3)$ and
$(\Phi_4,\Phi_5,\Phi_6)$ as $\vec Y=(Y_1,Y_2,Y_3)$. We sometimes
write $SO(3)_X$ and $SO(3)_Y$ for the two $SO(3)$ groups.

Though the $\16$ of $SO(1,9)$, in which the supersymmetries
transform, is irreducible, it is  reducible as a representation of
$W=SO(1,2)\times SO(3)_X\times SO(3)_Y$. Indeed, the action of $W$
commutes with the three operators
\begin{align}\notag B_0 &=\Gamma_{456789}\\
             \notag B_1&=\Gamma_{3456}\\
             B_2&=\Gamma_{3789}.\end{align}
They obey $B_0^2=-1$, $B_1^2=B_2^2=1$, and $B_0B_1=-B_1B_0=B_2$,
etc., and  generate an action of $SL(2,\R)$.  We can decompose the
$\16$ of $SO(1,9)$ as $V_8\otimes V_2$, where $V_8$ transforms in
the real irreducible representation $(\2,\2,\2)$ of $SO(1,2)\times
SO(3)_X\times SO(3)_Y$, and $V_2$ is a two-dimensional space in
which the $B_i$ are represented by
\begin{align}\notag\label{dorf}B_0&=\begin{pmatrix}0 &1\\ -1
&0\end{pmatrix}\\
\notag B_1&=\begin{pmatrix}0 &1\\ 1
&0\end{pmatrix}\\
B_2&=\begin{pmatrix}1 &0\\
0&-1\end{pmatrix}.\end{align}

A $W$-invariant embedding of the eight supercharges of
three-dimensional $\N=4$ supersymmetry into the four-dimensional
supersymmetry algebra can be obtained by putting a constraint on the
supersymmetry generators
\begin{equation}\label{seful}\left(\sin\psi B_1+\cos\psi
B_2\right)\varepsilon=\varepsilon,\end{equation} for some $\psi$.
The possible choices are rotated by the one-parameter group
generated by $B_0$. This is the outer automorphism group $F$. If
$\varepsilon$ and $\tilde\varepsilon$ obey (\ref{seful}) (with the
same value of $\psi$), then
\begin{equation}\label{useful}\bar\varepsilon\Gamma_3\tilde\varepsilon=0.
\end{equation}
The physical meaning of this is that, as a Janus configuration is
not invariant under translations of $y=x^3$, the anticommutator of
two fermionic symmetries of such a configuration never generates a
translation in the $y$ direction.

\subsection{Construction}\label{construction}

$\N=4$ super Yang-Mills theory has been generalized in \cite{DEG} to
allow a $y$-dependent coupling constant while preserving half the
supersymmetry.  We will extend this to include a varying $\theta$
angle.

We begin with the unperturbed ${\cal N}=4$ action
\begin{equation}\label{supac}I=\int \mathrm{d}^{4}x\frac{1}{e^2}
\,\Tr\,\left(\frac{1}{2}F_{IJ}F^{IJ}-i\bar\Psi
\Gamma^ID_I\Psi\right)\end{equation} and supersymmetry
transformations
\begin{align}\label{zupac}\delta A_I  &= i\bar\varepsilon\Gamma_I\Psi\\
\delta\Psi &=\frac{1}{2} \Gamma^{IJ}F_{IJ}\varepsilon.
\end{align}

We will perturb both the action and the supersymmetry
transformations to be $y$-dependent, while preserving half of the
supersymmetry.  The generators $\varepsilon$ of the unbroken
supersymmetries will themselves also be $y$-dependent.  (This fact
is perhaps the main novelty in our analysis here.)  However, the
$y$-dependence of $\varepsilon$ is rather special.
 The commutators of two
unbroken supersymmetries will be, of course, a translation in the
directions $x^0,x^1,x^2$, with $y$-independent coefficients (since
$y$-dependent translations of the other coordinates do not give
symmetries). This is tantamount to the condition
\begin{equation}\label{steful}\frac{\mathrm{d}}{\mathrm{d}y}\bar\varepsilon\Gamma^\mu\varepsilon=0,~~
\mu=0,1,2.\end{equation} For this to hold, the $y$-dependence of
$\varepsilon$ must be generated by the outer automorphism group $F$.
This result will emerge below from our explicit calculation (see
eqn. (\ref{ordog})).

Now we describe the corrections to the supersymmetry transformations
and to the action. Dimensional analysis permits us
 to add a correction to the
supersymmetry transformation of $\Psi$:
\begin{equation}\label{cortrans}\tilde \delta\Psi=
\frac{1}{2}\left(\Gamma\cdot X\left(s_1\Gamma_{456}
+s_2\Gamma_{789}\right) +\Gamma\cdot
Y\left(t_1\Gamma_{456}+t_2\Gamma_{789}\right)\right)\varepsilon,\end{equation}
where $\Gamma\cdot X$ and $\Gamma\cdot Y$ are abbreviations,
respectively, for $\sum_a\Gamma_a X^a$ and $\sum_p\Gamma_pY^p$. Here
$s_1,s_2,t_1$ and $t_2$ will be functions of $y=x^3$, along with
other parameters that appear momentarily.

To the action, we can add fermion bilinear terms:
\begin{equation}\label{acone}I'=\int \mathrm{d}^4x
\frac{i}{e^2}\Tr\,\bar\Psi\left(\alpha\Gamma_{012}+
\beta\Gamma_{456}+\gamma\Gamma_{789}\right)\Psi.
\end{equation}
This is the most general fermion bilinear that is gauge-invariant
and has $SO(1,2)\times SO(3)\times SO(3)$ symmetry. It is also
possible to add the following dimension 3 bosonic terms to the
action\footnote{The $\epsilon$ symbols that appear here are
antisymmetric tensors in the $012$, $456$, and $789$ subspaces,
normalized to $\epsilon^{012}=\epsilon^{456}=\epsilon^{789}=1$.}:
\begin{equation}\label{bcone} I''=\int \mathrm{d}^4x
\frac{1}{e^2}\,\left(u\epsilon^{\mu\nu\lambda}\Tr\,\left(A_\mu\partial_\nu
A_\lambda+\frac{2}{3}A_\mu A_\nu A_\lambda\right)+
\frac{v}{3}\epsilon^{abc}
\Tr\,X_a[X_b,X_c]+\frac{w}{3}\epsilon^{pqr}\Tr\,Y_p[Y_q,Y_r]\right).
\end{equation}
Finally, the action can have terms of dimension 2:
\begin{equation}\label{hormox}I'''=\int
d^4x\Tr\left(\frac{r}{2e^2}X_aX^a+\frac{\tilde
r}{2e^2}Y_pY^p\right).\end{equation}

It is convenient to define
\begin{equation}\label{qtype}q=e^2\frac{\mathrm{d}}{\mathrm{d}y}\frac{1}{e^2}.\end{equation}
We consider $q$  and similarly $\mathrm{d}\varepsilon/\mathrm{d}y$
and the parameters $\alpha,\beta,\gamma$, $s_i,t_j$, and $u,v,w$ to
be of first order, while $r$ and $\tilde r$, the second derivatives
of $e^2$ and $\varepsilon$, and the first derivatives of the other
parameters are second order. (Of course, homogeneous quadratic
expressions in first order quantities are second order also.)  We
already know that the zeroth order variation of $I$ vanishes, since
the pure ${\cal N}=4$ theory in four dimensions is supersymmetric.
We need to examine the supersymmetry of the first and second order
quantities.

\subsection{First Order Variations}\label{firstorder}

When we act with the unperturbed supersymmetry variation $\delta$ on
the unperturbed action $I$, the zeroth order terms vanish, as just
noted, but we do get first order terms involving the $y$ derivatives
of $e^2$ and $\varepsilon$:
\begin{align}\label{hope}\delta I|_1&=-i\,\Tr\int \mathrm{d}^4x
\frac{1}{e^2}\frac{\mathrm{d}\bar\varepsilon}{\mathrm{d}y} \Gamma^{KL}F_{KL}\Gamma_3\Psi\\
             &~~-\frac{i}{2}\,\Tr\int \mathrm{d}^4x \frac{q}{e^2}\bar\varepsilon
             \Gamma_{3}\Gamma^{KL}F_{KL}\Psi.\end{align}

First order variations come from several other places.  The
unperturbed supersymmetry variation $\delta$ acting on the perturbed
action $I'$ gives
\begin{align}\label{zope}\delta I'=-i\,\Tr\int \mathrm{d}^4x
\frac{1}{e^2}\bar\varepsilon\Gamma^{IJ}F_{IJ}\left(\alpha\Gamma_{012}+
\beta\Gamma_{456}+\gamma\Gamma_{789}\right)\Psi.
\end{align}
The perturbed supersymmetry variation $\tilde\delta$ acting on the
unperturbed action $I$ gives a first order contribution
\begin{align}\tilde\delta I|_1=i\,\Tr\int
d^4x\frac{1}{e^2}\bar\varepsilon\left((s_1\Gamma_{456}+s_2\Gamma_{789})\Gamma^{Ia}
D_IX_a+
(t_1\Gamma_{456}+t_2\Gamma_{789})\Gamma^{Ip}D_IY_p)\right)\Psi.\end{align}
 The remaining first order terms come from $\delta I''$:
\begin{equation}\label{helpdoc}\delta I''=i\Tr\,\int
d^4x\frac{1}{e^2}\left(u\,\epsilon^{\mu\nu\lambda}\bar\varepsilon\Gamma_\mu\Psi
F_{\nu\lambda}+v\epsilon^{abc}\bar\varepsilon\Gamma_a\Psi
[X_b,X_c]+w\epsilon^{pqr}\bar\varepsilon\Gamma_p\Psi[Y_q,Y_r]\right).
\end{equation}

Now let us give some samples of the use of the above formulas. First
we consider variations  proportional to $D_\mu X_a\Psi$, contracted
with $\bar\varepsilon$ and some gamma matrices.  The sum of such
contributions comes out to be
\begin{equation}\label{holsome}i\int
\mathrm{d}^4x\frac{1}{e^2}\left(-2\frac{\mathrm{d}\bar\varepsilon}{\mathrm{d}y}
-q\bar\varepsilon+2\bar\varepsilon(\alpha\Gamma_{0123}-\beta\Gamma_{3456}+
\gamma\Gamma_{3789})-\bar\varepsilon\left(s_1\Gamma_{3456}+s_2
\Gamma_{3789}\right)\right)\Tr\,D_\mu X_a\Gamma_{\mu a
3}\Psi.\end{equation} The condition for vanishing of terms of this
form therefore gives
\begin{equation}\label{firstone}
-2\frac{\mathrm{d}\bar\varepsilon}{\mathrm{d}y}
-q\bar\varepsilon+2\bar\varepsilon(\alpha\Gamma_{0123}-\beta\Gamma_{3456}+
\gamma\Gamma_{3789})-\bar\varepsilon\left(s_1\Gamma_{3456}+s_2
\Gamma_{3789}\right)=0.\end{equation}

A similar analysis of terms proportional to $D_3X_a \Psi$ gives a
very similar equation with a couple of signs reversed:
\begin{equation}\label{secondone}
2\frac{\mathrm{d}\bar\varepsilon}{\mathrm{d}y}
-q\bar\varepsilon+2\bar\varepsilon(-\alpha\Gamma_{0123}-\beta\Gamma_{3456}+
\gamma\Gamma_{3789})-\bar\varepsilon\left(s_1\Gamma_{3456}+s_2
\Gamma_{3789}\right)=0.\end{equation} By subtracting the last two
equations, we get an equation that determines the $y$-dependence of
$\bar \varepsilon$:
\begin{equation}\label{ordog}\frac{\mathrm{d}\bar\varepsilon}{\mathrm{d}y}=
\alpha\bar\varepsilon \Gamma_{0123}.\end{equation} (This shows that
$\varepsilon$ varies with $y$ by an element of the outer
automorphism group $F$, for reasons explained at the beginning of
section \ref{construction}). We will also need the transpose
\begin{equation}\label{ordogt}\frac{\mathrm{d}\varepsilon}{\mathrm{d}y}=
\alpha \Gamma_{0123}\varepsilon.\end{equation} Notice that as
expected \begin{equation} \frac{\mathrm{d}}{\mathrm{d}y} \bar
\varepsilon \Gamma^3 \tilde \varepsilon = \alpha \bar \varepsilon
\Gamma_{0123} \Gamma^3 \varepsilon+\alpha \bar \varepsilon \Gamma^3
\Gamma_{0123} \varepsilon=0 \end{equation} and \begin{equation}
\frac{\mathrm{d}}{\mathrm{d}y} \bar \varepsilon \Gamma^{\mu} \tilde
\varepsilon = \alpha \bar \varepsilon \Gamma_{0123} \Gamma^{\mu}
\varepsilon+\alpha \bar \varepsilon \Gamma^{\mu} \Gamma_{0123}
\varepsilon=0 \end{equation}

The sum of the two equations gives
\begin{equation}\label{nodog}0=\bar\varepsilon\left((s_1+2\beta)
\Gamma_{3456}+(s_2-2\gamma)\Gamma_{3789}+q\right).
\end{equation}

It is convenient to regard this as an equation that uniquely
determines $s_1$ and $s_2$  in terms of $\bar\varepsilon$, $\beta,$
$\gamma$, and $q$.  Indeed, if written out explicitly, eqn.
(\ref{nodog}) is equivalent to a pair of linear equations that have
a unique solution for the unknowns $s_1$, $s_2$. Explicitly solving
for $s_1$ and $s_2$ makes the formulas more complicated, and for now
it is more convenient to simply leave the equation for $s_1$ and
$s_2$ in the given form.

Upon exchanging $X$ and $Y$, and considering terms proportional to
$D_\mu Y_p\Psi$ or  $D_3Y_p\Psi$, we get two more similar equations.
One linear combination gives (\ref{secondone}) again, and the second
gives the counterpart of (\ref{nodog}):
\begin{equation}\label{modog}0=\bar\varepsilon\left((t_1-2\beta)\Gamma_{3456}+
(t_2+2\gamma)\Gamma_{3789}+q\right).
\end{equation}

A similar analysis of terms proportional to $[X_a,Y_p]\Psi$ gives
the following condition:
\begin{equation}\label{modo}-2\frac{\mathrm{d}\bar\varepsilon}{\mathrm{d}y}-q\bar\varepsilon
-2\bar\varepsilon(\alpha\Gamma_{0123}+\beta\Gamma_{3456}+\gamma\Gamma_{3789})
-\bar\varepsilon((s_1+t_1)\Gamma_{3456}+(s_2+t_2)
\Gamma_{3789})=0.\end{equation} With the aid of the above formulas,
this reduces to
\begin{equation}\label{odo}0=\bar\varepsilon\left(4\alpha\Gamma_{0123}
+2\beta\Gamma_{3456}+2\gamma\Gamma_{3789}-q\right).
\end{equation}
This equation determines $\beta$ and $\gamma$ in terms of
$\varepsilon$, $\alpha$, and $q$, and then eqns. (\ref{nodog}) and
(\ref{modog}) similarly determine $s_1,s_2,t_1$ and $t_2$ in terms
of the same variables.

The analysis of the remaining first order terms in the variation of
the action is similar.  The terms proportional to $D_3X^a\Psi$
vanish with the aid of the above formulas. The vanishing of terms
proportional to $F_{\mu\nu}\Psi$, $[X_a,X_b]\Psi$, and
$[Y_p,Y_q]\Psi$ serves, respectively, to determine the coefficients
$u,v,$ and $w$ in eqn (\ref{bcone}).

Let us  work out the terms $F_{\mu\nu}\Psi$.  In doing so, for
brevity we omit the usual factors $i\int \mathrm{d}^4x
\frac{1}{e^2}\,\Tr$, leaving the integration and the trace
understood.  From $\delta I|_1$, we get
\begin{equation}\label{firstcon}-\bar\varepsilon\alpha
\Gamma_{0123}\Gamma^{\mu\nu}F_{\mu\nu}\Gamma_3\Psi-\frac{1}{2}\bar\varepsilon
q\Gamma_3\Gamma^{\mu\nu}F_{\mu\nu}\Psi.\end{equation} And from
$\delta I'$, we get
\begin{equation}\label{secondcon}-\bar\varepsilon\Gamma^{\mu\nu}F_{\mu\nu}
\left(\alpha\Gamma_{012}+\beta\Gamma_{456}+\gamma\Gamma_{789}\right)\Psi.
\end{equation}
There is no  contribution from $\tilde\delta I|_1$. These
contributions add to
\begin{equation}\label{theirsm}-\bar\varepsilon
\left(2\alpha\Gamma_{0123}-\beta\Gamma_{3456}-\gamma\Gamma_{3789}
+\frac{q}{2}\right)
\Gamma^{\mu\nu}F_{\mu\nu}\Gamma_3\Psi.\end{equation} With the aid of
eqn. (\ref{odo}), this collapses to
\begin{equation}\label{eirsm}-4\alpha\bar\varepsilon\Gamma_{0123}
\Gamma^{\mu\nu}F_{\mu\nu}\Gamma_3\Psi=
4\alpha\bar\varepsilon\epsilon^{\mu\nu\lambda}F_{\mu\nu}\Gamma_\lambda.
\end{equation}
Comparing to (\ref{helpdoc}), we see that the remaining
supersymmetry variation $\delta I''$ will cancel this term precisely
if
\begin{equation}\label{irsm}u=-4\alpha.\end{equation}

A very similar analysis of the terms $[X_a,X_b]\Psi$ and
$[Y_p,Y_q]\Psi$ shows that these contributions to the supersymmetry
variation similarly cancel if
\begin{equation}\label{rsm}v=-4\beta,~~w=-4\gamma.\end{equation}

To summarize what we have obtained so far, we may begin with two
arbitrary functions $\alpha(y)$ and $q(y)$ and an arbitrary initial
value of $\bar \varepsilon(y)$ at, say, $y=y_0$.  The $y$-dependence
of $\bar \varepsilon$ is then determined from eqn. (\ref{ordog}),
and the other equations determine everything else in terms of
$\alpha$, $q$, and $\varepsilon$.   So far $\alpha(y)$ and $q(y)$
are arbitrary, but it turns out that vanishing of the second order
variations places a non-trivial restriction on these functions.

\subsection{Second Order Variations}\label{secondorder}

There are three sources of second order variations.

The supersymmetry variation of $I'''$, the part of the action that
is of dimension 2, is easily computed:
\begin{equation}\label{quicklee}\delta I'''=i\int \mathrm{d}^4x
\frac{1}{e^2}\Tr\,\bar\varepsilon\left(r\Gamma \cdot X+\tilde
r\Gamma\cdot Y\right)\Psi.\end{equation} It is of second order
simply because we consider $r$ and $\tilde r$ to be second order
quantities.

The modified supersymmetry variation $\tilde\delta$ acting on the
correction $I'$ to the action, is again not difficult to compute:
\begin{equation}\label{lessquick}\tilde\delta I'=-{i}\int
\mathrm{d}^4x\frac{1}{e^2}\Tr\,\bar\varepsilon\biggl(\bigl((s_1\Gamma_{456}+s_2\Gamma_{789})\Gamma\cdot
X+(t_1\Gamma_{456}+t_2\Gamma_{789}\Gamma\cdot
Y)\bigr)(\alpha\Gamma_{012}+\beta\Gamma_{456}+\gamma\Gamma_{789})\biggr)\Psi.\end{equation}
This is equivalent to
\begin{equation}\label{lessquick0}\tilde\delta I'=-{i}\int
\mathrm{d}^4x\frac{1}{e^2}\Tr\,\bar\varepsilon\biggl(\bigl((s_1\Gamma_{3456}+s_2\Gamma_{3789})\Gamma\cdot
X+(t_1\Gamma_{3456}+t_2\Gamma_{3789}\Gamma\cdot
Y)\bigr)(-\alpha\Gamma_{0123}+\beta\Gamma_{3456}+\gamma\Gamma_{3789})\biggr)\Psi.\end{equation}
This can be further simplified using eqns. (\ref{nodog}) and
(\ref{modog}).  The terms involving $X$ become
\begin{equation}\label{turkey}\tilde\delta I'_X=-i\int
\mathrm{d}^4x\frac{1}{e^2}\Tr\,\bar\varepsilon\biggl(\bigl(
2\beta^2+2\gamma^2+
(2\gamma\alpha+q\beta)\Gamma_{3456}+(2\beta\alpha-q\gamma)\Gamma_{3789}+q\alpha\Gamma_{0123}
\bigr)\Gamma\cdot X\Psi\biggr).\end{equation} The terms involving
$Y$ can be analyzed similarly, but one can also take a short cut
using symmetry, as we explain below.

The remaining second order terms are
\begin{align}\label{openquick}\notag\tilde \delta I|_2&=-i\int
\mathrm{d}^4x\Tr\,\biggl(\bigl(\frac{q}{2e^2}+\frac{1}{e^2}\frac{\mathrm{d}}{\mathrm{d}y}\bigr)
\biggl(\bar\varepsilon(s_1\Gamma_{3456}+s_2\Gamma_{3789} )\biggr)
\biggr)\Gamma\cdot X\Psi\\ & -i\int
\mathrm{d}^4x\Tr\,\biggl(\bigl(\frac{q}{2e^2}+\frac{1}{e^2}\frac{\mathrm{d}}{\mathrm{d}y}\bigr)
\biggl(\bar\varepsilon(t_1\Gamma_{3456}+t_2\Gamma_{3789} )\biggr)
\biggr)\Gamma\cdot Y\Psi.\end{align} This can again be simplified
using (\ref{nodog}) and (\ref{modog}). The terms containing $X$
become
\begin{align}\label{nquick}\notag\tilde\delta &I|_{2,X}=-i\int \mathrm{d}^4x\Tr\,
\biggl(\bigl(\frac{q}{2e^2}+\frac{1}{e^2}\frac{\mathrm{d}}{\mathrm{d}y}\bigr)\bigl(\bar\varepsilon
(-2\beta\Gamma_{3456}+2\gamma\Gamma_{3789}-q \bigr)\bigr)
\biggr)\Gamma\cdot X\Psi \\ = & -i\int
\mathrm{d}^4x\frac{1}{e^2}\Tr\,\bar\varepsilon\biggl((-2\beta'-\beta
q+2\gamma\alpha)\Gamma_{3456}+(2\gamma'+\gamma
q+2\beta\alpha)\Gamma_{3789}-q\alpha\Gamma_{0123}-\frac{q^2}{2}-q'\biggr)
\Gamma\cdot X\Psi
\end{align}

The sum of $\tilde\delta I'_X$ and $\tilde\delta I|_{2,X}$ is
\begin{equation}\label{penquick}-i\,\Tr\int \mathrm{d}^4x
\frac{1}{e^2}\bar\varepsilon\biggl(\bigl(-2\beta'+
4\gamma\alpha\bigr)\Gamma_{3456}+\bigl(2\gamma'
+4\beta\alpha\bigr)\Gamma_{3789}+\bigl(2\beta^2+2\gamma^2-\frac{q^2}
{2}-q') \biggr) \Gamma\cdot X\Psi.\end{equation} The terms
proportional to $\bar\varepsilon\Gamma_{0123}$ have canceled, but
the terms involving $\bar\varepsilon\Gamma_{3456}$ and
$\bar\varepsilon\Gamma_{3789}$ have not canceled.  As a result, it
is not in general possible to cancel (\ref{penquick}) with an
additional contribution of the form (\ref{quicklee}).  This is
possible if and only if $\bar\varepsilon$ is an eigenvector of the
matrix appearing in (\ref{penquick}).  We need
\begin{equation}\label{enquick}\bar\varepsilon\biggl(\bigl(-2\beta'
+4\gamma\alpha\bigr)\Gamma_{3456}+\bigl(2\gamma'
+4\beta\alpha\bigr)\Gamma_{3789}\biggr)=\bar\varepsilon
\lambda,\end{equation} where $\lambda$ is a multiple of the
identity.  This condition is equivalent to the expected one
(\ref{seful}), with $\psi$ now determined in terms of
$\alpha,\beta,\gamma$.

Now let $\varepsilon$ and $\tilde\varepsilon$ be any two generators
of the unbroken supersymmetry.  Then
$\bar\varepsilon\Gamma_3\tilde\varepsilon=0$, according to eqn.
(\ref{useful}). So contracting (\ref{enquick}) with
$\Gamma_3\tilde\varepsilon$ and expressing the result in terms of
$B_1=\Gamma_{3456}$, $B_2=\Gamma_{3789}$, we get
\begin{align}\label{lonely}0=\bar\varepsilon\Gamma_3
\biggl(\bigl(-\beta'+2\gamma\alpha\bigr)B_1+\bigl(\gamma'+2\beta\alpha\bigr)
B_2\biggr)\tilde\varepsilon.\end{align} Let us decompose that in
three terms
\begin{equation} 0=\bar\varepsilon\Gamma_3
\biggl(-\beta'B_1+\gamma'B_2\biggr)\tilde\varepsilon+\bar\varepsilon
\Gamma_{0123}\Gamma_3 \biggl(\gamma B_2 -\beta B_1 \biggr)\alpha
\tilde\varepsilon+\bar\varepsilon\Gamma_3 \biggl(\gamma B_2 -\beta
B_1 \biggr)\alpha \Gamma_{0123} \tilde\varepsilon.\end{equation}
This recombines to
\begin{equation}
\frac{\mathrm{d}}{\mathrm{d}y} \left( \bar \varepsilon \Gamma_3 (
\beta B_1 - \gamma B_2) \tilde\varepsilon\right) =0.
\end{equation}
So we can integrate to give
\begin{equation}\label{inte}
 \bar \varepsilon \Gamma_3 (\beta B_1 - \gamma B_2)
\tilde\varepsilon = C,
\end{equation}
with a constant $C$.

It is convenient to write the 16-dimensional space of positive
chirality spinors (in which $\varepsilon$ takes values) as
$V_8\otimes V_2$, where $V_8$ is an eight-dimensional space acted on
by a double cover of $W=SO(1,2)\times SO(3)_X\times SO(3)_Y$, and
$V_2$ is a two-dimensional space in which act the matrices
$B_0,B_1,$ and $B_2$ of eqn. (\ref{dorf}).  We take
$\varepsilon=v\otimes \varepsilon_0$, $\tilde\varepsilon =\tilde
v\otimes \tilde\varepsilon_0$, with $v,\tilde v\in V_8$,
$\varepsilon_0, \tilde\varepsilon_0\in V_2$. The quadratic form
$(\epsilon,\tilde\epsilon)=\bar\epsilon\Gamma_3\tilde\epsilon$ is
symmetric in $\epsilon,\tilde\epsilon$.  It can be decomposed as the
tensor product of an antisymmetric inner product in $V_8$ and an
antisymmetric inner product in $V_2$.  To write the inner product in
$V_2$, we write $\varepsilon_0$ as a column vector
\begin{equation}\label{hexplic}\varepsilon_0=\begin{pmatrix} a\\
b \end{pmatrix},\end{equation} and define $\bar\varepsilon_0$ as a
row vector:
\begin{equation}\label{xplic}\bar\varepsilon_0=\begin{pmatrix}-b &
a\end{pmatrix}.\end{equation} Then the antisymmetric inner product
in $V_2$ can be defined by
\begin{equation}\label{plic}\langle\varepsilon_0,\tilde\varepsilon_0\rangle=
\bar\varepsilon_0\tilde\varepsilon_0.\end{equation}

Equation (\ref{ordogt}) shows that the $y$ evolution of
$\varepsilon_0$ is just an $SO(2)$ rotation.  Let us work in a basis
in which $B_0,B_1,$ and $B_2$ act as in eqn. (\ref{dorf}), and
normalize $\varepsilon_0$ at some value of $y$ so that
\begin{equation}\label{nexplic}\varepsilon_0=\begin{pmatrix} \cos\,{\psi}/{2}\\
\sin{\psi}/{2} \end{pmatrix},\end{equation} for some $\psi$. Then
(\ref{ordogt}) implies this form is valid for all $y$ and moreover
\begin{equation}\psi'=2 \alpha.\end{equation}

We set $\tilde\varepsilon_0=\varepsilon_0$, and plug the expression
(\ref{nexplic})  into (\ref{inte}), with the result
\begin{equation}\label{intex}
\beta \cos \psi  + \gamma \sin \psi   = C
\end{equation}
Now we can eliminate $\beta$ and $\gamma$ from equations
(\ref{odo}), by contracting with $\Gamma_3\tilde\epsilon$ for a
conveniently chosen $\tilde\epsilon$.  If we take
$\tilde\epsilon=\tilde v\otimes \tilde\varepsilon_0$ with
\begin{equation}\label{explic}\tilde\varepsilon_0=
\begin{pmatrix} \cos\,{3 \psi}/{2}\\
-\sin\,{3\psi}/{2} \end{pmatrix},\end{equation} then (\ref{odo})
reduces simply to
\begin{align}\label{sorx}\notag
0&=2 \psi' \cos 2 \psi + 2 \beta \cos \psi + 2 \gamma \sin \psi - q
\sin 2 \psi\\ &=2 \psi' \cos 2 \psi  - q \sin 2 \psi+2C.
\end{align}

\subsubsection{Exchange of $\vec X$ and $\vec Y$}\label{exchange}

We can repeat this analysis with $\vec X$ and $\vec Y$ exchanged,
but it is  more illuminating to observe that the problem has a
symmetry that exchanges $\vec X$ and $\vec Y$.  As a transformation
of the underlying ten-dimensional spacetime, the relevant symmetry
acts by $x^{3+i}\leftrightarrow x^{6+i}$, $i=1,2,3$, together with a
reflection of one of the coordinates $x^0,x^1,x^2$ (so as to
preserve the overall orientation).  On the above variables, the
transformation exchanges $\beta$ with $\gamma$ and $B_1$ with $B_2$.
It also changes the sign of $\alpha$ and maps $\psi$ to
$\pi/2-\psi$. (This is implied by the relation $\alpha=d\psi/dy$ and
the fact that the symmetry exchanges eigenvectors of $B_1$ with
eigenvectors of $B_2$.) The formulas we obtain are symmetric in
$\vec X$ and $\vec Y$, even though this is not manifest in the
derivation. For example, the symmetry is present in (\ref{sorx}).

\subsection{Interpreting And Solving The
Equations}\label{interpreting}

According to (\ref{bcone}), the supersymmetric Lagrangian has a
three-dimensional Chern-Simons interaction, integrated in four
dimensions:
\begin{equation}\label{trypt}\int \mathrm{d}^4x
\frac{1}{e^2}\,u\epsilon^{\mu\nu\lambda}\Tr\,\left(A_\mu\partial_\nu
A_\lambda+\frac{2}{ 3}A_\mu A_\nu A_\lambda\right).\end{equation}

Let us compare this to the $\theta$-term of four-dimensional super
Yang-Mills theory. This usually takes the form
\begin{equation}\label{rypt}I_\theta=-\frac{1}{32\pi^2}
\int \mathrm{d}^4x \theta
\epsilon^{\mu\nu\alpha\beta}\,\Tr\,F_{\mu\nu}F_{\alpha\beta}.\end{equation}
Usually, one  assumes $\theta$ to be a constant and then the
integral is a topological invariant.  However, we wish to assume
that $\theta$ is a function of $y=x^3$. Then, after integration by
parts, we can write
\begin{equation}\label{ypt}I_\theta=\frac{1}{8\pi^2}\int \mathrm{d}^3x dy
\frac{\mathrm{d}\theta}{\mathrm{d}y}\epsilon^{\mu\nu\lambda}\Tr\,\left(A_\mu\partial_\nu
A_\lambda+\frac{2}{3}A_\mu A_\nu A_\lambda\right).  \end{equation}
We see that we can interpret the combination $u/e^2$ as
$\theta\,'/8\pi^2$.

On the other hand, in eqn. (\ref{irsm}), we concluded that
$u=-4\alpha$.  So we have a more direct interpretation of $\alpha$:
\begin{equation}\label{utp}\alpha=-\frac{e^2\theta\,'}{32\pi^2}.\end{equation}

The other key equation governing the $y$-dependence of the couplings
is (\ref{sorx}):
\begin{equation}\label{wefty}-2\psi' \cos 2\psi +q\sin
2\psi = C. \end{equation} Let us look for a domain wall solution
in which the $y$ coordinate extends over the whole real line and
the coupling parameters $e^2$ and $\theta$ are both constant for
$y\to\pm \infty$.  Then $\psi'$ and $q=-d\ln e^2/dy$ must vanish
for $y\to\pm\infty$. This being so, the integration constant $C$
must also vanish. So the equation reduces to
\begin{equation}\label{wufty}0=-2\psi' \cos 2\psi +q\sin
2\psi.\end{equation}

 Recalling that $\alpha=\psi'/2$,
$q=e^2d(1/e^2)/dy$, we can rewrite eqns. (\ref{utp}) and
(\ref{wufty}) in the form:
\begin{align}\notag\label{mufto}
\frac{\mathrm{d}\psi}{\mathrm{d}y}+\frac{e^2}{16\pi^2}\frac{\mathrm{d}\theta}{\mathrm{d}y}&=0\\
      -2\frac{\mathrm{d}\psi}{\mathrm{d}y}\cos 2\psi +e^2\sin 2\psi
      \frac{\mathrm{d}}{\mathrm{d}y}\frac{1}{e^2}& =0.\end{align}
These equations have the remarkable property of being invariant
under reparametrization of $y$.

Perhaps more to the point, we can solve them. (\ref{wufty}) is
equivalent to
\begin{equation}\label{ufty}\frac{\mathrm{d}}{\mathrm{d}y}\left(-\ln\sin
2\psi+\ln(1/e^2)\right)=0,\end{equation} so it says that
\begin{equation}\label{fty}\frac{1}{e^2}=D\sin 2\psi,\end{equation}
with some constant $D$.  Then we have
$\mathrm{d}\theta/\mathrm{d}y=-(16\pi^2/e^2)d\psi/dy=-16\pi^2 D \sin
2\psi d\psi/dy=d(8\pi^2 D\cos 2\psi)/dy$.  So we get
\begin{equation}\label{tify}\theta=2\pi a+8\pi^2 D\cos
2\psi,\end{equation} with another integration constant $a$.

The results for $\theta$ and $1/e^2$ are conveniently expressed in
terms of the usual $\tau$ parameter
\begin{equation}\label{rufty}\tau=\frac{\theta}{2\pi}+\frac{2\pi
i}{e^2},\end{equation} which takes values in the upper half plane.
We have
\begin{equation}\label{mufty}\tau=a+ 4\pi D(\cos 2\psi +i\sin
2\psi).\end{equation} Thus, $\tau$ takes values in a circle of
radius $4\pi D$, centered at the point $\tau=a$ on the real $\tau$
axis. (Just half of this circle is in the upper half plane.)
Curves of this type are precisely the geodesics on the upper
half-plane, with its standard $SL(2,\Bbb{R})$-invariant
metric.\footnote{A quick way to show this is to observe that the
line ${\rm Re}\,\tau=0$ is certainly a geodesic, since it is the
fixed line of the isometry $\tau\to-\bar\tau$. Every geodesic is
the image of this one under an $SL(2,\Bbb{R})$ transformation. On
the other hand, an $SL(2,\Bbb{R})$ transformation maps the line
${\rm Re}\,\tau=0$ to a semi-circle in the upper half-plane.} This
unexpected appearance of $SL(2,\Bbb{R})$ symmetry means that our
results are compatible with what is found in supergravity
\cite{BGH,CFKS,DEG2}, where $SL(2,\Bbb{R})$ symmetry is manifest.

Now we can classify half-BPS domain walls of this type.  We pick any
two points $\tau_-$ and $\tau_+$ in the upper half-plane and look
for a domain wall with the property that $\tau(y)\to \tau_\pm $ for
$y\to\pm \infty$. Any two points $\tau_+$ and $\tau_-$ in the upper
half plane are connected by a unique geodesic $L$, and the
trajectory $\tau(y)$ must lie on $L$ for all $y$.  We gain
absolutely no information about the function $\tau(y)$ except that
its image lies on $L$ and that the limits for $y\to \pm \infty$ are
$\tau_\pm$. In particular, since the equations are invariant under
reparametrization of $y$, there is no restriction on how the path
from $\tau_-$ to $\tau_+$ should be parametrized.

\subsubsection{Solving For The Remaining Variables}\label{solving}

From now on, we will keep $C=0$. We return to eqn. (\ref{odo}), and
contract with $\Gamma_3\varepsilon_0$.  The result is
\begin{equation} \psi' +  \beta \cos \psi -  \gamma \sin \psi=0.
\end{equation} Combining this with the $C=\beta \cos \psi +
\gamma \sin \psi=0$ we finally get $\beta, \gamma$:
\begin{align}\label{expsolve}\notag\beta & =- \frac{\psi'}{2 \cos \psi}
\\  \gamma& = \frac{\psi'}{2 \sin \psi} .\end{align}

From the ansatz (\ref{nexplic}) for $\varepsilon_0$ and the explicit
form of the matrices $B_1$ and $B_2$, we get
\begin{equation}\label{sefult}\bar \varepsilon \Gamma^3 \left(\sin\psi B_1+\cos\psi
B_2 +1\right)=0.\end{equation} By acting on (\ref{nodog}) with
$\Gamma^3$ and comparing to the last equation, we learn that
\begin{equation}s_1+2 \beta = -q \sin \psi = -2 \psi' \cos \psi +
\frac{\psi'}{\cos\psi}\end{equation} and
\begin{equation}s_2-2 \gamma = -q \cos \psi = + 2 \psi' \sin \psi
- \frac{\psi'}{\sin\psi}\end{equation} Hence
\begin{equation}\label{gome}s_1 = 2 \psi' \frac{\sin^2 \psi}{\cos
\psi},~~ s_2 = 2 \psi' \sin \psi.\end{equation} Similarly,
\begin{equation}\label{zome}t_1 = -2 \psi' \cos \psi, ~~t_2 = -2
\psi'\, \frac{\cos^2 \psi}{\sin \psi}.\end{equation}

The eigenvalue $\lambda$ in eqn. (\ref{enquick}) turns out to be
$\lambda = (d/dy)\left({\psi'}/{\sin \psi \cos \psi}\right).$
Finally, we can solve for $r$
\begin{equation}\label{zum} r = \lambda + 2\beta^2 + 2 \gamma^2 -
\frac{q^2}{2} - q' = 2\left( \psi' \tan \psi \right)' + 2 (\psi')^2
\end{equation} and by symmetry \begin{equation}\label{zzum}
 \tilde r = -2\left( \psi' \cot \psi \right)' + 2 (\psi')^2
\end{equation}

\subsubsection{Conformally Invariant Limit}\label{conformally}

Now (generalizing section 6 of \cite{DEG}) we would like to ask
whether, classically, it is possible to take a limit in which the
Janus configuration becomes conformally invariant.   As we have
presented it so far, this configuration involves an arbitrary
parametrization $\tau(y)$ of an arc in the upper half plane.  To
achieve conformal invariance (which acts by rescaling of $y$),
$\tau(y)$ should simply have a discontinuity, say $\tau(y)=\tau_-$
for $y<0$ and $\tau(y)=\tau_+$ for $y>0$.  When this is the case,
$q=(d/dy)\ln(1/e^2)$ and $\theta'$ have delta function
singularities; terms in the action linear in $q$ or $\theta'$ give
contributions to the action supported at the interface. After
integration by parts, the same is so for terms linear in  $q'$ or
$\theta''$. But contributions proportional to $q^2$ or
$(\theta')^2$ are divergent in the conformally invariant limit.
Our above formulas contain such terms, in view of the formulas for
$r$ and $\tilde r$.

In the absence of a varying $\theta$ angle, this problem can be
avoided \cite{DEG} by a position-dependent rescaling of scalar
fields. The same is possible in our case. After integration by
parts of the $\psi''$ term, the $r X^2$ part of the action becomes
\begin{equation}\label{melb}
\frac{1}{2 e^2}\Tr\, \left(- 4 \psi' \tan \psi \,X^a X'_a + 2
(\psi')^2 \tan^2 \psi \,X^a X_a \right). \end{equation} This
combines with the $(\partial_yX)^2$ term to a perfect square
\begin{equation}\label{elb} \frac{1}{e^2}\Tr\,(X' - \psi' \tan \psi X)^2
.\end{equation} Similarly for the $Y^2$ terms, one gets
\begin{equation} \label{kelm}\frac{1}{e^2}\Tr(Y' + \psi' \cot \psi
Y)^2.
\end{equation}

If we define new scalar fields $\tilde X = X \cos \psi$ and $\tilde
Y = Y \sin \psi$, the action simplifies. The terms just written
become simply
\begin{equation}\label{gelb}\frac{1}{e^2}\Tr\,\left(\frac{(d\tilde
X/dy)^2}{\cos^2\psi}
 +\frac{(d\tilde
Y/dy)^2}{\sin^2\psi}\right).\end{equation} Both $(\psi')^2$ and
$\psi''$ disappear from the action, which becomes linear in $\psi'$.
Hence the action has a well-defined limit to a localized
discontinuity in $\tau$.

Furthermore,  $s_1/s_2=t_1/t_2=\tan \psi$. This means that the
combination of gamma matrices which appears in the extra term
(\ref{cortrans}) in the supersymmetry transformation is proportional
to $B_1 \sin \psi + B_2 \cos \psi$, which leaves $\varepsilon$
invariant.  So the correction to the supersymmetry transformation is
\begin{equation}
\tilde \delta \Psi = -\Gamma^3 \Gamma \cdot X \psi' \tan \psi\,
\varepsilon + \Gamma^3 \Gamma \cdot Y \psi' \cot \psi\, \varepsilon.
\end{equation}
We can combine this with the similar term in the unperturbed
supersymmetry variation $\delta\Psi\sim \Gamma_3\Gamma\cdot (d
X/dy)\varepsilon+\Gamma_3 \Gamma\cdot (d Y/dy)\varepsilon+\dots$
to
\begin{equation}
\delta' \Psi = \Gamma^3\left(\frac{ \Gamma \cdot \tilde
X'}{\cos\psi}  + \frac{ \Gamma \cdot \tilde Y'}{\sin\psi}\right)
\varepsilon.
\end{equation}
The structure that we have just found will be more apparent from a
different viewpoint explained in section \ref{janagain}.

\section{$3d$ Superfield Method}

\subsection{Overview}\label{overview}

Our computation in section \ref{computation} was based on assuming
the relevant $R$-symmetry and adjusting the couplings to achieve
supersymmetry.  Here, we will follow a different approach, using
superfields to make manifest ${\cal N}=1$ supersymmetry and then
adjusting the couplings to achieve $R$-symmetry -- which then
implies the full ${\cal N}=4$ supersymmetry.

But instead of merely repeating the problem studied in section
\ref{computation} with a different approach, we will here study
several closely related problems.  So first we give an overview of
the contents of this section.

\subsubsection{A Three-Dimensional Problem}\label{threedim}

It is simplest to start with a purely three-dimensional problem.
From a three-dimensional point of view,  the generalized Janus
configuration proposed in the previous section contains a
Chern-Simons interaction as in eqn. (\ref{trypt}). Of course, this
configuration also has ${\cal N}=4$ supersymmetry in the
three-dimensional sense (eight supercharges, not counting
superconformal symmetries).  This suggests that our subject is
related to the problem of three-dimensional Chern-Simons
interactions with ${\cal N}=4$ supersymmetry, and that will turn out
to be the case.

As we recalled in the introduction, in three-dimensional
nonabelian gauge theory with a Chern-Simons term, it is difficult
to get past ${\cal N}=3$ supersymmetry if a conventional $F^2$
kinetic energy is present.\footnote{The standard argument for this
(for example, see \cite{Kao:1992ig}) is based on the structure of
the supermultiplets. In $2+1$ dimensions, the rotation group
$SO(2)$ is abelian, and the spin of a particle is an integer or
half-integer, either positive or negative.  In the presence of
both an $F^2$ term and a Chern-Simons interaction, the gauge
fields become massive \cite{DJT}, say with spin 1.  The ${\cal
N}=3$ supersymmetry algebra has three spin lowering operators, and
one can construct a supersymmetric theory of gauge fields,
scalars, and fermions in which the gauge field is contained in a
supermultiplet of states with spins $1,1/2,0,-1/2$. For ${\cal
N}=4$, one would need also states of spin $-1$, which would also
have to arise from gauge fields. Being in the same supermultiplet,
the spin 1 and spin $-1$ gauge fields would  have to transform the
same way under the gauge group, but this is not possible for gauge
fields in nonabelian gauge theory (gauge fields transform as
precisely one copy of the adjoint representation). In $U(1)\times
U(1)$ gauge theory, it is possible \cite{KS} to make an ${\cal
N}=4$ theory with $F^2$ and Chern-Simons interactions; one $U(1)$
gauge boson has spin 1 and the other has spin $-1$. This is
possible because, as $U(1)\times U(1)$ is abelian, the two gauge
bosons both transform trivially under the gauge group.} However,
it has been argued \cite{Schwarz} that one can achieve more
supersymmetry in the absence of the $F^2$ term, and an example
with ${\cal N}=8$ has been constructed \cite{BL}.

In section \ref{chernsimons}, making no {\it a priori} assumptions
about the appropriate gauge group or matter representations, we
describe a general ${\cal N}=4$ superconformal theory of this type.
Our method is to assume ${\cal N}=1$ superconformal symmetry, and
adjust the couplings to find an $SO(4)$ $R$-symmetry that ensures
that the model actually has ${\cal N}=4$ superconformal invariance.
We achieve a nice classification of models of this type.  They
correspond to supergroups in which the fermionic generators
transform in a  pseudoreal or symplectic representation of the
bosonic symmetries.  Apart from examples with abelian gauge group,
the main examples involve the classical supergroups $U(N|M)$ (and
its cousins $SU(N|M)$ and $PSU(N|N)$) and $OSp(N|M)$.

\subsubsection{Intersecting Branes}\label{intersecting}

\begin{figure}
  \begin{center}
    \includegraphics[width=3.5in]{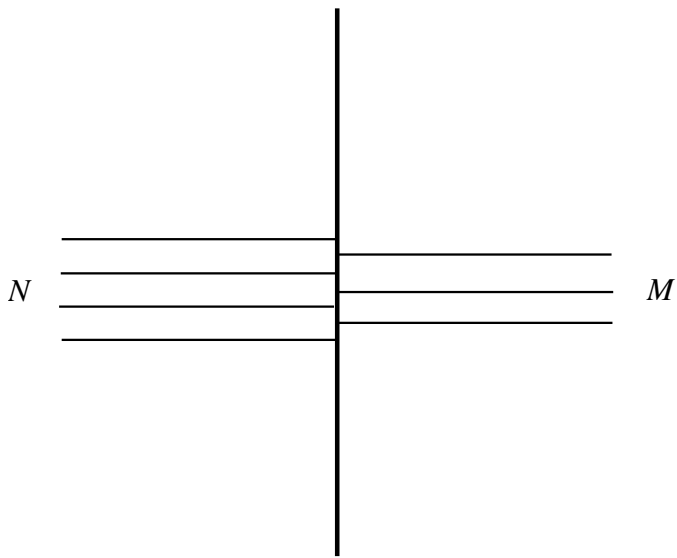}
  \end{center}
\caption{\small A configuration with $N$ D3-branes ending on an
NS5-brane from the left, while $M$ D3-branes end from the right. The
D3-brane worldvolumes span the 0123 directions, and those of the
NS5-branes span the 012456 directions. The horizontal direction in
the figure represents spacetime direction $x^3$, and the vertical
direction represents spacetime directions 456.}
  \label{pic1}
\end{figure}

These examples are related in an interesting way to a certain
familiar configuration in string theory.  Consider a system of
parallel D3-branes ending on an NS5-brane from left and right -- say
$N$ from the left and $M$ from the right, as in fig. \ref{pic1}. We
suppose that the D3-brane world-volumes are parametrized by
$x^0,x^1,x^2,x^3$, while the NS5-brane world-volume is parametrized
by $x^0,x^1,x^2$ and $x^4,x^5,x^6$. As usual, we set $y=x^3$. The
physics of this configuration is well-known.  For $y<0$, we have
$U(N)$ gauge theory; for $y>0$, we have $U(M)$ gauge theory. At
$y=0$, there are bifundamental hypermultiplets, transforming in the
representation $(N,\bar M)\oplus (\bar N,M)$ of $U(N)\times U(M)$.

This configuration is half-BPS, that is, it preserves eight
supercharges (enhanced to 16 in the infrared limit, where the gauge
theory becomes superconformal).  It remains half-BPS if one turns on
a ten-dimensional string theory axion field (the supersymmetric
partner of the dilaton), inducing a four-dimensional theta-angle.
However, it seems that in the literature, the low energy field
theory representing the configuration of fig. \ref{pic1} in the
presence of a theta-angle is not known.  As we will explain,
constructing this field theory is very similar to constructing the
supersymmetric Chern-Simons actions of section \ref{chernsimons}.

A theta-angle in four-dimensional gauge theory on a half-space is
equivalent, classically, to a Chern-Simons interaction on the
boundary, via a simple integration by parts:
\begin{equation}\label{zorypt}-\frac{\theta}{32\pi^2}
\int_{\eusm M_+} \mathrm{d}^4x\,
\epsilon^{\mu\nu\alpha\beta}\,\Tr\,F_{\mu\nu}F_{\alpha\beta}=
\frac{\theta}{8\pi^2}\int_{\partial \eusm M_+} \mathrm{d}^3x
\epsilon^{\mu\nu\lambda}\Tr\,\left(A_\mu\partial_\nu
A_\lambda+\frac{2}{3}A_\mu A_\nu A_\lambda\right).  \end{equation}
($\eusm M_+$ is the region $y>0$, and $\partial \eusm M_+$ is its
boundary.) So the brane configuration of fig. \ref{pic1} leads to
$U(N)\times U(M)$ gauge theory coupled to three-dimensional
bifundamental hypermultiplets, with  three-dimensional Chern-Simons
couplings for the $U(N)$ and $U(M)$ gauge fields. The Chern-Simons
couplings have opposite signs, coming from integration by parts from
left or right.

This is exactly the structure of one of the main examples of section
\ref{chernsimons}, the one related to the supergroup $U(N|M)$.  The
only difference is that the $U(N)$ and $U(M)$ gauge fields live in
four-dimensional half-spaces, while in section \ref{chernsimons}
they were purely three-dimensional.

The other main example from section \ref{chernsimons} is the
supergroup $OSp(N|M)$, which corresponds to gauge group $O(N)\times
Sp(M)$, again with bifundamental matter.  This example also arises
naturally from a brane construction.  In fact, one simply has to
modify the brane construction of fig. \ref{pic1} by including an
O3-plane, parallel to the D3-branes.  The gauge group of the
D3-branes is then orthogonal or symplectic, and jumps from one type
to the other in crossing the NS5-brane.   This gives $O(N)\times
Sp(M)$ gauge theory (with bifundamental hypermultiplets at the brane
intersection), as in the $OSp(N|M)$ example from section
\ref{chernsimons}.

The fact that the brane configurations give the same gauge groups
and matter representations as the Chern-Simons theories is too much
to be a coincidence, so it is reasonable to think that the relevant
supersymmetric field theories can be constructed in the same way.
Demonstrating this will be one of our goals.

\subsubsection{Back To Janus}\label{janusback}

But what does all this have to do with the generalized Janus
configurations of section 2?

The brane configuration of fig. \ref{pic1} has different
``branches'' of supersymmetric vacua.  If one displaces the
D3-branes incident from left and right in the $x^4,x^5,x^6$
directions (as is actually sketched in the figure), then the
bifundamental hypermultiplet fields $H$  become massive.  However,
it is also possible to give expectation values to the fields $H$.
This corresponds to displacing some of the D3-branes normal to the
NS5-brane by moving them in the directions $x^7,x^8,x^9$ -- or by
moving the NS5-brane in those directions. If $N=M$, the case we will
now focus on, then it is possible to detach the NS5-brane from all
D3-branes. The D3-branes that formerly ended from left or right on
the NS5-brane instead reconnect to each other, and at low energies
one is left with ${\cal N}=4$ super Yang-Mills theory with gauge
group $U(N)$.

The relation to Janus comes in because it is possible to modify
this process slightly.  Our field theory analysis of the
configuration of fig. \ref{pic1} will show that it is possible
while preserving supersymmetry  for the $U(M)$ theory on the right
of the NS5-brane and the $U(N)$ theory on the left to have
different four-dimensional gauge couplings. Moreover, it is
possible  to pick any embedding of three-dimensional ${\cal N}=4$
supersymmetry in the four-dimensional supersymmetry algebra, that
is, any value of the angle $\psi$ in eqn. (\ref{seful}), and
(roughly) any four-dimensional theta-angle $\theta$, to the left
of the NS5-brane. (The values of $\psi$ and $\theta$ on the right
are then uniquely determined.)

What happens, then, if we set $N=M$ and give an expectation value
to the hypermultiplet fields? We reduce to a four-dimensional
$U(N)$ gauge theory, but now with a coupling constant that
``jumps'' in crossing the hyperplane $y=0$.  There also are angles
$\psi$ or $\theta$ to the left of this hyperplane, which jump in
crossing the hyperplane in a way that turns out to be consistent
with eqn. (\ref{mufty}). In short, we reduce to precisely the
generalized Janus domain wall described in section
\ref{conformally}.

\begin{figure}
  \begin{center}
    \includegraphics[width=3.5in]{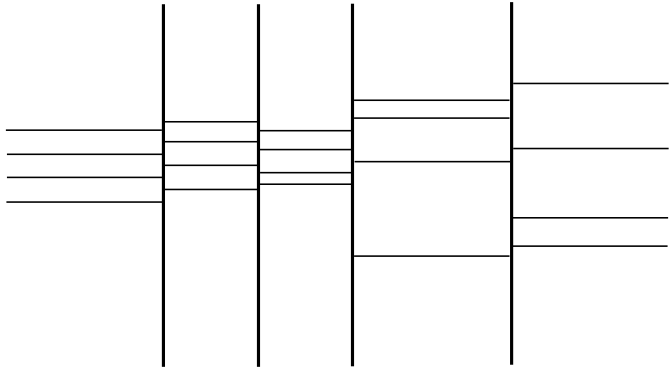}
  \end{center}
\caption{\small A system of $N$ parallel D3-branes intersecting
successive NS5-branes.  }
  \label{pic2}
\end{figure}

We can go farther in this direction.  We consider (fig. \ref{pic2})
a system of $N$ parallel D3-branes that intersect $k$ successive
NS5-branes.  From a field theory point of view, in each slab between
two NS5-branes (or each half-space to the left or right of all
branes) there is a four-dimensional $U(N)$ gauge theory.  At each
interface between two slabs (or half-spaces), there are
bifundamental hypermultiplets.  A supersymmetric configuration can
be constructed with any values of the angles $\psi$ and $\theta$ to
the left of all the NS5-branes, and any independently chosen
four-dimensional gauge couplings in each of the various segments.
(The values of $\psi$ and $\theta$ in the other regions are then
uniquely determined, via eqn. (\ref{mufty}).) After giving generic
expectation values to the hypermultiplet fields, we reduce to a
four-dimensional theory of $U(N)$ gauge couplings in the presence of
$k$ successive generalized Janus domain walls.

A generalized Janus configuration with an arbitrary $y$-dependence
of the gauge coupling, as described in \cite{DEG} at $\theta=0$ and
in section \ref{interpreting} in general, can be obtained as a limit
of this.  We take a suitable limit in which $k$ becomes large, the
NS5-branes are closely spaced, and the individual jumps in the gauge
coupling are small.

\subsubsection{Janus and Fivebranes}\label{janfive}
It is useful to make explicit the condition for a Janus
configuration to preserve the same supersymmetry as a defect or
boundary which is the field theory limit of a $(p,q)$ fivebrane.
The original system of D3-branes preserves $16$ out of $32$
supercharges of the type II string theory. Writing $\varepsilon_1$
and $\varepsilon_2$ for the supersymmetries of left-moving and
right-moving string modes, the supersymmetries left unbroken by
the D3-branes are characterized by
\begin{equation}\label{condone}
\varepsilon_2 = \Gamma_{0123} \varepsilon_1.\end{equation} A
$(p,q)$ (anti)fivebrane extended along the $012456$ directions
imposes a further constraint which depends on the appropriate
central charge $p \tau + q$.  In terms of $t=\arg(p\tau+q)$, the
condition is
\begin{equation}\label{condtwo} \varepsilon_1 =
-\Gamma_{012456}\left(\sin t\, \varepsilon_1 + \cos t\,
\varepsilon_2\right).\end{equation} In the presence of both types
of branes, we have \begin{equation}\label{bothtypes}\varepsilon_1
= \left(- B_1 \cos t + B_2 \sin t\right) \varepsilon_1.
\end{equation} Comparing this to the constraint (\ref{seful}) imposed by the
Janus configuration, we discover that they are compatible if $t =
\frac{\pi}{2} + \psi$. On the other hand, the Janus configuration
also prescribes that
\begin{equation}\label{ondtwo} \tau = a + 4 \pi D e^{2 i \psi}
\end{equation} with real constants $a,D$.
This condition defines a semicircle in the upper half plane, which
intersects the real axis at two points $a\pm 4\pi D$.

The condition $\arg (p \tau + q) = \frac{\pi}{2} + \psi$ is
equivalent to
\begin{equation}\label{ppp}\frac{\sin(\pi/2+\psi)}{\cos(\pi/2+\psi)}=\frac{{\rm
Im}\,(p\tau+q)}{{\rm Re}\,(p\tau+q)},\end{equation} or
\begin{equation}\label{kolf}-\frac{\cos\psi}{\sin\psi}=\frac{4\pi
D\,\sin 2\psi}{a+q/p+4\pi D\cos 2\psi}.\end{equation} This
condition is actually independent of $\psi$, and equivalent to
\begin{equation}\label{zox} a=-q/p-4\pi D.\end{equation}
So the rightmost intersection of the semicircle with the real axis
must be at $\tau=-q/p$.   Provided this condition is obeyed,
supersymmetry is preserved when a $(p,q)$-fivebrane is added to a
Janus configuration, regardless of the value of $\psi$ at the
location of the fivebrane.

We can repeat the same exercise for a $(p',q')$ fivebrane extended
along the $012789$ directions. (The symmetry exchanging directions
$456$ with $789$ is discussed in section \ref{exchange}.) The
requirement is then $t=\psi$ and the  leftmost intersection of the
semicircle with the real axis must be at $-{q'}/{p'}$.

Hence a Janus configuration in which $a$ and $D$ are rational
numbers can be combined with a $(p,q)$ fivebrane in a
supersymmetric fashion, for two different values of $p$ and $q$.
One compatible fivebrane runs in the 012456 directions, and has
$a+4\pi D=-q/p$.  The other one runs in the 012789 directions,
with $a-4\pi D=-q'/p'$.

The specific case of D5-branes, which corresponds to $p=0$,
requires a special treatment. In this case, $t$ is 0 or $\pi$
(depending on the sign $q$), and (\ref{bothtypes}) becomes
$\varepsilon_1=\pm \Gamma_{3456}\varepsilon_1$. This is the
supersymmetry of the original half-BPS Janus configuration
\cite{DEG}, in which the coupling constant varies but the
Yang-Mills theta-angle does not. This is the case that the
trajectory is a vertical line in the upper half plane,
corresponding to a limit of the semicircle in which $a,D\to\infty$
with $a-4\pi D$ (or $a+4\pi D$) fixed.

\subsubsection{Organization Of This Section}\label{organiz}

In section \ref{chernsimons}, we construct purely three-dimensional
theories with Chern-Simons couplings and ${\cal N}=4$ supersymmetry.
The technique that we will use is the one that we will follow
throughout this section: we use three-dimensional ${\cal N}=1$
superspace to construct a Chern-Simons theory with ${\cal N}=1$
supersymmetry, and then constrain the couplings so that an extra
global symmetry appears, promoting ${\cal N}=1$ to ${\cal N}=4$.

In section \ref{language}, as a prelude to some of the other
questions described above, we reformulate four-dimensional ${\cal
N}=4$ super Yang-Mills in terms of three-dimensional ${\cal N}=1$
superfields.  In section \ref{janagain}, we use this method to
recover the generalized Janus configuration of section
\ref{construction}. This involves a computation that is arguably
simpler than the one of section \ref{construction}.  In this
formulation, it is evident that the Janus configuration has a
conformally invariant limit, as found with greater effort in
section \ref{conformally}.   In section \ref{boundary}, we use
three-dimensional ${\cal N}=1$ superfields to analyze the low
energy field theories associated with the brane configurations of
figs. \ref{pic1} and \ref{pic2}, justifying some claims that were
made above.


\subsection{${\cal N}=4$ Chern-Simons Theory}\label{chernsimons}

For any gauge group $G$, and any hypermultiplet representation of
$G$, there is a unique classical theory with ${\cal N}=3$
Chern-Simons couplings.  Morally, we want to show that if the matter
content and gauge group are picked carefully, the resulting
classical theory will actually possess ${\cal N}=4$ supersymmetry,
or better $OSp(4|4)$  superconformal symmetry. It is known that the
${\cal N}=3$ theory is superconformal quantum mechanically as well,
and we expect the same to be true of the ${\cal N}=4$ theory.

In practice, because ${\cal N}=3$ superfields are not convenient,
we find it useful to start from an ${\cal N}=1$ Chern-Simons
theory coupled to matter and look for an enhancement to ${\cal
N}=4$. The hallmark of ${\cal N}=4$ is an $SO(4)$ $R$-symmetry
group under which the four supercharges transform in the vector
representation. The subgroup of $SO(4)$ that leaves fixed one of
the supercharges is therefore $SO(3)$.  So we start with an ${\cal
N}=1$ theory with an $SO(3)$ global symmetry.  We try to adjust
the couplings so that the $SO(3)$ is enhanced to an $SO(4)$ that
does {\it not} commute with ${\cal N}=1$ supersymmetry.  Instead,
the $SO(4)$ together with the ${\cal N}=1$ supersymmetry generate
a full ${\cal N}=4$ structure. In fact, the theory is really
conformal at the classical level, so the ${\cal N}=1$ theory we
start with  really has symmetry $OSp(1|4)$, and the enhancement is
to $OSp(4|4)$.

The mechanism for symmetry enhancement will be the following.  The
matter fields will be ${\cal N}=1$ superfields ${\cal
Q}^I_A=Q^I_A+i\theta^\alpha \lambda^I_{\alpha\,A}+\dots$, where
$\theta^\alpha$ are superspace coordinates, $Q$ and $\lambda$ are
bosonic and fermionic fields, and  the index $A=1,2$ transforms in
the two-dimensional representation of a group that we will call
$SU(2)_d$, the double cover of the global symmetry group $SO(3)$
mentioned in the last paragraph. The theory will have Yukawa
couplings that are schematically of the form $Q^2\lambda^2$.
Generically, these couplings are not invariant under separate
$SU(2)$ symmetries acting on only $Q$ or only $\lambda$.  Our
strategy is to adjust the superspace couplings so that such a
$SU(2)\times SU(2)$ symmetry appears; in view of the relation of
$SU(2)\times SU(2)$ to $SO(4)$, this will enhance ${\cal N}=1$
supersymmetry to ${\cal N}=4$.  (This $SU(2)\times SU(2)$ is a
cover of the group $SO(3)_X\times SO(3)_Y$ of section
\ref{computation}.)

As explained in the last paragraph, we will here make the assumption
that one factor of $SU(2)\times SU(2)$ acts only on $Q$, and the
other factor acts only on $\lambda$. This ansatz is too restrictive
to include the recently proposed Lagrangian \cite{BL} for an ${\cal
N}=8$  Chern-Simons theory with gauge group $SO(4)$.  It may also
omit interesting constructions with ${\cal N}=4$ supersymmetry.

\bigskip\noindent{\it Notation}

We generally follow the conventions  of \cite{Gates:1983nr},
chapter 2, for the ${\cal N}=1$ superspace in three dimensions. We
denote spinor indices as Greek indices $\alpha, \beta,\dots$. We
raise and lower indices with a matrix
\begin{equation} C_{\alpha \beta}= - C^{\alpha \beta} =
\begin{pmatrix}0 & -i \\ i & 0\end{pmatrix}
\end{equation} as \begin{equation} \lambda_{\alpha} =
\lambda^{\beta} C_{\beta \alpha} \qquad \lambda^{\alpha} = C^{\alpha
\beta} \lambda_{\beta} \qquad \lambda^2 = \frac{1}{2}
\lambda^{\alpha} \lambda_{\alpha} = i \lambda^+ \lambda^-
\end{equation}

Several useful relations are listed in appendix \ref{rel}. Superspace coordinates will be denoted as
$\theta^{\alpha}$. A basic real superfield is expanded as
\begin{equation}{\cal Q} = Q+ \theta^{\alpha} \lambda_{\alpha} -
\theta^2 F_Q \end{equation} For such a superfield, the kinetic term
is
\begin{equation} -\frac{1}{2} \int \mathrm{d}^2\theta\, (\partial_{\alpha} {\cal Q})^2
= - \frac{1}{2} \partial_{\mu} Q \partial^{\mu} Q +
\frac{1}{2}\lambda^{\alpha} (i \slashed{\partial})_{\alpha}^{\beta}
\lambda_{\beta}+\frac{1}{2} F_Q^2 .\end{equation} A real
superpotential may be added:
\begin{equation}\label{supint} \int \mathrm{d}^2\theta\, W({\cal Q}) = W''(Q)
\lambda^2 + W'(Q) F_Q.\end{equation}

 To describe $n$
hypermultiplets in terms of ${\cal N}=1$ superfields, we introduce
$4n$ superfields ${\cal Q}^I_A$, $I=1,\dots,2n$, $A=1,2$.  The group
$SU(2)_d$ acts on $A$, while $Sp(2n)$ acts on $I$.  The gauge group
$G$ will act via a homomorphism to $Sp(2n)$, so the hypermultiplets
form a quaternionic representation of $G$. The metric on the
hypermultiplet space will be $\epsilon^{AB} \omega_{IJ}$, where
$\epsilon^{AB}$ and $\omega_{IJ}$ are respectively $SU(2)$-invariant
and $Sp(2n)$-invariant antisymmetric tensors.  The structure
constants will have a quaternionic form $\tau^m_{IJ} =
T^{mK}_I \omega_{KJ}$, symmetric in $IJ$. (Here $m$ runs over a basis of the Lie
algebra $\frak g$ of $G$.) ${\cal Q}$ obeys the natural reality
condition ${\cal Q}^{\dagger A}_I = \epsilon^{AB}\omega_{IJ} {\cal
Q}^J_B$.

 The gauge multiplet consists of
a superconnection $\Gamma_{\alpha}$ and $\Gamma_{\mu}$ entering
supercovariant derivatives ${\cal D}_{\alpha} = \partial_{\alpha} -
i \Gamma_{\alpha}$ and ${\cal D}_{\mu}=\partial_{\mu} - i
\Gamma_{\mu}$. There is a constraint \begin{equation}\{ {\cal
D}_{\alpha},{\cal D}_{\beta} \} = 2 i \slashed{\cal D}_{\alpha \beta}
\end{equation} and a definition of field strength \begin{equation}
[{\cal D}_{\alpha},\slashed{\cal D}_{\beta \gamma} ] = C_{\alpha
(\beta}{\cal W}_{\gamma)}. \end{equation} In the Wess-Zumino gauge,
the only superpartner of the gauge field is
\begin{equation}{\cal
W}_{\alpha}|_{\theta=0}=\chi_{\alpha}.\end{equation}

The gauge-invariant extension of the matter kinetic energy is
obtained by replacing ordinary superspace derivatives by covariant
ones.  This gives
\begin{equation}\label{matact}
-\frac{1}{2}\int \mathrm{d}^2 \theta ({\cal D}_\alpha {\cal
Q}^I_A)^2 = \frac{1}{2}  \epsilon^{AB}\left(- \omega_{IJ}{\cal
D}_{\mu} Q^I_A {\cal D}^{\mu} Q^J_B + \omega_{IJ}\lambda^{I
\alpha}_A (i \slashed{\cal D})_{\alpha}^{\beta} \lambda^J_{B \beta}+
\omega_{IJ}F^I_{Q A} F^J_{Q B} + 2 \lambda^I_{A\alpha} \tau^m_{IJ}
\chi_m^{\alpha} Q^J_B\right).
\end{equation}

The standard kinetic term for the gauge fields is
\begin{equation}\label{lugo} \int \mathrm{d}^2\theta\, {\cal W}^2
=\frac{1}{2}\chi^{\alpha} (i \slashed{\cal D})_{\alpha}^{\beta}
\chi_{\beta}+ \frac{1}{4}F_{\mu \nu}F^{\mu \nu}. \end{equation} But
for the moment we are interested in gauge theories in which this
term is absent, and the gauge fields have only a Chern-Simons
action. The Chern-Simons term is essentially $\frac{1}{2}
\Gamma^{\alpha} {\cal W}_{\alpha} + \cdots$, where we omit some
extra terms cubic and quartic in $\Gamma^{\alpha}$. In the
Wess-Zumino gauge, this reduces to
\begin{equation}\label{gerto}
\frac{k^{mn}}{4 \pi} \left( A_m \wedge dA_n + \frac{2}{3}A_m\wedge
[A,A]_n - \chi^{\alpha}_m \chi_{\alpha n}\right),
\end{equation}
where $k^{mn}$ is an invariant quadratic form on the Lie algebra
$\frak g$ of $G$.  This quadratic form must obey a suitable
integrality condition in order for the quantum theory to be
well-defined.  (For example, if $G$ is a product of simple and
simply-connected factors, then $k$ is an integer multiple of the
Killing form for each factor.)

 In the absence of the
conventional kinetic term (\ref{lugo}), the part of the action
quadratic in $\chi$ is purely the mass term present in
(\ref{gerto}), so $\chi$ is an auxiliary field.  $\chi$ also enters
the Yukawa coupling $\chi Q\lambda$ in (\ref{matact}), so
integrating out $\chi$ gives a $Q^2\lambda^2$ coupling. This
coupling is not invariant under $SU(2)\times SU(2)$ (with the two
factors acting separately on $Q$ and $\lambda$). However, if the
superpotential $W({\cal Q})$ is homogeneous and quartic in $\cal Q$,
then a $Q^2\lambda^2$ term, also not invariant under $SU(2)\times
SU(2)$, arises when the auxiliary field $F_Q$ is integrated out of
(\ref{supint}).  Our procedure will be to choose $W$ so that the
$Q^2\lambda^2$ terms add up to be $SU(2)\times SU(2)$ invariant. The
rest of the action will be $SU(2)\times SU(2)$ invariant for any
choice of $W$.

The most general possible form of $W$, granted that it should be
homogeneous and quartic (for conformal invariance) and $G\times
SU(2)_d$ invariant, is as follows:
\begin{equation}\label{genform}
W({\cal Q})=\frac{\pi}{3} t_{IJ;KS}\epsilon^{AB} \epsilon^{CD} {\cal
Q}^I_A {\cal Q}^J_B {\cal Q}^K_C {\cal Q}^S_D.
\end{equation}
The tensor $t_{IJ;KS}$ is antisymmetric in the first two indices and
in the last two indices, and symmetric under exchange of the first
two with the last two. The full $Q^2\lambda^2$ interaction is
\begin{equation}\label{troublint}
\pi Q^I_A Q^J_B \lambda^{\alpha K}_{\dot C} \lambda^S_{\alpha \dot
D}\left(\epsilon^{A\dot C} \epsilon^{B \dot D} \tau^m_{IK}
\tau^n_{JS} k_{mn} + \frac{2}{3} t_{IJ;KS}\epsilon^{AB}
\epsilon^{\dot C \dot D} + \frac{4}{3} t_{IK;JS}\epsilon^{A\dot C}
\epsilon^{B \dot D}\right) .\end{equation} The first term comes from
integrating out the auxiliary fermion $\chi$, as described in the
last paragraph. The rest comes from the superpotential, via
(\ref{supint}).  If in (\ref{troublint}), we can antisymmetrize the
expression $Q^I_AQ^J_B$  in $A$ and $B$, then the result is
proportional to $\epsilon_{AB}$ and has the full $SU(2)\times SU(2)$
symmetry. Hence the condition for $SU(2) \times SU(2)$ invariance is
that the part proportional to $Q^{(I}_{(A}Q^{J)}_{B)}$ is zero:
\begin{equation} \label{necess}
\tau^m_{IK} \tau^n_{JS} k_{mn} +\tau^m_{JK} \tau^n_{IS} k_{mn} +
\frac{4}{3} t_{IK;JS}+ \frac{4}{3} t_{JK;IS}=0.
\end{equation}

By summing this equation over cyclic permutations of $IJK$, we can
eliminate $t$ and get a condition that involves $\tau$ only:
\begin{equation} \label{fund}
\tau^m_{(IJ} \tau^n_{K)S} k_{mn}=0.
\end{equation}
This fundamental identity is a strong requirement on the gauge group
and matter representation. Happily, it is possible to understand
this condition in detail, since it is equivalent to the Jacobi
identity for the following super Lie algebra:
\begin{align}\label{bizon}\notag
\left[M^m,M^n\right]&=f_s^{mn} M^s \\ \notag
\left[M^m,\lambda_I\right]&=\tau^m_{IJ} \omega^{JK} \lambda_K \\
 \{\lambda_I,\lambda_J\}&= \tau^m_{IJ} k_{mn} M^n.
\end{align}
Of these conditions, the first just says that $\frak g$ is a Lie
algebra, and the second is equivalent to the statement that the
hypermultiplets furnish a representation of this Lie algebra. The
interesting statement is the last one, which asserts that the Lie
algebra $\frak g$ can be extended to a super Lie algebra $\hat
{\frak g}$ by adjoining fermionic generators associated with the
hypermultiplet representation.  In verifying that eqn. (\ref{bizon})
does define a super Lie algebra, the only nontrivial condition to
verify is the $\lambda\lambda\lambda$ Jacobi identity:
\begin{equation}\label{holfo} [\lambda_I,\{\lambda_J,\lambda_K\}]+
[\lambda_J,\{\lambda_K,\lambda_I\}]+[\lambda_K,\{\lambda_I,\lambda_J\}]=0.
\end{equation}
A short calculation shows that this precisely coincides with
(\ref{fund}).

Moreover, we can now solve for $t$; each solution of the fundamental
identity (\ref{fund}) determines a solution to (\ref{necess}) as
well:
\begin{equation}
t_{IJ;KS} = \frac{1}{4} \tau^m_{IK} \tau^n_{JS} k_{mn}-
\frac{1}{4}\tau^m_{IS} \tau^n_{JK} k_{mn}
\end{equation}
The quadratic form $k^{mn}$ that controls the Chern-Simons couplings
in eqn. (\ref{gerto}) also appears in (\ref{fund}) and
(\ref{bizon}).  This means that the object $\hat
k=(k^{mn},\omega^{IJ})$ is an invariant (and nondegenerate)
quadratic form on the super Lie algebra $\hat {\frak g}$. Hence we
get a consistent $N=4$ Chern-Simons theory for each choice of a
supergroup whose fermionic generators form a (possibly reducible)
quaternionic representation of the bosonic subgroup, together with
an invariant nondegenerate quadratic form $\hat k$.  The
Chern-Simons couplings are determined by the restriction of $\hat k$
to the bosonic Lie algebra $\frak g$.

For our purposes, the prime examples are the classical supergroups
$U(N|M)$ (or their cousins $SU(N|M)$ and $PSU(N|N)$) and
$OSp(N|M)$. In each case, the gauge group is a product $U(N)\times
U(M)$ or $O(N)\times Sp(M)$, with equal and opposite Chern-Simons
couplings in the two factors.  The fermion fields are in
bifundamental hypermultiplets. The connection to a certain brane
construction was described in section \ref{intersecting}.

It will be useful to specialize the Lagrangian to the $U(N|M)$ case.
We write  $A_1$ and $A_2$ for the gauge fields of $U(N)$ and $U(M)$.
The hypermultiplets consist of a pair of $N\times M$ matrices ${\cal
Q}_A$, $A=1,2$, whose bosonic and fermionic components we denote
$Q_A, \psi_{\dot A}$. ${\cal Q}^\dagger_A$ will be the hermitian
adjoint of ${\cal Q}^A$, and similarly for $Q^A$. The structure
constants are most easily described by giving the moment maps
\begin{equation} \mu_{AB}^m = Q_A^I Q_B^J \tau_{IJ}^m \end{equation}
for the actions of $U(N)$ and $U(M)$: \begin{equation}
\mu^{(1)}_{AB} = Q^{\dagger}{}_{(A} Q_{B)} \qquad \qquad
\mu^{(2)}_{AB} = Q_{(A} Q^{\dagger}{}_{B)}.
\end{equation} The fundamental identity (\ref{fund}) is also easier
to understand when recast\footnote{We can equivalently write
$k_{mn} \mu^m_{(AB} \mu^n_{C)D} =0$; once this tensor is
symmetrized in three of the indices $ABCD$, it becomes
automatically symmetric in all four.} as
\begin{equation} \label{fundmu} k_{mn} \mu^m_{(AB} \mu^n_{CD)} =0 .\end{equation}
Indeed, due to the opposite signs of the Chern-Simons coefficients
of $U(N)$ and $U(M)$, the identity follows from the obvious
relation
\begin{equation} \Tr\, Q^{\dagger}{}_{(A} Q_{B}Q^{\dagger}{}_{C} Q_{D)} -
\Tr \,Q_{(A} Q^{\dagger}{}_{B}Q_{C} Q^{\dagger}{}_{D)}=0.
\end{equation}

The superpotential can be expressed in terms of the superfield
${\cal M}_{AB}^m = {\cal Q}_A^I {\cal Q}_B^J \tau_{IJ}^m$  whose
leading component is the moment map:
\begin{equation}\label{superp}W= \frac{\pi}{6} \epsilon^{AB}
\epsilon^{CD} {\cal Q}^I_A {\cal Q}^J_B {\cal Q}^K_C {\cal Q}^S_D
\tau^m_{IK} \tau^n_{JS}k_{mn} =\frac{\pi}{6}  \epsilon^{AB}
\epsilon^{CD} {\cal M}^m_{AC} {\cal M}^n_{BD}k_{mn}.
\end{equation} Informally, the superpotential is the square of the moment
map.  In the specific case of the $U(N|M)$ theory, this reads
\begin{equation}\label{keflo}
{\cal W}_4 = \frac{\pi k}{6} \Tr\, {\cal Q}^A {\cal Q}^{\dagger}{}_A
{\cal Q}^B {\cal Q}^{\dagger}{}_B - \frac{\pi k}{6}  \Tr\, {\cal
Q}^A {\cal Q}^{\dagger}{}_B {\cal Q}^B {\cal Q}^{\dagger}{}_A.
\end{equation}
The scalar potential is proportional to
\begin{equation}\label{zeflo}
\Tr\, Q^{[A} Q^{\dagger}{}_A Q^{B]} Q^{\dagger}{}_{[B} Q^C
Q^{\dagger}{}_{C]}.
\end{equation}
The equation for a critical point of the superpotential says that
\begin{equation}\label{opo} Q^A Q^\dagger_C Q^B=Q^B Q^\dagger_C
Q^A\end{equation} for all $A,B,C=1,2$, along with the hermitian
adjoint statement
\begin{equation}\label{zopo}Q^\dagger_A Q^C Q^\dagger_B=Q^\dagger_B
Q^C Q^\dagger_A.\end{equation} In particular, the bilinear
matrices $Q_A Q^\dagger_B$, which include the moment maps, commute
with each other. Their eigenvalues can be interpreted as points in
$\R^4$.  We will see in a later section that this is an important
consistency requirement for the brane picture.

There is one last calculation which we will find useful. Consider
the scalar potential in general \begin{equation} \frac{2
\pi^2}{9}\mu_{AB}^m \mu^{Bn}_C k_{ms} k_{np} Q^{AI}
\tau^s_{IJ}\omega^{JK}\tau^p_{KT} Q^{CT}. \end{equation} The part
symmetric in $AC$ involves a commutator of the gauge group structure
constants. Let us apply some transformations to half of this term.
First, separate the symmetric and antisymmetric parts: the symmetric
part is
\begin{equation}\frac{\pi^2}{9}\mu_{AB}^m \mu^{Bn}_C k_{ms} k_{np}
f^{sp}_q \mu^{qAC}. \end{equation} The antisymmetric part is
\begin{equation} \frac{\pi^2}{9}\mu_{AB}^m \mu^{ABn} k_{ms} k_{np}
Q^{I}_C \tau^s_{IJ}\omega^{JK}\tau^p_{KT} Q^{CT}. \end{equation} We
can use the fundamental identity to rearrange this to
\begin{equation} \frac{2\pi^2}{9} \mu_{AB}^m \mu^{Bn}_C k_{ms}
k_{np} Q^{IC} \tau^s_{IJ}\omega^{JK}\tau^p_{KT} Q^{AT}.
\end{equation} This combines symmetrically in $AC$ with the
remaining part of the original potential to give again the
commutator. The final result is
\begin{equation}\frac{\pi^2}{6}\mu_{AB}^m \mu^{Bn}_C k_{ms} k_{np}
f^{sp}_q \mu^{qAC}. \end{equation} It is clear that the potential
will be zero if the moment maps commute as elements of the Lie
algebra.

\subsubsection{ The Current Multiplet}\label{multiplet}

The conserved current that generates the gauge symmetry  is part
of a short multiplet of ${\cal N}=4$ supersymmetry or of the
superconformal symmetry $OSp(4|4)$. The lowest dimension operators
in this multiplet are the moment maps $\mu_{AB}$ (we suppress the
$\frak g$ index), which transform as  $\3 \otimes \1$ under
$SU(2)\times SU(2)$. In the conformal case, these fields satisfy a
BPS bound: their dimension is $1$ and equals the spin under the
$R$-symmetry group.  The first descendants, which we will call
$j_{A \dot B}$, are fermionic fields of dimension $3/2$
transforming as $\2 \otimes \2$. For the case that we have been
treating so far that the matter system consists of free
hypermultiplets, we have
\begin{equation} j_{A \dot B}^m = Q_A^I \lambda_{\dot B}^J
\tau_{IJ}^m. \end{equation} After some algebra, the ``Yukawa
coupling'' can be expressed in terms of this operator in a
manifestly $SU(2)\times SU(2)$-invariant form
\begin{equation}\label{yuka}
\pi Q^I_A Q^J_B \lambda^{\alpha K}_{ \dot C} \lambda^S_{\alpha
\dot D} \epsilon^{AB} \epsilon^{\dot C \dot D}\tau^m_{IS}
\tau^n_{JK}k_{mn}=\pi \epsilon^{AB} \epsilon^{\dot C \dot D}j_{A \dot C}^m
j_{B \dot D}^n k_{mn}.
\end{equation}

Let us reconsider in terms of these variables what happens when the
free hypermultiplets $\cal Q$ are coupled to gauge fields
represented by ${\cal N}=1$ supermultiplets.  (This is useful as
preparation for considering more general matter systems in section
\ref{general}.) The coupling of the auxiliary fermion $\chi$ in the
component Lagrangian is $\chi^{\alpha} j_{A \dot B \alpha} \epsilon^{A \dot B}$,
and integrating it away produces a $jj$ coupling that lacks
$R$-symmetry:
\begin{equation}\label{primo} \pi k_{mn} j_{A \dot B} \epsilon^{A \dot B}j_{C
\dot D} \epsilon^{C \dot D}.
\end{equation} The superpotential \begin{equation} \frac{\pi}{6}
\epsilon^{AB} \epsilon^{CD} {\cal M}^m_{AC} {\cal M}^n_{BD}k_{mn}
\end{equation} gives two contributions to the Yukawa couplings; the
two superderivatives can act on different factors of $\cal M$ or on
the same factor:
\begin{equation}\label{secondo} \frac{\pi}{6}  \epsilon^{AB} \epsilon^{CD}
j^m_{(A \dot C)} j^n_{(B\dot D)}k_{mn} + \frac{\pi}{3}
\epsilon^{AB} \epsilon^{CD} \mu^m_{A C} O^n_{\dot B \dot
D}k_{mn}.\end{equation} The operator $O_{\dot A \dot B}$ is a
fermion bilinear of dimension two and spin zero that is an
additional member of the ${\cal N}=4$ current multiplet for the
free fields:
\begin{equation} O_{\dot A \dot B}=\lambda^I_{\dot A}
\lambda^J_{\dot B} \tau_{IJ}.
\end{equation} The fundamental identity (\ref{fundmu}) can be
subjected to two superderivatives to give
\begin{equation}\label{helfox} \mu^{m AB} O^n_{\dot C \dot D}
k_{mn} + j^{m(A}_{\dot C}  j^{B) n}_{\dot D} k_{mn} =0
\end{equation} This identity allows one to rewrite the
$QQ\lambda\lambda$ ``Yukawa'' interaction, which is the sum of
(\ref{primo}) and (\ref{secondo}), as a bilinear expression in $j$
(we omit the factor $\frac{\pi}{3} k_{mn}$) . \begin{equation} 3j_{A
\dot B} \epsilon^{A \dot B}j_{C \dot D} \epsilon^{C \dot D}+
\epsilon^{AB} \epsilon^{CD} j^m_{A \dot C} j^n_{B\dot D}+
\epsilon^{AD} \epsilon^{CB} j^m_{A \dot C} j^n_{B\dot D}-
\epsilon^{AB} \epsilon^{CD} j^m_{A \dot B} j^n_{C \dot D}-
\epsilon^{AB} \epsilon^{CD} j^m_{A \dot D} j^n_{C \dot B}
.\end{equation} Rearranging the two indices of a current in the
third and fifth terms by the usual $U_{AB}=U_{BA} + \epsilon_{AB}
\epsilon^{CD} U_{CD}$ the non-$R$-symmetric terms drop and one is
left with a simple Yukawa: \begin{equation} \pi k_{mn} j^m_{A\dot B}
j^{n A \dot B}\end{equation}
\subsubsection{Chern-Simons
Coupling to General ${\cal N}=4$ Matter System}\label{general}

So far, we have constructed ${\cal N}=4$ Chern-Simons theories by
coupling a gauge field to an ${\cal N}=4$ matter system that
consists simply of some free hypermultiplets.  This can be
generalized to replace the free hypermultiplets with a sigma model
in which the target space is a hyper-Kahler manifold $X$.  An
important special case is that $X$ is the Higgs (or Coulomb)
branch of a superconformal field theory with $OSp(4|4)$ symmetry.
However, for much of the discussion, this is not required.

If $X$ is the Higgs branch of a CFT, then $X$ is conical and in
addition there is an $SU(2)$ symmetry acting on $X$ and rotating
the three complex structures.  However, in the following, we do
not need to assume the existence of such a symmetry. The only
symmetries that we will assume that act on scalar fields will be
the symmetries that make up the gauge group $G$.  The scalars will
be fields $Q^{IA}$ that are associated with local coordinates on
$X$, but in contrast to the discussion of the free
hypermultiplets, we do not assume any symmetry acting on the $A$
index.   We will simply look for an $SU(2)$ symmetry that acts
only on the fermion fields $\lambda_{I\dot A}$, transforming the
$\dot A$ index. This is enough to promote ${\cal N}=1$
supersymmetry to ${\cal N}=4$, or in the conformal case to promote
$OSp(1|4)$ to $OSp(4|4)$.  (In the conformal case, the $SU(2)$
that rotates the complex structures is part of $OSp(4|4)$.)  The
hyper-Kahler structure of $X$ can be described by the existence of
antisymmetric inner products $\omega_{IJ}$ and $\epsilon_{AB}$.

The moment maps $\mu^m{}_{AB}$ ($m$ being a $\frak g$ index) are
defined as follows.  Let $V^m$, $m=1,\dots,{\rm dim}\,\frak g$ be
the vector fields on $X$ generating the action of $\frak g$.  And
let $\Omega_{AB}$, $A,B=1,2$ be the three symplectic
forms\footnote{They can be defined by $\Omega_{AB}=dQ^{IC}\wedge
dQ^{JD}\omega_{IJ}\epsilon_{AC}\epsilon_{BD}$.} of the
hyper-Kahler manifold $X$. Then the functions $\mu^m{}_{AB}$ are
characterized as follows:
\begin{equation}\label{felx} d\mu^m{}_{AB} = i_{V^m}(\Omega_{AB}).\end{equation}
(Here $i_V$ is the operation of contraction with a vector field
$V$.) This condition plus $\frak g$-invariance determines
$\mu^m_{AB}$ uniquely if $G$ is semi-simple.  In general, there are
undetermined additive constants in $\mu$, which correspond
physically to the possibility of adding Fayet-Iliopoulos $D$-terms
for $U(1)$ gauge fields. The fundamental identity makes sense for
this class of models:
\begin{equation} \label{fundmunc} k_{mn} \mu^m_{(AB} \mu^n_{CD)} =0.\end{equation}
Here $k$ is some invariant and nondegenerate quadratic form on
$\frak g$.

Similarly, we can define ${\cal N}=4$ descendants of the fields
$\mu^m_{AB}$. The first descendant $j$, the fermionic current, is
proportional to $\lambda$ times a first derivative of $\mu$.  But
according to (\ref{felx}), the derivatives of $\mu$ are
essentially the vector fields $V^m$, so we can write $j$ in terms
of $V^m$:
\begin{equation} j^{m\,A \dot B} = V^{mIA} \lambda^{\dot BJ}
\omega_{IJ}. \end{equation}  The next descendant of $\mu$ is
constructed from the second derivative of $\mu$ or equivalently
from the first derivative of $V$.  In general, if $V$ is a Killing
vector field on a Riemannian manifold, one has $D_IV_J+D_JV_I=0$.
On a hyper-Kahler manifold, with $V$ assumed to preserve all three
complex structures, one has a stronger version of this statement:
\begin{equation}\label{kelx}D_{IA}V^m_{JB}=\tau^m_{IJ}\epsilon_{AB},\end{equation}
where $\tau^m_{IJ}$ is symmetric in $I$ and $J$.  In the case of
free hypermultiplets, the $\tau^m$ are simply constant matrices
(which generate the action of $G$ on the hypermultiplets), but in
general, they are tensor fields on $X$.

As in our study of the linear hypermultiplets, we construct the
action starting with ${\cal N}=1$ superfields and then looking for
an additional symmetry. Let us denote with ${\cal M}^{AB}$ again
the ${\cal N}=1$ superfield whose lowest component is $\mu^{AB}$.
The components of ${\cal M}^{AB}$ are
\begin{equation} {\cal D}_{\alpha} {\cal M}^{mAB}|_{\theta=0} =
j_\alpha^{AB}
\end{equation}
and
\begin{equation} {\cal D}^{\alpha}{\cal D}_{\alpha} {\cal M}^{AB}|_{\theta=0} =
\tau^{IJ} \lambda^{\alpha A}_I \lambda^B_{\alpha J}= O^{AB}.
\end{equation}
(In these formulas, we suppress the $\frak g$ index $m$.)

We can now repeat step by step the analysis done in the free field
case, and every step is formally identical. The potentially
non-$R$-invariant fermion bilinears are still given by
(\ref{primo}) and (\ref{secondo}), and they still add up to an
$R$-invariant sum (\ref{yuka})  if the moment map $\mu^m$ obeys
the fundamental identity:
\begin{equation} \label{fundmunox} k_{mn} \mu^m_{(AB} \mu^n_{CD)} =0.\end{equation}
In fact, we only need the weaker condition (\ref{helfox}), which is
a second descendant of the fundamental identity:
 \begin{equation}\label{helfoxt}
\mu^{m AB} O^n_{\dot C \dot D} k_{mn} + j^{m(A}_{\dot C}  j^{B)
n}_{\dot D} k_{mn} =0
\end{equation}
Elsewhere, we will show that there actually are ${\cal N}=4$ models
for which this weaker condition is obeyed (in fact, the first
descendant of the fundamental identity vanishes) but the fundamental
identity itself is not satisfied. However, in the superconformal
case, there are superconformal lowering operators and
(\ref{helfoxt}) actually implies (\ref{fundmunox}).

The superconformal transformations of the fundamental identity are
of interest. The current multiplet is a short multiplet of
$OSp(4|4)$, as the leading component $\mu^{AB}$ satisfies a BPS
bound: it has spin $(1,0)$ under the $R$-symmetry group and
dimension $1$. The fundamental identity has spin $(2,0)$ and
dimension $2$: it is the protected component of the product of two
current multiplets. This fact will probably play a useful role in a
quantum description of these theories.

\subsubsection{ Quivers}\label{quivers}

Thus, we have a general recipe to couple ${\cal N}=4$ Chern-Simons
gauge fields to any hyper-Kahler manifold $X$ that obeys the
fundamental identity.  Of course, this is only interesting if there
are examples beyond the ones associated with free hypermultiplets.
We will now describe a family of such examples. We begin with a
special case.

We start with a symmetry group $U(N_1)\times U(N_2)\times U(N_3)$
acting on the following free hypermultiplets: we include
hypermultiplets $Y$ that transform as $(N_1,\bar N_2,1)$ plus
complex conjugate, and  hypermultiplets $Z$ that transform as
$(1,N_2,\bar N_3)$ plus complex conjugate. (We use the same
symbols $Y$ and $Z$ to denote the hypermultiplets and the spaces
that they parametrize.) We write $\mu_Y$ and $ \mu'{}_Y$ for the
hyper-Kahler moment maps for the action of $U(N_1)$ and $U(N_2)$
on $Y$, and likewise $\mu_Z$ and $\mu'_Z$ for the hyper-Kahler
moment maps for the action of $U(N_2)$ and $U(N_3)$, respectively,
on $Z$.  The hyper-Kahler moment map for the action of
$U(N_1)\times U(N_2)\times U(N_3)$ on $Y\times Z$ is therefore
\begin{equation}\label{lurf}\mu_{Y\times Z}=(\mu_Y,\mu'{}_Y+\mu_Z,\mu'{}_Z),
\end{equation}
where the three components refer to the three factors. Now we let
$X$ denote the hyper-Kahler quotient $(Y\times Z)///U(N_2)$.  We
recall that the hyper-Kahler quotient is obtained by setting to
zero the moment map for $U(N_2)$, that is by imposing
\begin{equation}\label{urf}\mu'_Y+\mu_Z=0,\end{equation}
and dividing by $U(N_2)$.  On the quotient, we still have an action
of $U(N_1)\times U(N_3)$, and the moment map can be read off from
(\ref{lurf}):
\begin{equation}\label{kurf}\mu_X=(\mu_Y,\mu'{}_Z).\end{equation}

Now if $W$ is any hyper-Kahler manifold with action of a group
$U(N)$ (where $N$ may be $N_1$, $N_2$, or $N_3$) with moment map
$\mu$, we set $f_{(ABCD)}(\mu)=\sum_m \mu^m{}_{(AB}\mu^m{}_{CD)}$.
(The sum is taken in an orthonormal basis for the Killing form.)
Since $f$, whose subscripts we will suppress, is homogeneous and
quadratic, we have
\begin{equation}\label{uj}f(\mu)=f(-\mu).\end{equation}
 Since $Y$ obeys the
fundamental identity for the action of $U(N_1)\times U(N_2)$, we
have
\begin{equation}\label{olurf} f(\mu_Y)-f(\mu'{}_Y)=0.\end{equation}
(As usual, the minus sign reflects the structure of the invariant
quadratic form $\hat k$ of the supergroup $U(N_1|N_2)$.) Similarly,
$Z$ obeys the fundamental identity for the action of $U(N_2)\times
U(N_3)$, so
\begin{equation}\label{plurf}f(\mu_Z)-f(\mu'{}_Z)=0.\end{equation}
For the hyper-Kahler quotient $X$, we have the additional
condition (\ref{urf}), which by virtue of (\ref{uj}) implies that
$f(\mu'{}_Y)=f(\mu_Z)$.  Combining these results, we learn that
the hyper-Kahler moment map of $X$ obeys
\begin{equation}\label{okurf}f(\mu_Y)-f(\mu'{}_Z)=0.\end{equation}
This means that $X$ obeys the fundamental identity for the action of
$G=U(N_1)\times U(N_3)$, provided we take equal and opposite
Chern-Simons levels for the two factors in $G$.

\begin{figure}
  \begin{center}
    \includegraphics[width=3.5in]{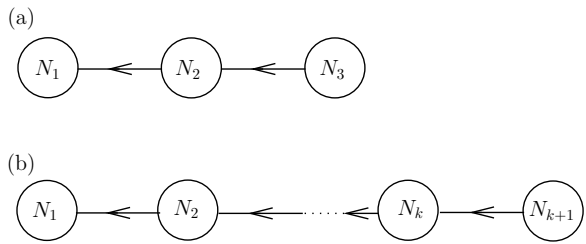}
  \end{center}
\caption{\small (a) A quiver associated with the first non-trivial
example of a hyper-Kahler manifold obeying the fundamental identity.
 (b)  More general linear quivers leading to solutions of the fundamental
 identity. }
  \label{pic3}
\end{figure}
Thus, we get our first example of a non-flat hyper-Kahler manifold
obeying the fundamental identity.  The construction can be
conveniently described via a simple quiver (fig. \ref{pic3}).  A
node in the quiver represents a unitary group $U(N)$ for some $N\geq
0$. A link connecting two nodes represents  bifundamental
hypermultiplets transforming under the given product of groups.  The
links are oriented, as indicated by the arrows.  We pick a nonzero
integer $r$ and assign to each node a Chern-Simons level which is
the product of $r$ times the number of arrows entering the node
minus the number of arrows leaving.  Thus, in the example of fig.
\ref{pic3}a, there are three nodes, with levels $r,0,-r$.  We take
the hyper-Kahler quotient (of the space parametrized by the
hypermultiplets) by the product of all groups associated with nodes
of level 0. The result is a hyper-Kahler manifold $X$. It is acted
on by a group $G$ that is the product of the factors associated with
nonzero levels.

The explicit example based on the product $U(N_1)\times
U(N_2)\times U(N_3)$ is associated to the quiver of fig.
\ref{pic3}a.   The general case of a linear quiver, as in fig.
\ref{pic3}b, is similar.  The gauge group $G$ is a product of two
unitary groups, associated with the ends of the quiver; these are
the only nodes with nonzero labels. The fundamental identity is
obeyed for the action of $G$ on $X$, by virtue of essentially the
same argument that we used for the quiver of fig. \ref{pic3}a.
Moreover, essentially the same argument works for orthosymplectic
quivers, in which the gauge groups are alternatively orthogonal or
symplectic along the chain. Here one uses at each step the
solution of the fundamental identity associated to $OSp(N|M)$.

Because of their origin as hyper-Kahler quotients of linear spaces,
the spaces $X$ obtained this way are actually conical, and have an
$SU(2)$ action rotating the three complex structures.  In fact,
these examples are associated to superconformal field theories,
which we will study in detail elsewhere.

\subsection{$4d$ ${\cal N}=4$ Super Yang-Mills in a $3d$ Language}\label{language}

Given the simplicity of the Chern-Simons calculation, it is
natural to wonder if a similar method can be applied to the Janus
configuration, or to other modifications of ${\cal N}=4$ super
Yang-Mills that preserve ${\cal N}=4$ supersymmetry in the
three-dimensional sense.

As preparation, we will describe the undeformed ${\cal N}=4$
Lagrangian in the 3d ${\cal N}=1$ language. It will be then very
simple to deform this to allow for various kinds of ``defects,''
including $y$-dependent couplings (section \ref{janagain}), and
interfaces between two possibly different gauge theories with
bifundamental matter living at the interface (section
\ref{boundary}). It is also possible to include defects that support
extra hypermultiplets coupled to the bulk gauge fields [REFERENCE].
Any of these defects can coexist, provided that they preserve the
same $8$ supersymmetries.

From a three-dimensional point of view, the gauge field $A_\mu$,
$\mu=0,\dots,3$, splits up as a three-dimensional gauge field
$A_\mu, $ $\mu=0,1,2$, and an adjoint-valued scalar $A_3$. In
terms of $3d$ ${\cal N}=1$ superfields, the three-dimensional
gauge field  is part of  a standard superconnection
$\Gamma_{\alpha}$; this multiplet, which already appeared in
section \ref{chernsimons}, also describes a fermion $\sigma_{A
\alpha}$. (In the purely three-dimensional discussion, the
analogous field was an auxiliary field and was called $\chi$.) On
the other hand, $A_3$ is the leading component of a real
superfield
\begin{equation}\label{zito}{\cal A}_3 = A_3 + \theta^{\alpha} \sigma_{3 \alpha} +
\theta^2 F_3^a\end{equation} that transforms inhomogeneously under
gauge transformations:
\begin{equation}\label{interm}
{\cal A}_3^g = g^{-1} {\cal A}_3 g - g^{-1} \partial_3 g .
\end{equation}  The quantity that transforms most simply
 is the covariant derivative $\partial_3+{\cal A}_3$. The inhomogeneous term
 in (\ref{interm})  also
changes the expression for the gauge-covariant superderivative
\begin{equation} {\cal D}_{\alpha} {\cal A}_3 = D_{\alpha} {\cal
A}_3 + \{\Gamma_{\alpha},{\cal A}_3 \} - \partial_3
\Gamma_{\alpha} .\end{equation}  Accordingly, the component
expansion of ${\cal D}_{\alpha} {\cal A}_3$  contains not the
naive non-gauge invariant derivative $\partial_{\mu} A_3$ or
covariant derivative $D_\mu A_3$, but the whole field strength
$F_{\mu 3}$. Meanwhile, the scalar fields
 $X^a$ are the leading components of  real superfields
\begin{equation}{\cal X}^a = X^a + \theta^{\alpha} \rho^a_{1 \alpha}
+ \theta^2 F_X^a \end{equation} which transform in the adjoint
representation of the gauge group and as a $\3$ of the diagonal
flavor symmetry group $SU(2)_d$. The same is true of $Y^p$:
\begin{equation}\label{het}{\cal Y}^p = Y^p + \theta^{\alpha} \rho^p_{2 \alpha}
+ \theta^2 F_Y^a .\end{equation}

The fermions as described so far transform as $\3\oplus\3\oplus
\1\oplus \1$ under $SU(2)_d$.  To prove $R$-symmetry, we will have
to reorganize the fermions as the sum of two copies of the
representation $\2\otimes \2$ of the full $R$-symmetry group
$SU(2)\times SU(2)$.  Each $\3$ is related by $SU(2)\times SU(2)$
to a linear combination of the two $\1$'s. Which linear
combination appears will depend on how three-dimensional ${\cal
N}=4$ supersymmetry is embedded in four-dimensional ${\cal N}=4$,
that is, it will depend on the angle $\psi$ in eqn. (\ref{seful}).
Of course, as long as we consider the pure ${\cal N}=4$ theory (as
opposed to the generalizations that we introduce starting in
section \ref{janagain}), supersymmetry will hold simultaneously
for all values of $\psi$.

We will now write the four-dimensional ${\cal N}=4$ theory in
terms of these three-dimensional superfields.  In this formalism,
the part of the kinetic energy that involves derivatives in the
$012$ directions will come from what would usually be called
kinetic energy terms in three dimensions.  Terms involving
derivatives in the $x^3$ direction will arise from two different
sources, the peculiar gauge-covariant derivatives of ${\cal A}_3$
and some carefully chosen terms in the superpotential involving
covariant derivatives in the $x^3$ direction.

Altogether, the kinetic terms in the action come from a superspace
interaction
\begin{equation}\label{kinetic}
\frac{1}{e^2}\int \mathrm{d}^2\theta\,\Tr \left(  -{\cal W}^2 +
(D_\alpha {\cal A}_3)^2+ (D_\alpha {\cal X}^a)^2+(D_\alpha {\cal
Y}^a)^2- 2 {\cal Y}^a [\partial_3+{\cal A}_3,{\cal X}^a]\right).
\end{equation}
Expanding this in components, the 3d gauge action is
\begin{equation}\label{gaugac}
-\frac{1}{e^2} \int \mathrm{d}^2 \theta \,\Tr\, {\cal W}^2 =
\frac{1}{e^2}\Tr\left(\frac{1}{2}F_{\mu \nu} F^{\mu \nu} -i
\sigma_A^{\alpha} \slashed{D}_{\alpha}^{\beta} \sigma_{A
\beta}\right).
\end{equation}
 The kinetic energy for $A_3$ is
\begin{equation}\label{kina}
\frac{1}{e^2} \int \mathrm{d}^2 \theta\, \Tr(D_\alpha {\cal A}_3)^2
= \frac{1}{e^2}\Tr\left(F_{\mu 3}F^{\mu 3} -i \sigma_3^{\alpha}
\slashed{D}_{\alpha}^{\beta} \sigma_{3 \beta}- (F_A)^2 - 2
\sigma_3^{\alpha} D_3 \sigma_{A \alpha}\right).
\end{equation}
 The three-dimensional part of the kinetic energy for ${\cal X}^a$ reads
\begin{equation}\label{kinx}
\frac{1}{e^2} \int \mathrm{d}^2 \theta\, \Tr(D_\alpha {\cal X}^a)^2
=\frac{1}{e^2} \Tr\left((D_\mu X^a) (D^\mu X^a) -i \rho_1^{a \alpha}
\slashed{D}_{\alpha}^{\beta} \rho_{1 \beta}^a- (F^a_X)^2 - 2
\rho_1^{a \alpha} [\sigma_{A \alpha},X^a]\right)
\end{equation}
and similarly for ${\cal Y}$
\begin{equation}\label{kiny}
\frac{1}{e^2}  \int \mathrm{d}^2 \theta \,\Tr(D_\alpha {\cal Y}^a)^2
= \frac{1}{e^2}\Tr\left((D_\mu Y^a) (D^\mu Y^a) -i \rho_2^{a \alpha}
\slashed{D}_{\alpha}^{\beta} \rho_{2 \beta}^a- (F^a_Y)^2 - 2
\rho_2^{a \alpha} [\sigma_{A \alpha},Y^a]\right).
\end{equation}
And the part of the kinetic energy involving $x^3$ derivatives of
$X$ and $Y$ comes from
\begin{align}\label{polix}
-\frac{2}{e^2}  \int \mathrm{d}^2\theta \,\Tr \,{\cal Y}^a
[\partial_3+{\cal A}_3,{\cal X}^a]= -\frac{2}{e^2}\Tr \left(D_3
F_X^a Y^a+  X^a [F_A,Y^a]+D_3 X^a F_Y^a +\right. \notag \\ \left.+
Y^a [\sigma_3^{\alpha},\rho^a_{1 \alpha}]+\rho_2^{a \alpha}
[\sigma_{3 \alpha},X^a]+ \rho_2^{a \alpha} D_3\rho_{1 \alpha}^a
\right).
\end{align}
The sum of all of these terms gives the conventional
four-dimensional kinetic energy for all fields.

In addition to eqn. (\ref{polix}), which may be regarded as a
superpotential interaction from a three-dimensional point of view,
we need a conventional cubic superpotential:
\begin{equation}
{\cal W}_3=\frac{\epsilon_{abc}}{e^2} \Tr \left(-\cos \psi
\left(\frac{1}{3}{\cal X}^a[{\cal X}^b,{\cal X}^c] - {\cal
X}^a[{\cal Y}^b,{\cal Y}^c]\right)+ \sin \psi \left(\frac{1}{3}{\cal
Y}^a[{\cal Y}^b,{\cal Y}^c] - {\cal Y}^a[{\cal X}^b,{\cal
X}^c]\right) \right)
\end{equation}
The angle $\psi$ will ultimately coincide with the angle that
appears in eqn. (\ref{seful}) characterizing the embedding of
three-dimensional supersymmetry in four dimensions.  For the
moment, the main point is that we can get the same
four-dimensional ${\cal N}=4$ theory for any choice of $\psi$.
Finally, the four-dimensional theta-angle comes from
\begin{equation}
\frac{\theta}{4 \pi^2} \int \mathrm{d}^2 \theta \,\Tr\, D^\alpha
{\cal A}_3 W^\alpha = \frac{\theta}{4 \pi^2}\Tr \left(F \wedge F +i
\slashed{D}^{\alpha \beta}(\sigma_{A \alpha} \sigma_{3 \beta}) +
\frac{1}{2}D_3( \sigma_A^{\alpha} \sigma_{A \alpha})\right)
\end{equation}
(The terms other than $\Tr\,F\wedge F$ are total derivatives of
gauge-invariant quantities.)

Let us now check that the sum of the above terms reproduces the
standard $\N=4$ Lagrangian. We need to verify that by integrating
away the auxiliary fields $F$,  the quartic  scalar potential and
the $x^3$ part of the kinetic energy are reproduced, and that the
correct Yukawa couplings arise as well. The superpotential in the
three-dimensional sense is \begin{equation}\label{needthis}{\cal
W}=\int \mathrm{d}x^3\left(-\frac{2}{e^2}\Tr \,{\cal Y}^a
[\partial_3+{\cal A}_3,{\cal X}^a]+{\cal W}_3 \right).\end{equation}
Its derivatives evaluated at $\theta=0$ are
\begin{align} \notag\label{grad}
-e^2 \frac{\partial {\cal W}}{\partial {\cal X}^a} &= - 2 D_3 Y^a  +
\cos \psi \epsilon_{abc}([X^a, X^b] - [Y^a, Y^b]) + 2\sin \psi
\epsilon_{abc} [X^b, Y^c]  \\ -\notag e^2 \frac{\partial {\cal
W}}{\partial {\cal Y}^a} &=  2 D_3 X^a  - 2\cos \psi
\epsilon_{abc}[X^a,
Y^b] + \sin \psi \epsilon_{abc} ([X^b, X^c] - [Y^b, Y^c]) \\
e^2 \frac{\partial {\cal W}}{\partial {\cal A}_3} &= 2[X^a,Y^a].
\end{align}

The quartic scalar potential and some parts of the kinetic energy
arise by squaring these expressions, adding, and integrating over
$x^3$.  This process generates terms quartic in $X$ and $Y$ that
are independent of $\psi$, and can easily be rearranged into the
standard $R$-symmetric ${\cal N}=4$ quartic potential. There are
dangerous non-$R$-symmetric cubic terms, but they are total
derivatives
\begin{equation}
- \frac{1}{e^2}\cos \psi\, D_3 \left(\epsilon_{abc} \Tr [X^a ,X^b]
Y^c\right)- \frac{1}{e^2} \sin \psi \, D_3\left( \epsilon_{abc} \Tr
X^a [Y^b, Y^c]\right).\end{equation} Finally, the quadratic terms
give the contributions $(D_3X)^2$ and $(D_3Y)^2$ to the kinetic
energy.

Next we can look at the Yukawa couplings. Dropping unnecessary
indices for clarity, $-\frac{1}{e^2} X^a$ couples to
 \begin{equation} 2[\rho_1^{a \alpha},\sigma_{A\alpha}] +2
[\rho_2^{a \alpha},\sigma_{3\alpha}] + \cos \psi \epsilon_{abc}
\left([\rho^{\alpha b}_1,\rho^c_{1\alpha}]- [\rho^{\alpha
b}_2,\rho^c_{2\alpha}]\right) + 2\sin \psi\epsilon_{abc}
[\rho^{\alpha b}_1,\rho^c_{2\alpha}],\end{equation} while
$-\frac{1}{e^2} Y^a$ couples to
 \begin{equation} 2[\rho_2^{a \alpha},\sigma_{A\alpha}] -2
[\rho_1^{a \alpha},\sigma_{3\alpha}] + \sin \psi \epsilon_{abc}
\left([\rho^{\alpha b}_1,\rho^c_{1\alpha}]-
[\rho^{\alpha b}_2,\rho^c_{2\alpha}]\right) - 2\cos \psi\epsilon_{abc}
[\rho^{\alpha b}_1,\rho^c_{2\alpha}].\end{equation}

We want to pair up $\rho_1$ and $\rho_2$ with linear combinations
of $\sigma_A$ and $\sigma_3$ in order to restore the full
$R$-symmetry. The kinetic terms of the fermions are
\begin{equation}- \frac{1}{e^2} \left(i \sigma_A^{\alpha}
\slashed{D}_{\alpha}^{\beta} \sigma_{A \beta}+i \sigma_3^{\alpha}
\slashed{D}_{\alpha}^{\beta} \sigma_{3 \beta}+i \rho_1^{a \alpha}
\slashed{D}_{\alpha}^{\beta} \rho_{1 \beta}^a+i \rho_2^{a \alpha}
\slashed{D}_{\alpha}^{\beta} \rho_{2 \beta}^a
\right)\end{equation} and clearly constrain the possible linear
combinations to be an $SO(2)$ rotation. The combinations will be
\begin{align} \sqrt{2}\Psi_1^{A \dot B} &= \rho_1^{(A B)} +
\epsilon^{AB}\left(\cos \psi \sigma_A - \sin \psi
\sigma_3\right) \\
\sqrt{2}\Psi_2^{A \dot B} &= \rho_2^{(A B)} + \epsilon^{AB}\left(\sin \psi
\sigma_A + \cos \psi \sigma_3\right).
\end{align} The Yukawa couplings of $X$ and $Y$, namely
\begin{equation}
-\frac{1}{e^2} \Tr X^{AB} \left(\cos \psi \left([\Psi_1^{A \dot
C},\Psi^B_{1\dot C}]- [\Psi_2^{A \dot C},\Psi^B_{2\dot C}]\right) +
2 \sin \psi [\Psi_1^{A \dot C},\Psi^B_{2\dot C}]
\right)\end{equation} and \begin{equation} -\frac{1}{e^2} \Tr
Y^{\dot C \dot D} \left(\sin \psi \left([\Psi_1^{A \dot
C},\Psi^B_{1\dot C}]- [\Psi_2^{A \dot C},\Psi^B_{2\dot C}]\right) -
2\cos \psi [\Psi_1^{A \dot C},\Psi^B_{2\dot C}]\right),
\end{equation}
are $R$-symmetric with this choice.  If we further identify
\begin{equation}\Psi = B_1 \varepsilon_0 \otimes \Psi_1 + B_2
\varepsilon_0 \otimes \Psi_2\end{equation} we reproduce the
standard kinetic terms and Yukawa couplings of ${\cal N}=4$ super
Yang-Mills. (Appendix \ref{norm} contains some additional
conventions and details.)

The remaining $D_3$ part of the fermion kinetic terms should pair up
$\Psi_1$ and $\Psi_2$. Indeed there is a $-\frac{1}{e^2}\rho_2 D_3
\rho_1$ term in eqn. (\ref{polix}), but the $\sigma$ terms appear to
be missing. To make things $R$-symmetric, we need an additional
coupling $-(\sin \psi \sigma_A + \cos \psi \sigma_3) D_3 (\cos \psi
\sigma_A - \sin \psi \sigma_3)$, but what we see in (\ref{kina}) is
$-\sigma_3 D_3 \sigma_A$. Luckily, the last is equal to the former
plus a total derivative $-\sin \psi\, D_3 (\cos \psi
\frac{1}{2}(\sigma_3\sigma_3-\sigma_A\sigma_A) + \sin \psi \sigma_3
\sigma_A)$. The term proportional to the theta-angle is also a total
derivative, clearly.

We have reproduced the standard ${\cal N}=4$ Lagrangian. As an
extra check, notice that the only $\psi$ dependence in the
$R$-symmetric component action is in the total derivative terms,
in agreement with the fact that the same ${\cal N}=4$ theory is
invariant under each of the different $OSp(4|4)$ supergroups.

The various ${\cal N}=1$ fermionic superpartners are packaged
together in $R$-symmetric combinations in a way which may appear
quite obscure. It is useful to consider the opposite point of
view: start from the ${\cal N}=4$ gauge multiplet and look at it
as an $OSp(4|4)$ multiplet. Some tedious computations collected in
the appendix \ref{norm} show that the fields are organized into
two mirror $OSp(4|4)$ multiplets, which have the same general
structure as a $3d$ current multiplet. One multiplet contains
$Y^{A B}, \Psi_2, \cos \psi F_{3\mu} - \frac{1}{2} \sin \psi
\epsilon_{\mu \nu \rho} F^{\nu \rho}, \partial_3 X^{\dot A \dot
B}$ while the other contains $X^{\dot A \dot B}, \Psi_1, \sin \psi
F_{3\mu} + \frac{1}{2} \cos \psi \epsilon_{\mu \nu \rho} F^{\nu
\rho},\partial_3 Y^{AB}$. Further reduction from ${\cal N}=4$ to
${\cal N}=1$ reproduces the detailed structure of the
$R$-symmetric fermion combinations.

\subsection{Generalized Janus, Again}\label{janagain}

Now it is straightforward to apply this method to build again the
generalized Janus configuration of section \ref{computation}.

A key step in verifying $R$-symmetry in the previous section was to
integrate by parts to remove non-$R$-symmetric total derivatives. If
we make the couplings\footnote{We henceforth write $\theta_{Y\neg
M}$ for the gauge theory theta-angle to avoid confusion with the odd
superspace coordinates.} $e^2$ and $\theta_{Y\neg M}$ functions of
$x^3$ and repeat the calculation, then $SU(2)\times SU(2)$
invariance is broken down to a diagonal subgroup $SU(2)_d$ by terms
proportional to ${de^2}/{dx^3}$ and ${d\theta_{Y\neg M}}/{dx^3}$
that arise when we integrate by parts. Let us try to correct the
Lagrangian to restore the symmetry. By dimensional reasoning and
gauge invariance, the only possibility is to add a superpotential
term that is bilinear in the scalar fields. Moreover, this term must
be $SU(2)_d$-invariant. A bit of inspection shows that adding terms
proportional to ${\cal X}^2$ or ${\cal Y}^2$ will do irreparable
damage to $R$-symmetry. With some hindsight, and inspired by the
results of section \ref{computation}, we will add the following term
to the Lagrangian:
\begin{equation}\label{needagain}
\frac{2}{e^2}  \int \mathrm{d}^2\theta \left( \psi' \frac{\sin
\psi}{\cos \psi}\right) \Tr{\cal Y}^a {\cal X}^a=
\frac{2}{e^2}\left( \psi' \frac{\sin \psi}{\cos \psi}\right) \Tr(
F_X^a Y^a+ X^a F_Y^a + \rho_2^{a \alpha} \rho_{1 \alpha}^a ).
\end{equation}
We will also assume the coupling dependence deduced in the earlier
computation; we suppose that $\tau=\theta_{Y\neg M}/2\pi+4\pi i/e^2$
takes the form
\begin{equation}\label{gufty}
\tau = a + 4 \pi D e^{2 i \psi} \end{equation} with real constants
$a$ and $D$. (Alternatively, instead of building in our prior
knowledge of this, we could use ${\cal N}=1$ superfields to give a
new derivation of this result.)

The superpotential is now the sum of (\ref{needthis}) and the
correction term of eqn. (\ref{needagain}):
\begin{equation}\label{needmore}{\cal W}=\int \mathrm{d}x^3\left(-\frac{2}{e^2}\Tr\, {\cal
Y}^a D_3{\cal X}^a+{\cal W}_3+\frac{2}{e^2} \left( \psi' \frac{\sin
\psi}{\cos \psi}\right) \Tr{\cal Y}^a {\cal X}^a
\right).\end{equation} When we vary this with respect to $X$, we
must integrate by parts the $YD_3X$ term.  We encounter a derivative
$\mathrm{d}(1/e^2)/\mathrm{d}x^3$, which we express in terms of
$\psi'=d\psi/dx^3$ using (\ref{mufto}). The gradient of the
superpotential becomes
\begin{align}\label{oxford}
-e^2 \frac{\partial {\cal W}}{\partial {\cal X}^a} &= - 2D_3 Y^a -
2\psi' \frac{\cos \psi}{\sin \psi} Y^a + \cos \psi
\epsilon_{abc}([X^a, X^b] - [Y^a, Y^b]) + 2\sin \psi \epsilon_{abc}
[X^b, Y^c]
\notag\\
-e^2 \frac{\partial {\cal W}}{\partial {\cal Y}^a} &=  2D_3 X^a -
2\psi' \frac{\sin \psi}{\cos \psi} X^a - 2\cos \psi
\epsilon_{abc}[X^a, Y^b] + \sin \psi \epsilon_{abc} ([X^b,
X^c] - [Y^b, Y^c]) \notag \\
e^2 \frac{\partial W}{\partial {\cal A}_3^a} &= 2[X^a,Y^a]
\end{align}

Notice that the combinations $D_3 Y^a + \psi' \frac{\cos
\psi}{\sin \psi} Y^a$ and $D_3 X^a - \psi' \frac{\sin \psi}{\cos
\psi} X^a$ appear here, as in (\ref{elb}) and (\ref{kelm}). It is
useful to remember the results of section \ref{computation}: in
that formalism, it was possible to reabsorb both the $X^2$ and
$Y^2$ terms in the component Lagrangian and the extra $\gamma
\cdot X \epsilon$ terms in the supersymmetry transformations by
rescaling the scalar fields as $\tilde X = X \cos \psi$ and
$\tilde Y = Y \sin \psi$. Similarly here, the choice of the extra
superpotential term  is such that the  terms in the gradient of
the superpotential that are linear in $X$ and $Y$ can be expressed
as ${D_3 \tilde X^a}/{\cos \psi}$ and $-{D_3 \tilde Y^a}/{\sin
\psi}$.

The quartic potential for $X$ and $Y$ is unchanged. The dangerous
non-$R$-symmetric cubic terms receive several contributions, but
these add up to  true total derivatives:
\begin{equation}\label{ame}-\partial_3
\left(\frac{1}{e^2} \cos \psi \epsilon_{abc} \Tr [X^a ,X^b]
Y^c\right)\end{equation} and \begin{equation}\label{bme}
-\partial_3\left(\frac{1}{e^2} \sin \psi\, \epsilon_{abc} \Tr X^a
[Y^b, Y^c]\right).\end{equation} The $R$-symmetric terms cubic in
$X$ are
\begin{equation}-\frac{2}{3 e^2}\frac{\psi'}{\cos \psi}
\epsilon_{abc} \Tr X^a [X^b, X^c]+ \frac{1}{3} \partial_3
\left(\frac{1}{e^2} \sin \psi \Tr X^a [X^b,
X^c]\right).\end{equation} The $R$-symmetric terms cubic in $Y$ are
\begin{equation}\frac{2}{3 e^2}\frac{\psi'}{\sin \psi}
\epsilon_{abc} \Tr Y^a [Y^b, Y^c]+ \frac{1}{3} \partial_3\left(
\frac{1}{e^2} \cos \psi \Tr Y^a [Y^b, Y^c]\right).\end{equation} We
see that the bosonic Lagrangian indeed agrees with the one in
section \ref{computation} up to total derivatives.

The fermion bilinear terms are also $R$-symmetric up to a total
derivative. The $\rho$ bilinears are:
\begin{equation}-\frac{2}{e^2}\rho_2^{a \alpha} D_3 \rho^a_{1\alpha} + 2 \psi'
\frac{1}{e^2} \frac{\sin \psi}{\cos \psi} \rho_2^{a\alpha}
\rho_{1\alpha}^a.\end{equation} The $\sigma$ bilinears are
\begin{equation}\label{golfx}-\frac{2}{e^2} \sigma_3^{\alpha} D_3 \sigma_{A\alpha}
+\frac{2}{e^2} \psi' \sigma_A^{\alpha}
\sigma_{A\alpha}.\end{equation} The last term is the non-trivial
term provided by the variable theta-angle $\theta_{Y\neg M}=2\pi
a+8\pi^2 D\cos 2\psi$
\begin{equation}
\frac{\theta}{4 \pi^2} \int \mathrm{d}^2 \theta \,\Tr\, D^\alpha
{\cal A}_3 W^\alpha = \frac{\theta}{4 \pi^2}\Tr \left(F \wedge F +i
\slashed{D}^{\alpha \beta}(\sigma_{A \alpha} \sigma_{3 \beta}) +
\frac{1}{2}D_3( \sigma_A^{\alpha} \sigma_{A \alpha})\right),
\end{equation}
after integration by parts.  From now on we will occasionally drop
the spinor indices for clarity. To verify $R$-symmetry, it is useful
to integrate the $\rho$ bilinear by parts to make the derivative
antisymmetric; this gives\footnote{Here and in (\ref{relt}), we use
$D=1/e^2\sin 2\psi$.}
\begin{equation}-\frac{1}{e^2}\rho_2^a \overleftrightarrow {D_3} \rho_1^a
+ 2D \psi' \rho_2^a \rho_1^a.\end{equation} We must verify that the
$\sigma$ bilinear can be written in the same form, with $\rho_i$
replaced by the appropriate linear combinations of $\sigma_3,
\sigma_A$. The derivative may act both on $\sigma_3, \sigma_A$ and
on the coefficients of the linear combination, and after some
rearrangements the form we need to find is
\begin{equation}\label{relt}-\frac{1}{e^2}\sigma_3 \overleftrightarrow {D_3} \sigma_A
+ \frac{\psi'}{ e^2}(\sigma_A \sigma_A + \sigma_3 \sigma_3) + D
\psi' \sin 2 \psi (\sigma_A \sigma_A - \sigma_3 \sigma_3) + 2D \psi'
\cos 2 \psi \sigma_A \sigma_3.\end{equation} Let us rearrange this a
little more, by integrating the antisymmetric derivative of $\sigma$
back to the form $\sigma_3 D_3 \sigma_A$ which appears in the
Lagrangian (\ref{golfx}). The result of the integration by parts
cancels the term $2D \psi'\cos 2 \psi \sigma_A \sigma_3$ and the
rest combines to $ \frac{2\psi'}{ e^2}\sigma_A \sigma_A$, as in
(\ref{golfx}).
 Hence we have proved $R$-symmetry of
the whole component Lagrangian.

Let us collect all the various non $R$-symmetric total derivatives
that appeared in this calculation from the cubic bosonic terms,
from the integration by parts of fermion bilinears, and from the
theta-angle.  They add up to
\begin{align}\label{theylo}\notag
\frac{\mathrm{d}}{\mathrm{d}x^3}\left(-\frac{1}{e^2}\right.& \cos
\psi \,\epsilon_{abc} \Tr\, [X^a, X^b] Y^c-\frac{1}{e^2} \sin \psi
\,\epsilon_{abc} \Tr\, X^a [Y^b, Y^c] - \frac{1}{e^2} \Tr\,
\sigma_A^{\alpha} \sigma_{3 \alpha}
\\&\left. - \frac{1}{e^2}\Tr\, \rho^{a \alpha}_1\rho^a_{2 \alpha} +
\frac{\theta_{Y\neg M}}{8 \pi^2} \Tr\, \sigma_A^{\alpha} \sigma_{A
\alpha}\right).\end{align} This formula will be useful at the next
step, when adding a boundary to the theory.

\subsection{Bifundamental Defect}\label{boundary}

We are going to apply what we have learned to a problem described in
section \ref{intersecting} and in fig. \ref{pic1}  -- an NS5-brane
with $N$ D3-branes ending from the left and $M$ from the right.
This system has been much-studied at $\theta_{Y\neg M}=0$, and the
resulting low energy physics is well-known.  There is an $\N=4$
theory with gauge group $U(N)$ in the half-space $x^3\leq 0$,
another $\N=4$ theory with gauge group $U(M)$ in the half-space
$x^3\geq 0$, and there are bifundamental hypermultiplets supported
on the hyperplane $x^3=0$ and interacting with the gauge fields on
both sides.  The problem also has a variant with an orientifold
threeplane parallel to the D3-branes; the gauge group is then
$SO(N)\times Sp(M)$, still with bifundamental hypermultiplets
supported at $x^3=0$.

As far as we know, the low energy effective action describing this
system at $\theta_{Y\neg M}\not=0$ has not been elucidated in the
literature.  It is easy to explain why. The formula
\begin{equation}\label{zorypto}-\frac{\theta_{Y\neg M}}{32\pi^2}
\int_{M_+} \mathrm{d}^4x\,
\epsilon^{\mu\nu\alpha\beta}\,\Tr\,F_{\mu\nu}F_{\alpha\beta}=
\frac{\theta_{Y\neg M}}{8\pi^2}\int_{\partial M_+} \mathrm{d}^3x
\epsilon^{\mu\nu\lambda}\Tr\,\left(A_\mu\partial_\nu
A_\lambda+\frac{2}{3}A_\mu A_\nu A_\lambda\right)  \end{equation}
(where $M_+$ is a half-space, and $\partial M_+$ is its boundary)
shows that supersymmetrizing the interaction $\Tr\,F\wedge F$ in
four-dimensional  gauge theory on a half-space is very similar to
supersymmetrizing the Chern-Simons interaction in three dimensions.
How to do this in a theory with the equivalent of three-dimensional
${\cal N}=4$ supersymmetry has not been clear.

However, in section \ref{chernsimons}, we constructed ${\cal N}=4$
Chern-Simons couplings for precisely the relevant cases --
$U(N)\times U(M)$ or $SO(N)\times Sp(M)$ gauge theory with
bifundamental hypermultiplets.  As we will see, the problem in
which the two factors of the gauge group live on four-dimensional
half-spaces $x^3\leq 0$ and $x^3\geq 0$ can be treated very
similarly to the purely three-dimensional problem of section
\ref{chernsimons}.

If $\theta_{Y\neg M}$ is of the form $2\pi q$, $q\in \Z$, then by an
$S$-duality transformation, one can set $\theta_{Y\neg M}$ to zero,
at the cost of replacing the NS5-brane with a $(1,q)$-fivebrane.  So
our analysis also governs a system of D3-branes ending from left and
right on a $(1,q)$-fivebrane.

We will not assume {\it a priori} that the gauge groups and matter
representations are the particular ones appropriate to the D3-NS5
system. But since we will find the same constraints as in section
(\ref{chernsimons}), this will turn out to be the case

\subsubsection{Gauge Fields In A Half-Space}\label{calcul}

We closely follow the logic of section \ref{janagain}, constructing
$\N=4$ super Yang-Mills theory in terms of three-dimensional $\N=1$
superfields.  Now, however, our gauge fields are defined only in a
half-space, say $x^3\geq 0$.  It is most simple to consider first a
one-sided problem with gauge fields in only one half-space and with
no hypermultiplets.\footnote{The case with constant $e^2$ and
$\theta_{Y\neg M}$ and no hypermultiplets will be treated more
directly elsewhere. The present approach has the advantages of
letting $e^2$ and $\theta_{Y\neg M}$ vary, and of extending to the
two-sided case.} This corresponds, in terms of branes, to having
D3-branes on only one side of an NS5-brane. Then in section
\ref{twosided}, we generalize to include hypermultiplets.  From the
standpoint of branes, the generalization is relevant to the
two-sided case with different gauge groups on the two sides, and
hypermultiplets supported at $x^3=0$.

The main difference from the previous analysis is that the various
non-$R$-symmetric total derivatives in the component Lagrangian
cannot be discarded.  They give boundary contributions at $x^3=0$.
These boundary contributions will play a role similar to the terms
that in section \ref{chernsimons} were found by integrating out
the auxiliary field $\chi$; they combine with terms coming from
the $\N=1$ superpotential to give an $R$-symmetric action.

In the following analysis, one can permit $e^2$ and $\theta_{Y\neg
M}$ to be $x^3$-dependent, as long as they are constrained by eqn.
(\ref{gufty}).  The boundary contributions that we focus on here do
not involve derivatives of $e^2$ and $\theta_{Y\neg M}$, so it
simply does not matter whether $e^2$ and $\theta_{Y\neg M}$ are
constant. These boundary terms can be read off from (\ref{theylo}),
and are
\begin{equation}\label{himt}
- \frac{1}{e^2} \cos \psi \epsilon_{abc} \Tr [X^a, X^b] Y^c -
\frac{1}{e^2} \sin \psi \epsilon_{abc} \Tr X^a [Y^b, Y^c]
\end{equation}
and
\begin{equation}\label{glimt}-\frac{1}{e^2} \Tr \sigma_A^{\alpha} \sigma_{3
\alpha} - \frac{1}{e^2}\Tr \rho^{a \alpha}_1\rho^a_{2 \alpha} +
\frac{\theta_{Y\neg M}}{8 \pi^2} \Tr \sigma_A^{\alpha} \sigma_{A
\alpha}.
\end{equation}

As usual, we want to make the $\rho_1 \rho_2$ term part of an
$R$-symmetric interaction $\Psi_1 \Psi_2$ by combining it with the
appropriate bilinear in $\sigma_3,\sigma_A$. After doing this, the
remaining truly non-$R$-symmetric terms are
\begin{equation}\label{fimt} -\frac{1}{e^2}\sin
\psi \cos \psi\Tr (\sigma_3\sigma_3-\sigma_A \sigma_A) -
\frac{2}{e^2}\sin^2 \psi \Tr \sigma_3 \sigma_A+\frac{ \theta_{Y\neg
M}}{8 \pi^2} \Tr \sigma_A \sigma_A.
\end{equation}

In the presence of a boundary, the computation of the derivatives of
the superpotential also needs to be re-examined.  Formula
(\ref{oxford}) for $\partial {\cal W}/\partial{\cal X}^a$ receives
an extra delta function contribution by integration by parts of the
$YD_3X$ contribution to ${\cal W}$.  So one now has
\begin{equation}\label{gelm}
-e^2 \frac{\partial {\cal W}}{\partial {\cal X}^a} = - 2D_3 Y^a -
2\psi' \frac{\cos \psi}{\sin \psi} Y^a + \cos \psi
\epsilon_{abc}([X^a, X^b] - [Y^a, Y^b]) + 2\sin \psi \epsilon_{abc}
[X^b, Y^c] +Y^a\delta(x^3).\end{equation} The action will contain a
term $\int \mathrm{d}x^3 |\partial {\cal W}/\partial{\cal X}|^2$,
and as we do not want  a term proportional to $\int \mathrm{d}x^3
\delta(x^3)^2$, we conclude that the boundary condition must be
$\vec Y=0$.

This argument is a little disingenuous, since the underlying theory
has a complete symmetry between $\vec Y$ and $\vec X$. Instead of
including in ${\cal W}$ a term $-\int \mathrm{d}x^3 \,\Tr \,{\cal
Y}D_3{\cal X}$, we could have integrated by parts and included a
term $\int \mathrm{d}x^3\,\Tr\,(D_3{\cal Y}) {\cal X}$.  This change
would not have affected the reasoning in section \ref{janagain}, but
an argument similar to the above\footnote{More generally, we could
take ${\cal W}$ to be a more generic linear combination of the two
expressions. Then to cancel delta function terms, we would be led to
impose $\vec X=\vec Y=0$ on the boundary. This, however, is
incompatible with preserving one-half of the supersymmetry.  We will
show this in more detail elsewhere, but a quick argument is as
follows. Given that $\vec Y=0$ on the boundary, we will deduce below
from supersymmetry that $\vec X$ must obey modified Neumann boundary
conditions (\ref{reft}).  It is therefore not possible for $\vec X$
to obey Dirichlet boundary conditions.} would now lead us to a
boundary condition $\vec X=0$. Actually, the boundary condition
$\vec Y=0$ is very natural for describing the D3-NS5 system of fig.
\ref{pic1}. If the NS5-brane is characterized by $x^7=x^8=x^9=0$,
then, as the scalar fields $Y^p$ parametrize the position of the
D3-branes in those directions, the boundary condition $\vec Y=0$ is
natural.  The boundary condition $\vec X=0$ is the one we want if
the NS5-brane is characterized by $x^4=x^5=x^6=0$.

The boundary condition $\vec Y=0$ is extended by ${\cal N}=1$
supersymmetry to a superspace boundary condition $\vec{\cal Y}=0$.
Using the superspace expansion of eqn. (\ref{het}), this amounts
to
\begin{equation}\label{peft}0 = Y^a=\rho_2^a=F_Y^a.\end{equation}
On the other hand, $F_Y={\partial \cal W}/\partial Y$ has been
computed in (\ref{oxford}).  Given the boundary condition $\vec
Y=0$, the vanishing of $F_Y$ at the boundary gives a boundary
condition for $\vec X$:
\begin{equation}\label{reft}0=D_3
X^a-\psi'\frac{\sin\psi}{\cos\psi}X^a=\frac{D_3\tilde
X^a}{\cos\psi},\end{equation} with $\tilde X^a=X^a\,\cos\psi$.
This boundary condition must of course also be extended to a
modified Neumann boundary condition on the superfield $\vec{\cal
X}$.

Since we want ${\cal N}=4$ supersymmetry, not just ${\cal N}=1$
supersymmetry, we must extend (\ref{peft}) to a set of boundary
conditions with the full $SU(2)\times SU(2)$ $R$-symmetry.  In
particular, the $SU(2)\times SU(2)$-symmetric extension of the
boundary condition $\rho_2^a=0$ is to require also that
\begin{equation}\label{lokey} \sin \psi \sigma_A +
\cos \psi \sigma_3=0 \end{equation} at the boundary. ${\cal N}=1$
supersymmetry then extends this to a further boundary condition
\begin{equation}\label{hokey} F_{3 \mu}= \tan \psi \frac{1}{2}\epsilon_{\mu \nu\rho}
F^{\nu \rho}.\end{equation}

Setting $Y=0$ automatically sets to zero the non-$R$-invariant
bosonic terms in (\ref{himt}). What about the fermionic terms in
(\ref{fimt})?  Precisely if
\begin{equation}\label{torgy}\frac{\theta_{Y\neg M}}{2 \pi} =-\frac{4 \pi}{e^2}
\frac{\sin \psi }{\cos \psi}\end{equation} at $x^3=0$,
(\ref{fimt}) becomes a ``perfect square''
\begin{equation}\label{perfsq}
 \frac{1}{e^2} \tan \psi (\cos \psi \sigma_3 + \sin \psi
\sigma_A)^2,\end{equation} and vanishes by virtue of the boundary
condition (\ref{lokey}). The condition (\ref{torgy}) is necessary
for this result, since it was needed to cancel a $\sigma_A^2$
coupling that is not $R$-symmetric.

Thus, we have established the full $R$-symmetry and hence $\N=4$
supersymmetry in the absence of hypermultiplets supported at
$x^3=0$.  Before going on to the more general case in section
\ref{twosided}, we pause to interpret the relation (\ref{torgy})
that was needed for this result.

Consider four-dimensional gauge fields with the action
\begin{equation}\label{gork}I=\int \mathrm{d}^4x \left(\frac{1}{
2e^2}F_{IJ}F^{IJ} -\frac{\theta_{Y\neg
M}}{32\pi^2}\epsilon^{IJKL}\Tr F_{IJ}F_{KL}\right)\end{equation}
(here we take indices $I,J,\dots=0,\dots,3$).  In this theory,
``free'' boundary conditions, in which the variation of the
connection is unconstrained on the boundary, read
\begin{equation}\label{ork}F_{3\mu}+\frac{e^2\theta_{Y\neg M}}{8\pi^2}\frac{1}{2}
\epsilon_{\mu\nu\rho} F^{\nu\rho}=0.\end{equation} Comparing this to
(\ref{hokey}), we see that the condition (\ref{torgy}) on
$\theta_{Y\neg M}$ is the condition under which the boundary
conditions derived from the action are compatible with
supersymmetry.\footnote{The action (\ref{gork}) also allows
Dirichlet boundary conditions, in which the connection and its
variation vanish on the boundary, but this is not compatible with
supersymmetry.}

The condition (\ref{torgy}) is essentially one that we have
already seen.  In eqn. (\ref{ppp}), we write the condition for
supersymmetry of a D3-brane plus a $(p,q)$-fivebrane.  For an
NS-fivebrane, that is for $p=1$, $q=0$, the condition is
\begin{equation}\label{kelp}-\frac{\cos \psi}{\sin\psi}=\frac{{\rm
Im}\,\tau}{{\rm Re}\,\tau}=\frac{4\pi/e^2}{\theta_{Y\neg
M}/2\pi},\end{equation} and this is equivalent to (\ref{torgy}).  So
this is really the expected condition for the supersymmetry of a
D3-brane ending on an NS5-brane.

The physics should be unchanged if we replace $\theta_{Y\neg M}$ by
$\theta_{Y\neg M}+2\pi q$ and replace the NS5-brane with a
$(1,q)$-fivebrane.  Shifting $\theta_{Y\neg M}$ by $2\pi q$ adds to
the action a bulk ``topological'' term with that coefficient.  To
ensure that the results are independent of $q$, it must be that
replacing the NS5-brane with a $(1,q)$-fivebrane has the effect of
adding to the effective action for D3-branes ending on the fivebrane
a boundary Chern-Simons coupling with coefficient $-q$.

\subsubsection{Including Hypermultiplets}\label{twosided}

Now we want to add three-dimensional hypermultiplets supported at
$x^3=0$ and transforming in some pseudoreal representation of the
gauge group. As in section \ref{chernsimons}, we represent them by
${\cal N}=1$ superfields ${\cal
Q}^I_A=Q^I_A+\theta^\alpha\lambda_{\alpha\,A}^I+\dots$; moreover, we
construct ${\cal N}=1$ couplings that have $SU(2)_d$ symmetry acting
on these superfields and adjust those couplings so that the symmetry
group is enlarged to  $SU(2)\times SU(2)$, with one factor acting on
$Q$ and one on $\lambda$.  How this $SU(2)\times SU(2)$ relates to
the $SO(3)_X\times SO(3)_Y$ symmetry that rotates the fields $\vec
X$ and $\vec Y$ of the bulk theory will become clear momentarily.

Once we include hypermultiplets, it is important to consider the
case that the gauge group $G$  is a product of simple factors $G_i$.
(The non-trivial examples generally have more than one factor.) Each
Lie algebra ${\frak g}_i$ has a quadratic form $(~,~)_i$, which  we
denote by $(a,b)_i=-\Tr\,ab$. Each factor has its own gauge coupling
$e_i$, its own theta-angle $\theta_{Y\neg M,i}$, and its own
supersymmetry angle $\psi_i$. They each obey (\ref{torgy})
\begin{equation}\label{quelm}\frac{\theta_{Y\neg M,i}}{2\pi}=-
\frac{4\pi}{e_i^2}\frac{\sin \psi_i}{\cos \psi_i}.\end{equation}
The derivation of this formula (either by canceling a $\sigma_A^2$
coupling or by considering the boundary condition obey by $F$) is
unaffected by the existence of hypermultiplets.  In addition, if
we set $D=1/e^2\sin\,2\psi$, it will turn out that all factors of
the gauge group have the same value of $D$.

Each factor $G_i$ of the gauge group will be localized either at
$x^3\geq 0 $ or at $x^3\leq 0$. Until it is necessary to combine the
different semisimple factors of the gauge group, we will proceed as
if there were just one factor $G$.

\bigskip\noindent{\it The  Hypermultiplet Action}

Now we consider the part of the action that involves the
hypermultiplets. The gauge-covariant kinetic energy of ${\cal Q}$
is familiar:
\begin{equation}\label{kinetico}
-\frac{1}{2}\int \mathrm{d}^2 \theta ({\cal D}_\alpha {\cal
Q}^I_A)^2 = \frac{1}{2}  \epsilon^{AB}\left(- \omega_{IJ}D_{\mu}
Q^I_A D^{\mu} Q^J_B + \omega_{IJ}\lambda^{I \alpha}_A (i
\slashed{D})_{\alpha}^{\beta} \lambda^J_{B \beta}+ \omega_{IJ}F^I_{Q
A} F^J_{Q B} + 2 \lambda^{\alpha I}_A \tau^m_{IJ} \sigma_{Am \alpha}
Q^J_B\right).
\end{equation}
As in the purely three-dimensional problem, since ${\cal Q}$ has
dimension $1/2$,   a general quartic superpotential for the matter
theory will preserve superconformal invariance at least
classically:
\begin{equation}\label{genquart}
\int \mathrm{d}^2\theta W_4({\cal Q}).
\end{equation}
Unlike the purely three-dimensional case, we now have dimension 1
bulk superfields ${\cal X}$ and ${\cal Y}$, so we can preserve
conformal symmetry with a superpotential coupling ${\cal X}{\cal
Q}^2$ or ${\cal Y}{\cal Q}^2$, where here of course ${\cal X}$ and
${\cal Y}$ are evaluated at $x^3=0$.  Actually, in the unperturbed
problem, the boundary condition on ${\cal Y}$ was simply ${\cal
Y}=0$, so a boundary coupling involving ${\cal Y}$ does not add
anything.\footnote{In the combined system, the boundary condition
${\cal  Y}=0$ is modified to ${\cal Y}\sim {\cal Q}^2$, as we see
momentarily.  Still this means that a ${\cal Y}{\cal Q}^2$
superpotential interaction is equivalent to a modification of the
quartic superpotential $W_4({\cal Q})$.} As we will see,
$R$-symmetry requires a ${\cal X}{\cal Q}^2$ term in the
superpotential.  It turns  out that ${\cal X}$ should couple
precisely to the moment map superfield ${\cal M}^m_{AB}={\cal
Q}^I_A{\cal Q}^J_B\tau^m_{IJ}$. In order to facilitate the
calculations, we will replace in the remainder of this section the
vector indices with symmetric pairs of doublet indices: for example
the ${\cal X}{\cal Q}^2$ interaction term is
\begin{equation}\label{xqq}
c\int \mathrm{d}^2\theta {\cal X}_m^{AB}{\cal Q}^I_A {\cal Q}^J_B
\tau^m_{IJ} = c \tau^m_{IJ} (F_{xm}^{AB}Q^I_A Q^J_B+2 X_m^{AB}
F_{qA}^I Q^J_B+2 \rho_{1m}^{AB \alpha}\lambda^I_{A\alpha}
Q^J_B+X_m^{AB}\lambda^I_A \lambda^J_B )
\end{equation}

Basic formulae about this replacement are collected in appendix
\ref{norm}. An important one is that $X^{AB} X_{AB} = - 2 X^a X^a$.
The $\delta(x^3)$ term in the ${\cal X}$ gradient of the
superpotential becomes
\begin{equation} \frac{1}{e^2} Y_{AB}^m +c \tau^m_{IJ} Q^I_A Q^J_B.
\end{equation} The vanishing of the $\delta(x^3)^2$ term in the
action now tells us that the Dirichlet boundary condition on $Y$
must be modified to
\begin{equation}\label{moddir}Y^{m a} =-c e^2 \tau^m_{IJ} (Q^I \sigma^a
Q^J).\end{equation} Again, this boundary condition will imply a set
of boundary conditions for the other fields. By ${\cal N}=1$
supersymmetry,
 \begin{equation}\label{moddirf}\rho^{m }_{\alpha AB} = -c e^2
 \tau^m_{IJ} Q^I_{(A}\lambda^J_{B)\alpha}.\end{equation} Extending this by $R$-symmetry, we get the most
interesting relation: \begin{equation}\label{moddirfo}\cos \psi
\sigma_{3\alpha} + \sin \psi \sigma_{A \alpha}= -c e^2 \tau^m_{IJ}
Q^{I A} \lambda^J_{A \alpha}.\end{equation} One further application
of ${\cal N}=1$ supersymmetry gives another interesting relation:
\begin{equation}\label{girevel}F_{3\mu} = \tan \psi \frac{1}{2}
\epsilon_{\mu \nu \rho} F^{\nu \rho} - \frac{c e^2}{\cos\psi}
J^{\mu}.\end{equation} (Appendix \ref{none} is useful to understand
the precise normalization.)

This should be compared to free boundary conditions on a gauge
field with a theta-angle and a boundary coupling:
\begin{equation}\label{ogork}I=\int_{M} \mathrm{d}^4x \left(\frac{1}{
2e^2}F_{IJ}F^{IJ} -\frac{\theta_{Y\neg
M}}{32\pi^2}\epsilon^{IJKL}\Tr F_{IJ}F_{KL}\right)+\int_{\partial
M}I'.\end{equation} Requiring that $I$ should be stationary with no
restriction on the variation of $A$ at the boundary, we get a
boundary condition on $A$ that coincides with (\ref{girevel}), if
the current of the boundary fields is defined as usual by
$J^\mu=\delta I'/\delta A_\mu$ and if in addition
\begin{equation}\label{zogork}c=\frac{1}{2} \cos\psi.\end{equation}
The same restriction on $c$ will appear in a moment from the
$R$-symmetry analysis. The boundary condition on $Y$ can be written
in terms of the moment map
\begin{equation}\label{moddirg}Y^{m a} =-\frac{1}{2}e^2\cos\psi \mu^{m a}.\end{equation}
For this equation to be $R$-symmetric, the $SU(2)$ factor in the
$R$-symmetry group that acts on $Y$ must be the same as the one
that acts on $Q$.
 So we
aim for a construction in which one factor in the  $SU(2)\times
SU(2)$ $R$-symmetry group rotates $Q$ and $\vec Y$, while the other
rotates $\vec X$ and $\lambda$.

The non-$R$-symmetric bosonic boundary terms of eqn. (\ref{himt}),
namely
\begin{equation}-\epsilon_{abc}\delta(x_3) \Tr \left(\cos \psi
\frac{1}{e^2} Y^a [X^b, X^c]+ \sin\psi \frac{1}{e^2} X^a [Y^b,
Y^c]\right),\end{equation}  were dismissed in section \ref{calcul}
because of the boundary condition $\vec Y=0$.   These terms are now
equivalent to $X^2Q^2$ and $XQ^4$ boundary couplings, which will in
general break $R$-symmetry.  The existence of separate $SU(2)$
groups rotating $\vec X$ and $Q$ strongly constrains $X^2Q^2$
interactions and means that $XQ^4$ interactions should be absent. We
can rewrite the preceding formula in terms of doublet indices
\begin{equation}\label{nrsym}-i \delta(x_3) \Tr \left(\cos \psi
\frac{1}{2e^2} Y^A_B [X^B_C, X^C_A]+ \sin\psi \frac{1}{2e^2} X^A_B
[Y^B_C, Y^C_A]\right).\end{equation}

There are further non-$R$-symmetric terms arising from the
superpotential.  Its derivative with respect to $Q$ is
\begin{equation}\label{nexc}
2 c \tau^m_{IJ} X_m^{AB} Q^J_B + \partial^A_I W_4 \\
\end{equation}
and the square of this gives interactions
\begin{equation}2 c^2 \tau^m_{IJ}\tau^n_{KT} X_m^{AB}X_n^{CD} Q^J_B Q^T_D
\omega_{IK} \epsilon_{AC} + 2 c \tau^m_{IJ} X_m^{AB} Q^J_B
\partial_A^I W_4.\end{equation}

Now we have several kind of terms to play against each other. If the
terms proportional to $X^aX^b$ are symmetrized in $a$ and $b$, the
result is actually proportional to $\vec X\cdot \vec X$, and thus is
$R$-symmetric, that is, invariant under separate rotations of $X$
and $Q$. However, the part antisymmetric in $a$ and $b$, which
contracts $\epsilon_{abc} X^b X^c$ with an expression bilinear in
$Q$, is non-$R$-symmetric. It is \begin{equation}c^2
\left(\tau^m_{IJ}\omega_{IK}\tau^n_{KT}-\tau^n_{IJ}\omega_{IK}\tau^m_{KT}\right)
X_m^{AB}X_n^{CD} Q^J_B Q^T_D \epsilon_{AC}=c^2
f^{mnp}X_m^{AB}X_n^{CD} \epsilon_{AC} Q^J_B \tau_{p
JT}Q^T_D.\end{equation} On the other hand, we can apply the $Y$
boundary condition to the first term in (\ref{nrsym}) and replace
the trace $\Tr Y[X,X]$ with an explicit sum over the gauge group
structure constants:\begin{equation}\label{nrsymx }\frac{1}{2}\cos
\psi c f^{mnp} X^{B}_{mC} X^{C}_{nA} \tau_{pIJ} Q^{IA}
Q^J_B.\end{equation} The $QQXX$ terms cancel against each other due
to the boundary conditions on $Y$ if $c = \frac{1}{2} \cos \psi$. We
will deal with the $XQ^4$ interaction momentarily.

\bigskip\noindent{\it The Yukawa Couplings}

Now we come to what is in a sense the main point: for the
configuration to be supersymmetric, we require just the same
condition on the gauge group and hypermultiplet representation as
in section \ref{chernsimons}.  As before, this result will come
from ensuring $R$-symmetry of the ``Yukawa couplings''
$Q^2\lambda^2$. This is the only point in the derivation at which
we sum over all simple factors $G_i$ in the gauge group.

There are two sources of $Q^2 \lambda^2$ couplings. One is
(\ref{perfsq}); the second comes from the $R$-symmetrization of the
coupling \begin{equation} \cos \psi \rho^{AB\alpha}_{1m}
\lambda^I_{A \alpha} \tau^m_{IJ} Q^J_B. \end{equation} This requires
as usual a term \begin{equation} \cos \psi (\cos \psi \sigma_{Am} -
\sin \psi \sigma_{3m}) \lambda^I_{A \alpha} \tau^m_{IJ} Q^{JB}
\end{equation} but only
\begin{equation} \lambda^I_A \tau^m_{IJ} Q^{JA} \sigma_{Am} \end{equation}
is present in the Lagrangian. As a result, after completing the
$\rho$ coupling to an $R$-symmetric coupling, one is left with
\begin{equation}\sin \psi (\sin \psi \sigma_{Am} + \cos \psi
\sigma_{3m}) \lambda^I_{A \alpha} \tau^m_{IJ} Q^{JB} \end{equation}

If we start from this term and from (\ref{perfsq}) and apply the
fermion boundary conditions (\ref{moddirfo}), we get the following
key boundary Yukawa coupling:
\begin{equation}\label{noldo} \pi Q^I_A Q^J_B \epsilon^{\alpha \beta}
\lambda^K_{\alpha \dot C} \lambda^S_{\beta \dot D}\epsilon^{A\dot C}
\epsilon^{B \dot D} \tau^m_{IK} \tau^n_{JS} \tilde k_{mn}.
\end{equation}  The interaction (\ref{noldo}) involves hypermultiplet
fields only, and receives contributions from every factor $G_i$.  In
this formula, $\tilde k^{mn}$ is a quadratic form on the Lie algebra
of $G$ that is defined as follows.  On $\frak g_i$, the quadratic
form equals $\pm 8\pi  (~,~)_i/e_i^2 \sin 2\psi_i$, where the sign
is $+$ or $-$ according to whether the group $G_i$ is supported for
$x^3<0$ or for $x^3>0$.

Eqn. (\ref{noldo}) is not $R$-symmetric, but it is identical in form
to the first term in eqn. (\ref{troublint}) -- the term which in the
purely three-dimensional derivation came from integrating out the
auxiliary field $\chi$. So the cure is the same as in section
\ref{chernsimons}. After picking the same superpotential ${\cal W}_4
= \frac{\pi}{6} \tilde k_{mn} \mu^{m\,AB}\mu^n{}_{AB}$ as in the
purely three-dimensional case, we can combine two kinds of Yukawa
couplings into an $R$-symmetric combination. In doing so, we have to
obey the same constraint on the matter and gauge content as for the
Chern-Simons theory.  So $G$ must be the bosonic part of a
supergroup $\hat G$, whose Lie algebra has an invariant,
nondegenerate quadratic form whose restriction to $\frak g$ is
$\tilde k$.

\bigskip\noindent{\it Application To The D3-NS5 System}

Let us specialize this to the D3-NS5 system, with $N$ D3-branes on
one side of an NS5-brane, and $M$ on the other.  The supergroup is
$U(N|M)$, and the gauge group is $G=U(N)\times U(M)$.  The usual
invariant quadratic form $k$ on $\frak g$ is equal to the trace
$\Tr$ on one summand of $\frak g$ and $-\Tr$ on the other.  The
form $\tilde k$ found above must be a multiple of this (since it
must be the restriction to $\frak g$ of an invariant quadratic
form on the super Lie algebra).  Hence, writing $e_1, e_2$ and
$\psi_1,\psi_2$ for the values in the two factors, we have
\begin{equation}\label{gorme} e_1^2\sin 2\psi_1=e_2^2\sin
2\psi_2.\end{equation}

This result along with (\ref{torgy}) has a simple interpretation. It
says that the points $(e_1,\theta_{Y\neg M,1},\psi_1)$ and
$(e_2,\theta_{Y\neg M,2},\psi_2)$ obey a relation of the specific
form $\tau=4\pi D(\exp(2i\psi) -1)$. As $\psi$ varies, this defines
a semicircle in the upper half plane whose rightmost intersection
with the real axis is at $\tau=0$. We explained in section
\ref{janfive} that a Janus configuration of precisely that type
preserves the same supersymmetry as an NS5-brane.  Hence, there
should be a low energy supersymmetric action describing an NS5-brane
interacting with such a Janus configuration.  This is what we have
found, at least for the case that the couplings change only by
jumping in crossing the fivebrane. The extension to the general case
is immediate, since, as in section \ref{calcul}, even if we let
$e^2$ and $\theta_{Y\neg M}$ vary, the boundary terms do not depend
on their derivatives.

Actually, as explained in section \ref{janusback}, the existence of
the general Janus configuration can be inferred from the properties
of the D3-NS5 system. We simply consider, as in fig. \ref{pic2}, a
system of D3-branes interacting with NS5-branes located at different
values of $x^3$, with constant couplings $e_i$ and $\theta_{Y\neg
M,i}$ and supersymmetry parameters $\psi_i$ in between the
NS5-branes.
 We take the number of D3-branes to everywhere equal $N$. We
describe each interface by the above construction, with jumps in
couplings that are constrained by (\ref{torgy}) and by the fact that
$D=1/e_i^2\sin 2\psi_i$ must be constant.  Thus the couplings all
take values in the usual semicircle.

Finally, we remove  the NS5-branes by displacing them in the
$x^7-x^8-x^9$ directions.  This causes the various $U(N)$ gauge
groups (in the half-spaces and slabs separated by NS5-branes) to
recombine into a single four-dimensional gauge group. The couplings,
however, jump in a discrete version of the Janus configuration,
which can approximate a continuous Janus configuration when the
number of NS5-branes is very large.

The
 process of displacing the NS5-branes so that they do not meet
the D3-branes corresponds in field theory to giving expectation
values to the hypermultiplet fields $Q$. (This is familiar in the
absence of the theta-angle.)  As we explained in analyzing eqn.
(\ref{zopo}), setting the potential energy to zero requires that the
matrices $Q_AQ^\dagger{}_B$ commute, and likewise the matrices
$Q^\dagger{}_BQ_A$. Generic expectation values for these matrices
break the $U(N)\times U(N)$ symmetry to a diagonal $U(N)$ or a
subgroup thereof.

If we symmetrize in $A$ and $B$, the matrices
$QQ^\dagger{}_{(AB)}$ and $Q^\dagger Q_{(AB)}$ become the moment
maps for the two factors of the gauge group, and according to
(\ref{moddirg}), they are proportional to  $\vec Y$ at $x^3=0$.
The matrices $QQ^\dagger{}_{(AB})$ and $Q^\dagger Q_{(AB)}$ have
the same eigenvalues, so the boundaries values of $\vec Y$ in one
$U(N)$ group are conjugate to those in the other, consistent with
the claim that the symmetry breaking combines the two $\vec Y$
fields into a single such field of a single $U(N)$ gauge symmetry.
If the matrices $QQ^\dagger{}_{(AB)}$ and $Q^\dagger Q_{(AB)}$ are
(large) multiples of the identity, the $U(N)$ gauge symmetry is
unbroken, the NS5-branes decouple, and we reduce to a discrete
Janus configuration.

\bigskip\noindent{\it The $XQ^4$ Terms}

We still have to check the vanishing of the $XQ^4$ term in the
action. One contribution from the $XYY$ bulk total derivative is simply
\begin{equation}\label{lb1}-i
\frac{e^2}{4}\sin\psi \cos^2 \psi f^{mnp} X^A_{mB} \mu^{B}_{C\,n}
\mu^{C}_{A\,p}.
\end{equation}
 The second comes from the ${\cal Q}$ derivative of the superpotential
 and is more complex: \begin{equation}\frac{2\pi}{3 }
 \tilde k_{mn}\cos \psi \mu^{AB m}
 Q^I_A (\tau^m \tau^n)_{IJ} Q^{J C} X_{BC n}.\end{equation}
These terms better cancel each other out. We need to rearrange the
second term quite a bit. We can use an antisymmetrization of the
$AC$ upper indices first to transform it to \begin{equation}\pi
\tilde k_{mn}\cos \psi \left(\mu^{AB m} Q^I_A (\tau^m \tau^n)_{IJ}
Q^{J C} X_{BC n}-\frac{1}{3 } \mu^{BC m} Q^I_A (\tau^m \tau^n)_{IJ}
Q^{J A} X_{BC n}+\frac{1}{3 } \mu^{AB m} Q^{I C} (\tau^m
\tau^n)_{IJ} Q^{J}_A X_{BC n}\right).\end{equation} Now we can use
the (\ref{fund}) identity to combine the last two terms into
\begin{equation}\pi \tilde k_{mn}\cos \psi \left(\mu^{AB m} Q^I_A
(\tau^m \tau^n)_{IJ} Q^{J C} X_{BC n}+ \mu^{AB m} Q^{I C} (\tau^m
\tau^n)_{IJ} Q^{J}_A X_{BC n}\right).\end{equation} Finally, these
two terms differ only by the order of $\tau^m$ and $\tau^n$, and one
can use the Jacobi identity on the gauge generators to finally
recast it in the same form as \ref{lb1}, but with opposite sign
\begin{equation}\pi \tilde k_{mn}\cos \psi \mu^{AB m} Q^I_A \tau^k_{IJ}
Q^{J C} X^n_{BC} f_{mnk}.\end{equation}

\bigskip\noindent{\it Coupling to General CFT}

As in the purely three-dimensional case of section
\ref{chernsimons}, it is natural to express the various ingredients
in this construction in terms of the current supermultiplet, with an
eye towards a generalization involving  a generic supersymmetric
model that satisfies the fundamental identity. For example, the
boundary coupling to $X$ can be expressed in terms of the $\N=1$
supermultiplet $\cal M$ whose lowest term is the moment map:
\begin{equation}\label{bocs}
\frac{1}{2} \cos \psi \int \mathrm{d}^2\theta {\cal X}_m^{AB} {\cal
M}^m_{AB}.
\end{equation}
The boundary conditions on the bulk fields have the various members
of the ${\cal N}=4$ current supermultiplet on the right hand side,
and are immediately generalized to the hyper-Kahler sigma model. The
proof of $R$-symmetry of the Yukawa couplings is identical to the
Chern-Simons case, and leads one to hyper-Kahler manifolds with
moment maps satisfying the fundamental identity. The only new
ingredient is to prove that the non-$R$-symmetric terms of the form
$X^2Q^2$ and $XQ^4$ do cancel as a consequence of the fundamental
identity as well. As the superpotential is given in terms of the
moment map superfield, the bosonic terms will involve the gradient
squared of moment maps. On the other hand the terms which come from
the total derivatives $\partial_3 \Tr XXY$ and $\partial_3 \Tr XYY$
involve the moment maps directly. To be able to compare the two
terms, we will need a simple but useful relation from symplectic
geometry between the Poisson bracket of moment maps and the moment
map of the Lie bracket of the corresponding vector fields:
\begin{equation} i_{V^n}\left( i_{V^m}(\omega_{AB})\right) = f_{mnp} \mu^p_{AB}\end{equation}
Here is a useful consequence of this relation: \begin{equation}
\label{lb2}  \frac{1}{2}(d\mu^{[m}_{AB},d\mu^{n]}_{CD}) =
\frac{1}{2}(i_{V^{[m}} \omega_{AB},i_{V^{n]}} \omega_{CD}) =
\epsilon_{BC} i_{V^n}\left(
i_{V^m}(\omega_{AD})\right)+\mathit{sym}= \epsilon_{BC}f^{mn}_p
\mu^p_{AD}+\mathit{sym}\end{equation} We used the fact that the
three complex structures have the same algebra as the unit
quaternions, so that up to an appropriate constant $\omega^1_{ij}
\omega^2_{kt} g^{jk} = \omega^3_{it}$ and $\omega^1_{ij}
\omega^1_{kt} g^{jk} = g_{it}$ (here $i,j,k,t$  are indices in the
tangent or cotangent bundle). In particular it is also true that
\begin{equation} \label{lb3}\frac{1}{2}(d\mu^{(m}_{AB},d\mu^{n)}_{CD}) =
\frac{1}{2}(i_{V^{(m}} \omega_{AB},i_{V^{n)}} \omega_{CD}) =
\epsilon_{BC} \epsilon_{AD}(V^n,V^m)+\mathit{sym}\end{equation}

Let us put these two relations to work to cancel the non R-symmetric
$X^2\mu$ and $X\mu^2$ terms in the Lagrangian. In particular
\ref{lb2} will play the role that the Jacobi identity for $\tau$
played in the free hypermultiplet case. The $XX\mu$
non-$R$-symmetric term from the superpotential is
\begin{equation} \frac{1}{8} \cos^2 \psi X_m^{AB} X_n^{CD}
(d\mu^{m}_{AB},d\mu^{n}_{CD})\end{equation} More precisely the part
symmetric in $(AB)$ and $(CD)$ is proportional again to $X^a X^a$ by
\ref{lb3} and is $R$-symmetric, while the antisymmetric part is
\begin{equation} \frac{1}{2} \cos^2 \psi X_m^{AB} X_n^{CD}
\epsilon_{BC}f^{mn}_p \mu^p_{AD}\end{equation} by \ref{lb2} and
cancels against the $\Tr XXY$ total derivative. The analysis of the
terms linear in $X$ proceeds quite smoothly as well: we start with
\begin{equation} \frac{\pi}{6}\cos \psi \tilde k_{mp}\mu^p_{AB}
X_n^{CD} (d\mu^{m}_{AB},d\mu^{n}_{CD})  \end{equation} and proceed
in complete parallelism to the free field computation.

The quiver construction of section \ref{quivers} gives examples of
three-dimensional superconformal field theories that obey the
fundamental identity, so that the above analysis is applicable.
Actually, we can now motivate this quiver construction. In fig.
\ref{pic2} of section \ref{quivers}, the slabs between two
NS5-branes are  macroscopically  only three-dimensional (as the
$x^3$ coordinate is bounded between two branes), so  at low energies
one can use an effective three-dimensional description of the gauge
fields that live in the slabs. In this description, the
three-dimensional gauge fields in the slabs have no Chern-Simons
couplings, since the contributions to those couplings cancel at the
two ends of the slab. Hence these gauge fields only have ordinary
$F^2$ kinetic energy. In the limit that the slabs are thin, the
three-dimensional gauge fields become strongly coupled. We can
integrate them out of the low energy analysis by taking a
hyper-Kahler quotient of the hypermultiplets that they couple to.
This leads to the quiver construction of section \ref{quivers}. The
theories that arise from the quiver construction must obey the
fundamental identity; indeed, we know from the analysis of the
D3-NS5 system that the brane configuration of fig. \ref{pic2} can be
coupled to bulk gauge fields with $\theta_{Y\neg M}\not=0$ (even
before taking the limit that the slabs become thin). We verified the
fundamental identity directly in section \ref{quivers}.

\appendix
\section{Some useful relations about spin indices}\label{rel}
The obvious identity \begin{equation} A_{[\alpha} B_{\beta]} =
-C_{\alpha \beta} A^{\gamma} B_{\gamma} \end{equation} will be often
useful, along with the similar formula \begin{equation} A^{\alpha}
B_{\beta} - A_{\beta} B^{\alpha} = \delta^{\alpha} _{\beta}
A^{\gamma} B_{\gamma}.
\end{equation} A vector is represented in  spinor notation as a
matrix
\begin{equation} \slashed{V}_{\alpha \beta} = \begin{pmatrix}V_0 + V_1
&V_2\\
 V_2 & V_0 - V_1\end{pmatrix}.\end{equation} The norm is
\begin{equation} \slashed{V}^{\alpha \beta} \slashed{W}_{\alpha
\beta} = 2 V^{\mu} W_{\mu}.\end{equation} in signature $-++$. The
exterior product is computed by \begin{equation} \slashed{V}_{\alpha
\beta} \slashed{W}_{\gamma \delta} C^{\beta \gamma} +
\slashed{V}_{\delta \beta} \slashed{W}_{\gamma \alpha} C^{\beta
\gamma} = 2 i \slashed{S}_{\alpha \delta}  \qquad S = V \wedge W
\end{equation} Moreover  \begin{equation} \slashed{V}_{\alpha \beta}
\slashed{W}_{\gamma \delta} \slashed{S}_{\rho \lambda} C^{\beta
\gamma} C^{\delta \rho} C^{\lambda \alpha}= -2 i \epsilon^{\mu \nu
\sigma} V_{\mu} W_{\nu} S_{\sigma} \end{equation}

Similar formulae are valid for $SU(2)$ indices, raised and lowered
by the conventional alternating tensor $\epsilon_{12} =
\epsilon^{12}=1$. A vector is represented in  spinor notation
as\begin{equation} \slashed{V}_{AB} = \begin{pmatrix}i V_2 + V_1&
V_3 \\ V_3 &i V_2 - V_1\end{pmatrix}.\end{equation} The norm is
\begin{equation} \slashed{V}^{AB} \slashed{W}_{AB} = -2 V^{\mu} W_{\mu}.\end{equation} The exterior product is
computed by \begin{equation} \slashed{V}_{AB} \slashed{W}_{CD}
\epsilon^{BC} + \slashed{V}_{DB} \slashed{W}_{CA} \epsilon^{BC} = 2
i \slashed{S}_{AD}  \qquad S = V \wedge W
\end{equation} Moreover  \begin{equation} \slashed{V}_{AB}
\slashed{W}_{CD} \slashed{S}_{EF} \epsilon^{BC} \epsilon^{DF}
\epsilon^{FA}= 2 i \epsilon^{abc} V_{a} W_{b} S_{c}
\end{equation}

\section{Some Miscellanea About ${\cal N}=1$}\label{none}
The content of this appendix is mostly trivial, but it is a useful
gymnastic in preparation to the next appendix, where the closure of
the ${\cal N}=4$ SUSY algebra is explicitly checked for our
Chern-Simons theory. Let us do a bit of ${\cal N}=1$ SUSY to check
the superfield lagrangians. Basic real superfield
\begin{align}
\delta \phi&= \varepsilon^{\alpha} \psi_{\alpha} \notag \\ \delta
\psi_{\alpha}&= \varepsilon^{\beta} i \slashed{\partial}_{\alpha
\beta} \phi - \varepsilon_{\alpha} F \notag \\ \delta F &=
\varepsilon^{\alpha} i \slashed{\partial}_{\alpha}^{\beta}
\psi_{\beta}
\end{align}

\begin{align}
\delta^2 \phi&= \varepsilon^{\alpha} \varepsilon^{\beta} i
\slashed{\partial}_{\alpha \beta} \phi \notag \\ \delta^2
\psi_{\alpha}&= \varepsilon^{\beta} \varepsilon^{\gamma}i
\slashed{\partial}_{\alpha \beta} \psi_{\gamma} -
\varepsilon_{\alpha} \varepsilon^{\beta} i
\slashed{\partial}_{\beta}^{\gamma} \psi_{\gamma}=
\varepsilon^{\beta} \varepsilon^{\gamma}i
\slashed{\partial}_{\gamma \beta} \psi_{\alpha}\notag \\
\delta^2 F &= \varepsilon^{\alpha} \varepsilon^{\gamma}i
\slashed{\partial}_{\alpha}^{\beta} i \slashed{\partial}_{\gamma
\beta} \phi-\varepsilon^{\alpha} \varepsilon_{\beta} i
\slashed{\partial}_{\alpha}^{\beta} F = \varepsilon^{\alpha}
\varepsilon^{\beta} i \slashed{\partial}_{\alpha \beta}F\end{align}

Kinetic terms: \begin{align} \delta \left(\frac{1}{2} F^2\right) &=
\varepsilon^{\alpha} F i \slashed{\partial}_{\alpha}^{\beta}
\psi_{\beta} \notag \\ \delta \left(\frac{1}{2} \psi^{\alpha} i
\slashed{\partial}_{\alpha}^{\beta} \psi_{\beta}\right)  &=
\varepsilon^{\gamma} i \slashed{\partial}_{\gamma}^{\alpha} \phi i
\slashed{\partial}_{\alpha}^{\beta} \psi_{\beta}-
\varepsilon^{\alpha} F i \slashed{\partial}_{\alpha}^{\beta}
\psi_{\beta}= -\frac{1}{2} \varepsilon^{\gamma}i
\slashed{\partial}_{\alpha}^{\beta} i
\slashed{\partial}_{\beta}^{\alpha} \phi \psi_{\gamma}-
\varepsilon^{\alpha}
F i \slashed{\partial}_{\alpha}^{\beta} \psi_{\beta} \notag \\
\delta \left(-\frac{1}{4} i \partial^{\alpha}_{\beta} \phi i
\partial^{\beta}_{\alpha} \phi \right)&= -\frac{1}{2} \varepsilon^{\gamma} i \partial^{\alpha}_{\beta} \phi i
\partial^{\beta}_{\alpha} \psi_{\gamma}= \frac{1}{2}\varepsilon^{\gamma} i
\slashed{\partial}^{\beta}_{\alpha} i
\slashed{\partial}^{\alpha}_{\beta} \phi \psi_{\gamma}
\end{align}

Superpotentials: \begin{align} \delta\left( W(\phi)' F\right)&=W''
\varepsilon^{\alpha} \psi_{\alpha} F + W'\varepsilon^{\alpha} i
\slashed{\partial}_{\alpha}^{\beta} \psi_{\beta} \notag \\
\delta \left(W(\phi)'' \frac{1}{2} \psi^{\alpha}
\psi_{\alpha}\right)& = \frac{1}{2} W''' \varepsilon^{\beta}
\psi_{\beta} \psi^{\alpha} \psi_{\alpha} + W'' \varepsilon^{\beta} i
\slashed{\partial}^{\alpha}_{ \beta} \phi \psi_{\alpha} -W''
\varepsilon^{\alpha} F \psi_{\alpha}
\end{align}

For the gauge multiplet
\begin{align}
\delta \slashed{A}_{\alpha \beta} &= \varepsilon_{\alpha}
\chi_{\beta} +
\varepsilon_{\beta} \chi_{\alpha} \notag \\
\delta \chi_{\alpha} &= \varepsilon^{\beta} f_{\alpha \beta}
\end{align}

\begin{align}
\delta^2 \slashed{A}_{\alpha \beta} &=
\varepsilon_{\alpha}\varepsilon^{\gamma} f_{\beta \gamma} +
\varepsilon_{\beta}\varepsilon^{\gamma} f_{\alpha \gamma} \notag \\
\delta^2 \chi_{\alpha} &= \varepsilon^{\beta} \delta f_{\alpha
\beta}
\end{align}

We need $\delta^2 \slashed{A}_{\alpha \beta} =  \varepsilon^{\gamma}
\varepsilon^{\delta} i \slashed{F}_{\gamma \delta;\alpha \beta}$ and
we learn
\begin{equation} i \slashed{F}_{\gamma \delta;\alpha \beta} =
\frac{1}{2} \epsilon_{\delta \alpha} f_{\beta \gamma} +\frac{1}{2}
\epsilon_{\gamma \alpha} f_{\beta \delta} +\frac{1}{2}
\epsilon_{\delta \beta} f_{\alpha \gamma} +\frac{1}{2}
\epsilon_{\gamma \beta} f_{\alpha \delta}
\end{equation} and
\begin{equation} f_{\alpha \beta} = \frac{1}{2} i
\slashed{F}_{\alpha \gamma;\beta}^{\gamma}\end{equation} Hence
\begin{equation} \varepsilon^{\beta} \delta f_{\alpha \beta} = \frac{1}{2} \left(\varepsilon^{\beta}
\varepsilon^{\gamma}i\slashed{\partial}_{\alpha \gamma} \chi_{\beta}
+\varepsilon^{\beta} \varepsilon_{\beta}i\slashed{\partial}_{\alpha
\gamma} \chi^{\gamma}-\varepsilon^{\beta}
\varepsilon_{\alpha}i\slashed{\partial}_{\beta}^{\gamma}
\chi_{\gamma}-\varepsilon^{\beta}
\varepsilon_{\gamma}i\slashed{\partial}_{\beta}^{\gamma}
\chi_{\alpha}\right)\end{equation} and works.

Kinetic terms
\begin{align}
\delta \frac{1}{2} \chi^{\alpha} i
\slashed{\partial}_{\alpha}^{\beta} \chi_{\beta} &=
\varepsilon^{\gamma} f_{\gamma}^{\alpha}i
\slashed{\partial}_{\alpha}^{\beta} \chi_{\beta} \notag \\
\delta -\frac{1}{4} f^{\alpha \beta} f_{\alpha \beta} &=-\frac{1}{2}
\left(f^{\alpha \beta}
\varepsilon^{\gamma}i\slashed{\partial}_{\alpha \gamma} \chi_{\beta}
+f^{\alpha \beta} \varepsilon_{\beta}i\slashed{\partial}_{\alpha
\gamma} \chi^{\gamma}\right)=-\frac{1}{2} \left(f^{\alpha \gamma}
\varepsilon^{\beta}i\slashed{\partial}_{\alpha \gamma}
\chi_{\beta}+2f^{\alpha \beta} \varepsilon_{\beta}
i\slashed{\partial}_{\alpha \gamma} \chi^{\gamma}\right) \end{align}
Remember $\slashed{\partial}^{\alpha \beta} f_{\alpha \beta}=0$.
More conventionally, \begin{equation} -\frac{1}{4} f^{\alpha \beta}
f_{\alpha \beta}= \frac{1}{16}\slashed{F}^{\alpha;\beta
\gamma}_{\gamma} \slashed{F}_{\alpha \delta;\beta}^{\delta}
=\frac{1}{16}\slashed{F}^{\alpha;\beta \delta}_{\gamma}
\slashed{F}_{\alpha
\delta;\beta}^{\gamma}+\frac{1}{16}\slashed{F}^{\alpha\gamma;\beta
\delta} \slashed{F}_{\alpha \gamma;\beta \delta} = \frac{1}{4}F^{\mu
\nu}F_{\mu \nu} \end{equation} The first term drops by symmetries.

 Chern Simons term
\begin{align}
\delta \frac{1}{2} \chi^{\alpha} \chi_{\alpha} &=
\varepsilon^{\gamma}
f_{\gamma}^{\alpha} \chi_{\alpha} \notag \\
\delta \frac{1}{4} \slashed{A}_{\alpha \beta} f^{\alpha \beta} &=
f^{\alpha \beta}\varepsilon_{\alpha} \chi_{\beta}
 \end{align}
More conventionally, \begin{equation} \frac{1}{4}
\slashed{A}^{\alpha \beta} f_{\alpha \beta} = \frac{1}{8}
\slashed{A}^{\alpha \beta} \slashed{F}_{\alpha
\gamma;\beta}^{\gamma} = \frac{1}{8} C^{\alpha \delta} C^{\beta
\sigma} C^{\gamma \tau} \slashed{A}_{\delta \sigma}
\slashed{F}_{\alpha \gamma;\beta \tau} = \frac{1}{4} \epsilon^{\mu
\nu \rho} A_{\mu} F_{\nu \rho} \end{equation}

Consider some real multiplets in a real representation of the gauge
group, with covariant derivative $i D_{\mu} = i \partial_{\mu}
\phi^i+ A_{\mu} T^i_j \phi^j$. The coupling to gauge fields is
\begin{equation}\frac{1}{2} \slashed{J}^{\alpha \beta} \slashed{A}_{\alpha \beta}
= \frac{1}{2} \psi^{\alpha} \slashed{A}_{\alpha}^{\beta}
T\psi_{\beta}-\frac{1}{2} \slashed{A}^{\alpha}_{\beta} T\phi i
\slashed{\partial}^{\beta}_{\alpha} \phi \end{equation} The current
\begin{equation}-\psi^{\alpha}T\psi^{\beta}-\phi T i
\slashed{\partial}^{\alpha \beta} \phi \end{equation} Notice that if
$j^{\alpha} = \phi T \psi^{\alpha}$ \begin{equation} \delta
j^{\alpha} = \varepsilon^{\beta}\psi_{\beta} T \psi^{\alpha}+
\varepsilon^{\beta} \phi T i \slashed{\partial}_{ \beta}^{\alpha}
\phi + \varepsilon^{\alpha} \phi T F =
\varepsilon_{\beta}\slashed{J}^{\alpha \beta} + \varepsilon^{\alpha}
\phi T F
\end{equation}

Hence \begin{align} \delta \frac{1}{2} \slashed{J}^{\alpha \beta}
\slashed{A}_{\alpha \beta} = \slashed{J}^{\alpha \beta}
\varepsilon_{\alpha} \chi_{\beta} + \cdots \notag \\
\delta -j^{\alpha} \chi_{\alpha} =-
\varepsilon_{\beta}\slashed{J}^{\alpha \beta} \chi_{\alpha}
\end{align}

\section{Closure of the Chern-Simons Supersymmetry Algebra}
\label{cherclo} The supersymmetry transformations are
\begin{align}
\delta Q^I_A &= \varepsilon_A^{\dot B \alpha} \lambda^I_{\dot B \alpha} \notag \\
\delta \lambda^I_{\dot A \alpha} &= \varepsilon_{\dot A}^{B \beta} i
\slashed{D}_{\alpha \beta} Q^I_B + \frac{1}{3} \varepsilon^B_{\dot A
\alpha} T^{mI}_J Q^{JC} Q^K_C \tau^n_{KT} Q^T_B k_{mn}\notag\\
\delta \slashed{A}_{m \alpha \beta} &= k_{m n} \varepsilon^{A \dot
B}_{(\alpha} \lambda^I_{\beta) \dot B} \tau^n_{IJ} Q^J_A\end{align}

For readability, we will denote $ T^{mI}_J Q^{JC}$ as $(TQ)^{IC}$,
$Q^K_C \tau^n_{KT} Q^T_B$ as $(Q\tau Q)_{CB}$, $\lambda^I_{\beta)
\dot B} \tau^n_{IJ} Q^J_A$ as $(\lambda \tau Q)_{\beta \dot B A}$
and leave $k_{mn}$ implicit, with a single exception for a formula
where two $k_{mn}$ appear.
\begin{align}
\delta Q^I_A &= \varepsilon_A^{\dot B \alpha} \lambda^I_{\dot B \alpha} \notag \\
\delta \lambda^I_{\dot A \alpha} &= \varepsilon_{\dot A}^{B \beta} i
\slashed{D}_{\alpha \beta} Q^I_B + \frac{1}{3} \varepsilon^B_{\dot A
\alpha} (TQ)^{IC} (Q\tau Q)_{CB}\notag\\
\delta \slashed{A}_{\alpha \beta} &= \varepsilon^{A \dot B}_{\alpha}
(\lambda \tau Q)_{\beta \dot B A}+\varepsilon^{A \dot B}_{\beta}
(\lambda \tau Q)_{\alpha \dot B A}\end{align}

The second variation of $Q$ is
\begin{equation} \delta^2 Q^I_A=\varepsilon_A^{\dot B \alpha}
\varepsilon_{\dot B}^{C \beta} i \slashed{D}_{\alpha \beta} Q^I_C +
\frac{1}{3} \varepsilon_A^{\dot B \alpha} \varepsilon^D_{\dot B
\alpha} (T Q)^{IC} (Q\tau Q)_{CD} k_{mn}
\end{equation}
In the first term the $\varepsilon^2$ part is antisymmetric in $A$
and $C$ and can be rewritten as
\begin{equation} \frac{1}{2} \varepsilon_C^{\dot B \alpha} \varepsilon_{\dot B}^{C
\beta} i \slashed{D}_{\alpha \beta} Q^I_A \end{equation} and is the
conventional gauge covariant translation. The second term needs some
rearrangements
\begin{equation}
-\frac{1}{3} \varepsilon_A^{\dot B \alpha} \varepsilon^D_{\dot B
\alpha} (TQ)^I_C (Q\tau Q)^C_D
\end{equation}
Antisymmetrizing on $AC$ gives
\begin{equation}
\frac{1}{3} \varepsilon_C^{\dot B \alpha} \varepsilon_{\dot B \alpha
D} (TQ)^{IC} (Q\tau Q)^D_A-\frac{1}{3} \varepsilon_C^{\dot B \alpha}
\varepsilon^D_{\dot B \alpha} (TQ)^I_A (Q\tau Q)^C_D
\end{equation}
Application of the fundamental identity to the first term by
cyclically permuting the $JKT$ indices in $T^{mI}_J \tau^n_{KT}
k_{mn}$ gives finally
\begin{equation} \frac{1}{2} \varepsilon^{\dot B \alpha C}
\varepsilon^D_{\dot B \alpha} \mu_{CD}^n k_{mn} T^{mI}_J Q^J_A.
\end{equation}
This is a gauge transformation by a parameter $\frac{1}{2}
\varepsilon^{\dot B \alpha C} \varepsilon^D_{\dot B \alpha}
\mu_{CD}^n k_{mn}$

Let us now look at the fermion supersymmetry transformation:
\begin{align}
\delta^2 \lambda^I_{\dot A \alpha} &= \varepsilon_{\dot A}^{B \beta}
\varepsilon_B^{\dot C \gamma} i \slashed{D}_{\alpha \beta}
\lambda^I_{\dot C \gamma}+\varepsilon_{\dot A}^{B \beta}
\varepsilon^{C \dot D}_{\alpha}(TQ)^I_B (\lambda \tau Q)_{\beta \dot
D C}+\varepsilon_{\dot A}^{B \alpha} \varepsilon^{C \dot
D}_{\beta}(TQ)^I_B (\lambda \tau Q)_{\beta \dot D
C} + \notag \\
& \frac{1}{3} \varepsilon^B_{\dot A \alpha} \varepsilon^{C \dot D
\beta} (T\lambda)^I_{\dot D \beta} (Q \tau Q)_{CB}+\frac{1}{3}
\varepsilon^B_{\dot A \alpha} \varepsilon^{\dot D \beta}_C (TQ)^{IC}
(\lambda \tau Q)_{\dot D \beta B}+\frac{1}{3} \varepsilon^B_{\dot A
\alpha} \varepsilon^{\dot D \beta}_B (TQ)^{IC} (Q \tau \lambda)_{C
\dot D \beta}
\end{align}

In the first term we just need to antisymmetrize the spin indices
$\alpha \gamma$
\begin{equation} \varepsilon_{\dot A}^{B \beta}
\varepsilon_B^{\dot C \gamma} i \slashed{D}_{\beta \gamma}
\lambda^I_{\dot C \alpha} + \varepsilon_{\dot A}^{B \beta}
\varepsilon_{B\alpha}^{\dot C} i \slashed{D}_{\beta}^{\gamma}
\lambda^I_{\dot C \gamma} \end{equation} The first part is the usual
translation, while the second part will go to the equations of
motion.

The last three terms can be rearranged through the fundamental
identity and recombined together. The total $QQ\lambda$ part is
\begin{equation} \varepsilon_{\dot A}^{B \beta}\varepsilon^{C \dot
D}_{\alpha}
 (TQ)^I_B (\lambda \tau
Q)_{\beta \dot D C}+\varepsilon_{\dot A}^{B \beta} \varepsilon^{C
\dot D}_{\beta}
 (T Q)^I_B k_{m n} (\lambda \tau Q)_{\alpha \dot D C}+
 \varepsilon^B_{\dot A \alpha} \varepsilon^{\dot D \beta}_C (T
Q)^{IC} (Q \tau \lambda)_{B \dot D \beta} \end{equation}. We expect
to find the same gauge transformation as before:
\begin{equation}
\frac{1}{2} \varepsilon^{\dot B \alpha C} \varepsilon^D_{\dot B
\alpha} \mu_{CD}^n k_{mn} T^{mI}_J \lambda^J_{\dot A}
\end{equation}
We saw from the derivative term that the equations of motions
instead should contain only the contraction $\varepsilon_{\dot A}^{B
\beta} \varepsilon_{B\alpha}^{\dot C}$. The residual symmetric part,
proportional to $\varepsilon_{\dot A \alpha}^{(B} \varepsilon_{\dot
D \beta}^{C)}$ should cancel.
\begin{align} \varepsilon_{\dot A}^{B \beta}\varepsilon^{C \dot
D}_{\alpha}
 (TQ)^I_B (\lambda \tau Q)_{\beta \dot D C}+\varepsilon_{\dot A}^{C \beta}\varepsilon^{B \dot D}_{\alpha}
 (TQ)^I_B (\lambda \tau Q)_{\beta \dot D C}+\varepsilon_{\dot A}^{B \beta} \varepsilon^{C \dot D}_{\beta}
 (TQ)^I_B (\lambda \tau Q)_{\alpha \dot D C}+ \notag \\ \varepsilon_{\dot A}^{C \beta} \varepsilon^{B \dot D}_{\beta}
 (TQ)^I_B (\lambda \tau Q)_{\alpha \dot D C}-
 \varepsilon^B_{\dot A \alpha} \varepsilon^{\dot D \beta C} (TQ)^I_C (Q\tau \lambda)_{B \dot D \beta}-
\varepsilon^C_{\dot A \alpha} \varepsilon^{\dot D \beta B} (TQ)^I_C
(Q \tau \lambda)_{B \dot D \beta}\notag \\- \varepsilon^{\dot B
\beta C} \varepsilon^D_{\dot B \beta} (Q \tau Q)_{CD}
(T\lambda)^I_{\dot A \alpha}
\end{align}
The first, second, third, fourth,fifth and sixth terms all come
together and the expression simplifies to
\begin{equation}-2\varepsilon^{\dot B \beta C}\varepsilon^D_{\beta \dot B}(TQ)^I_D (Q\tau \lambda)_{C \alpha \dot A}- \varepsilon^{\dot B \beta C}
\varepsilon^D_{\dot B \beta} (Q \tau Q)_{CD} (T\lambda)^I_{\dot A
\alpha}
\end{equation}

This is zero by the fundamental identity. The remaining terms
proportional to $\varepsilon_{\dot A}^{B \beta}
\varepsilon_{B\alpha}^{\dot C}$ are
\begin{equation} \frac{1}{2}\varepsilon_{\dot A B}^{\beta}\varepsilon^{B \dot
D}_{\alpha}
 (TQ)^I_C (\lambda \tau Q)^C_{\beta \dot D}+ \frac{1}{2}\varepsilon_{\dot A B}^{\beta} \varepsilon^{B \dot
D}_{\beta}
 (TQ)^I_C (\lambda \tau Q)^C_{\alpha \dot D}-
 \frac{1}{2} \varepsilon_{\dot A \alpha B} \varepsilon^{\dot D \beta B} (TQ)^{IC} (Q \tau \lambda)_{C \dot D \beta} \end{equation} It
is straightforward to rearrange the spin indices and recombine
everything to \begin{equation}\varepsilon_{\dot A
B}^{\beta}\varepsilon^{B \dot D}_{\alpha}
 (TQ)^I_C k_{m n}(\lambda \tau
Q)^C_{\beta \dot D} \end{equation}
 The fermionic equations of motion are
\begin{equation}\varepsilon_{\dot A}^{B \beta}
\varepsilon_{B\alpha}^{\dot C}\left( i \slashed{D}_{\beta}^{\gamma}
\lambda^I_{\dot C \gamma} -
 T^{mI}_T Q^T_D k_{m n} \tau^n_{KJ}
 Q^{JD} \lambda^K_{\beta \dot C}\right)=0\end{equation}

 Finally we want to look at the supersymmetry variations of the gauge fields
 \begin{equation}\delta^2 \slashed{A}_{m \alpha \beta} = \varepsilon^{A \dot
B}_{(\alpha} \varepsilon_A^{\dot C \gamma}(\lambda_{\dot C \gamma}
\tau \lambda_{\beta) \dot B}) + \varepsilon^{A \dot B}_{(\alpha}
\varepsilon_{\dot B}^{C \gamma}(Q_A \tau i \slashed{D}_{\beta)
\gamma} Q_C)+\frac{1}{3} \varepsilon^{A \dot B}_{(\alpha}
\varepsilon^D_{\beta)\dot B } k_{mn} (Q_A \tau^n
 T^o Q^C) k_{op} (Q \tau^p Q)_{CD}
\end{equation}

The first term is easy to discuss: the part of the $\varepsilon$
bilinear which is symmetric in $\dot B\dot C$ drops out and the rest
becomes
\begin{equation} \frac{1}{2}\left(\varepsilon^{\dot C \gamma}_A \varepsilon^A_{\dot C
(\alpha}\right)\left(\lambda^{\dot B}_{\beta)}\tau \lambda_{\dot B
\gamma} \right) =\frac{1}{2}\left(\varepsilon^{\dot C \gamma}_A
\varepsilon^A_{\dot C (\alpha}\right)k_{mn} \slashed{J}^{\lambda
n}_{\beta) \gamma}
\end{equation}

The second term written in full is \begin{equation} \varepsilon^{A
\dot B}_{\alpha}\varepsilon_{\dot B}^{C \gamma} (Q_A \tau i
\slashed{D}_{\beta \gamma} Q_C)+ \varepsilon^{A \dot B}_{\beta}
\varepsilon_{\dot B}^{C \gamma} (Q_A \tau i \slashed{D}_{\alpha
\gamma} Q)_C\end{equation} The part symmetric in $AC$ is the usual
gauge transformation \begin{equation} - i D_{\alpha \beta}
(\frac{1}{2} \varepsilon^{\dot B \alpha C} \varepsilon^D_{\dot B
\alpha} \mu_{CD}^n k_{mn} \end{equation}

The antisymmetric part is \begin{equation}
\frac{1}{2}\left(\varepsilon^{\dot C \gamma}_A \varepsilon^A_{\dot C
(\alpha}\right) \left(Q_B \tau i \slashed{D}_{\beta)\gamma}
Q^B\right)=\frac{1}{2}\left(\varepsilon^{\dot C \gamma}_A
\varepsilon^A_{\dot C (\alpha}\right)k_{mn}\slashed{J}^{Q n}_{\beta)
\gamma} \end{equation}

The third term is written in full as \begin{equation}\frac{1}{3}
\varepsilon^{A \dot B}_{\alpha} \varepsilon^D_{\beta\dot B }k_{m n}
(Q_A \tau^n T^o Q^C) k_{op} (Q \tau Q)_{CD} +\frac{1}{3}
\varepsilon^{A \dot B}_{\beta} \varepsilon^D_{\alpha\dot B } k_{m n}
(Q_A \tau^n T^o Q^C)k_{op} (Q \tau^p Q)_{CD}  \end{equation} The
$\varepsilon$ bilinears are actually antisymmetric in $AD$, hence we
can simplify a bit
\begin{equation}\frac{1}{3}
\varepsilon^{\dot B}_{D \alpha} \varepsilon^D_{\beta\dot B } k_{m
n}(Q_A \tau^n T^o Q^C)k_{op} (Q \tau^p Q)^A_C
\end{equation}

The $Q^J_A \tau^n_{IJ}  T^{oI}_S Q^{SC}$ multiplies a term symmetric
in $AC$. From the relation $T^I_J \omega_{IK} = \tau_{JK}$ it is
possible to see that the product of structure constants is made into
a commutator by symmetrizing $AC$, and the term simplifies to
something proportional to $\mu^p_{AC} \mu^{qAC} f_{pqm}$ which is
zero by complete antisymmetry of the structure constants.

Hence we learn that \begin{equation}\delta^2 \slashed{A}_{m \alpha
\beta}=\frac{1}{2}\left(\varepsilon^{\dot C \gamma}_A
\varepsilon^A_{\dot C (\alpha}\right)k_{mn}\slashed{J}^{n}_{\beta)
\gamma}
\end{equation}
Comparison with the expected result gives the equations of motion
for the gauge fields: \begin{equation} f_{m \alpha
\beta}=\slashed{J}^{n}_{\alpha \beta}\end{equation} Comparison with
appendix \ref{none} gives the normalization of the Chern Simons
term: $\frac{k^{mn}}{2} A\wedge dA$, hence to get a canonical
normalization we need to replace $k^{mn}\to \frac{k^{mn}}{2\pi}$.

The equations of motion for the fermions become \begin{equation}i
\slashed{D}_{\beta}^{\gamma} \lambda^I_{\dot C \gamma} - 2\pi k_{m
n} T^{mI}_T Q^T_D \tau^n_{KJ} Q^{JD} \lambda^K_{\beta \dot
C}.\end{equation} The Yukawa couplings must be $- \pi
(\lambda^{\beta \dot C}\tau^m Q_D)k_{m n}(Q^D\tau^n \lambda_{\beta
\dot C})$

Another useful normalization is the comparison with the ${\cal N}=1$
formalism. Just set $\varepsilon^A_{\dot B \alpha} = \delta^A_{\dot
B} \varepsilon_{\alpha}$ in the SUSY transformations to verify the
values of the auxiliary fields: (we also introduce the factor of
$2\pi$)
\begin{align}
\delta Q^I_A &= \varepsilon^{\alpha} \lambda^I_{A \alpha} \notag \\
\delta \lambda^I_{A \alpha} &= \varepsilon^{ \beta} i
\slashed{D}_{\alpha \beta} Q^I_A + \frac{2\pi}{3} \varepsilon_{\alpha} T^{mI}_J Q^{JC} Q^K_C \tau^n_{KT} Q^T_A k_{mn}\notag\\
\delta \slashed{A}_{m \alpha \beta} &= 2 \pi k_{m n}
\varepsilon_{(\alpha} \lambda^{A I}_{\beta)} \tau^n_{IJ}
Q^J_A\end{align}

We learn \begin{equation}F^I_A = -\frac{2 \pi}{3} T^{mI}_J Q^{JC}
Q^K_C \tau^n_{KT} Q^T_A k_{mn}\end{equation} and
\begin{equation}\chi_{\alpha} =2 \pi k_{m n}
 \lambda^{A I}_{\alpha} \tau^n_{IJ} Q^J_A\end{equation} as expected.

\section{Relating $4d$ and $3d$ Expressions}\label{norm}
Let us review again the ${\cal N}=4$ current multiplet in $3d$. For
free hypermultiplets the gauge current is
\begin{equation} J_{\beta \gamma} = \lambda^{\dot B}_{\beta}\tau
\lambda_{\dot B \gamma} + Q_B \tau i \slashed{D}_{\beta \gamma} Q^B
\end{equation} The supersymmetry variation of the moment map defines
the superpartner of the gauge current:
\begin{equation} \delta \mu_{AB} =\varepsilon_{(A}^{\dot C \alpha} \lambda^I_{\dot C \alpha} \tau_{IJ} Q^J_{B)}
= \varepsilon_{(A}^{\dot C \alpha} j_{B) \dot C \alpha} \quad j_{A
\dot B \alpha} = \lambda^I_{\dot B \alpha} \tau_{IJ} Q^J_A
\end{equation} The supersymmetry variation of the current
superpartner is then
\begin{equation}\delta j_{A \dot B \alpha} = \varepsilon_A^{\dot C \beta} \lambda^J_{\dot C \beta} \lambda^I_{\dot B \alpha} \tau_{IJ} +Q^J_A \varepsilon_{\dot B}^{C \beta}
i\slashed{D}_{\alpha \beta} Q^I_C \tau_{IJ} \end{equation} We can
separate various components by taking symmetric and antisymmetric
parts in the bosons and fermions. The part symmetric in the two $Q$
is \begin{equation}\frac{1}{2}\varepsilon_{\dot B}^{C
\beta}i\slashed{D}_{\alpha \beta} \mu_{AC} \end{equation}. The
fermion bilinear which is a spacetime scalar is \begin{equation}
\frac{1}{2}\varepsilon_{A\alpha}^{\dot C} \lambda^J_{\dot C \beta}
\lambda^{I \beta}_{\dot B}  \tau_{IJ}= \varepsilon_{A\alpha}^{\dot
C} O_{\dot C \dot B} \quad O_{\dot C \dot B} = \frac{1}{2}
\lambda^J_{\dot C \beta} \lambda^{I \beta}_{\dot B}  \tau_{IJ}
\end{equation} and the remaining part is \begin{equation}
\frac{1}{2}\varepsilon_{A \dot B}^{\beta} \lambda^J_{\dot E \beta}
\lambda^{I \dot E}_{\alpha} \tau_{IJ} -\frac{1}{2} \epsilon^{DE}
Q^J_D \varepsilon_{A \dot B }^{\beta} i \slashed{D}_{\alpha \beta}
Q^I_E \tau_{IJ} = - \frac{1}{2} \varepsilon_{A \dot B }^{\beta}
\slashed{J}_{\alpha \beta} \end{equation}

In the paper we use boundary conditions of the form $Y_{AB} = c
\mu_{AB}$. This equation makes sense because both $Y^{AB}$ and the
moment map are the leading components of two identical $OSp(4|4)$
supermultiplets. As a result, under 3d ${\cal N}=4$ supersymmetry
variation one gets a multiplet of boundary conditions equating
corresponding members of the supermultiplets. Indeed, the fields of
the ${\cal N}=4$ gauge multiplet in four dimensions decompose under
the $3d$ ${\cal N}=4$ supergroup into two multiplets which are
identical in quantum numbers to the current supermultiplet or to the
mirror current supermultiplet respectively ($Y$ has spin $(1,0)$ and
$X$ has spin $(0,1)$ under the $R$-symmetry group). The precise
decomposition depends on the value of $\psi$. The supersymmetry
variation
\begin{equation} \delta Y^a = i \bar \varepsilon \Gamma^a \Psi =-
\frac{i}{2} \epsilon^{abc} \bar \varepsilon \Gamma_{bc} \Gamma^3
B_1 \Psi \end{equation} As the generators $\Gamma_{bc}$ act on
$V_8$ only, the supersymmetry variation involves the projection of
$\Psi$ on a specific vector in $V_2$. If we use the decomposition
\begin{equation}\Psi = B_1 \varepsilon_0 \otimes \Psi_1 + B_2
\varepsilon_0 \otimes \Psi_2\end{equation} we can rewrite the
supersymmetry variation of $Y$  \begin{equation} \delta Y^a = -
\frac{i}{2} \epsilon^{abc} (\bar \varepsilon \Gamma^3 B_0
\varepsilon_0) \Gamma_{bc} \Psi_1 \end{equation} Now that everything
happens in $V_8$ we can reintroduce the $SO(2,1)\times SO(3) \times
SO(3)$ indices: \begin{equation} \delta Y_{AB} =
\varepsilon_{(A}^{\dot C \alpha}  \Psi_{1 B) \dot C \alpha}
\end{equation} and discover the boundary condition $ \Psi_{1 B \dot
C \alpha}=c  j_{B \dot C \alpha}.$ For the next step, we need
\begin{equation} \delta \Psi_1 = \bar \varepsilon_0 B_2 \Psi =
\frac{1}{2} \bar \varepsilon_0 B_2 \Gamma^{IJ} F_{IJ} \varepsilon.
\end{equation}There are several contributions. The gamma matrix
bilinears can be rewritten as various $B_i$ times generators of
$SO(2,1)\times SO(3) \times SO(3)$. These $B_i$ matrices are
sandwiched between $\bar \varepsilon_0$ and $\varepsilon =
\varepsilon_0 \otimes \eta$, and we can compute the inner products
in $V_2$ right away: if the matrix is $B_2$ the term drops out, as
$\bar \varepsilon_0 \varepsilon=0$, if the matrix is $B_1$ the
result is $\psi$ independent, if it is $1$ the result is
proportional to $-\sin \psi$, if it is $B_0$ it is proportional to
$- \cos \psi$. The $D_3 Y^p$ has $B_2$ and corresponds to the
derivative of the moment map in the $\delta j$. The gauge field
strengths come as $\frac{1}{2} F_{\mu \nu} \epsilon^{\mu \nu \rho}
\sin \psi + F^{\rho3} \cos \psi$, and should be equal to $c
J^{\rho}$. Physically, we know that
\begin{equation} F^{\rho3} = \frac{1}{2}J^{\rho}-\frac{1}{2} F_{\mu \nu}
\epsilon^{\mu \nu \rho} \tan \psi, \end{equation} where the second
piece on the right hand side is the current induced by the Chern
Simons term a the boundary or by the theta-angle. The factor of two
in front of the current is due to the slightly non-standard
normalization of the gauge field kinetic term. Hence we learn that
$c=\frac{1}{2}\cos \psi$ and $\frac{g^2 \theta_{Y\neg M}}{8 \pi^2}
=\frac{\sin \psi }{\cos \psi}$.

Finally, there are terms as $D_3 X_{\dot A \dot B}$ and $- \sin \psi
[X,X]_{\dot A \dot B}$ whose sum equals $c O_{\dot A \dot B}$.

Along similar lines we could compare the Yukawa couplings computed
with the $3d$ formalism to the conventional ${\cal N}=4$ ones. If we
plug
\begin{equation}\Psi = B_1 \varepsilon_0 \otimes \Psi_1 + B_2
\varepsilon_0 \otimes \Psi_2\end{equation} in the fermion kinetic
terms in ${\cal N}=4$ super Yang-Mills we get an expression in $V_8$
\begin{equation}-\frac{i}{e^2}\bar\Psi \Gamma^{\mu} D_{\mu}\Psi = -
\frac{i}{2e^2} \epsilon^{\mu \nu \rho} \sum_i \bar \Psi_i
\Gamma_{\nu \rho} D_{\mu} \Psi_i.\end{equation} we can reintroduce
the $SO(2,1)\otimes SO(3) \otimes SO(3)$ indices
\begin{equation} \frac{1}{e^2} \sum_i \Psi^{A \dot B \alpha}_i
\slashed{D}_{\alpha}^{\beta} \Psi_{i A \dot B \beta}
\end{equation} The $X$ Yukawa couplings in ${\cal N}=4$ super
Yang-Mills are \begin{equation}- \frac{i}{e^2} \bar \Psi \Gamma^a
[X^a,\Psi] =\frac{i}{2 e^2} \epsilon^{abc} \bar \Psi \Gamma_{bc}
\Gamma^3 B_1 [X^a,\Psi] \end{equation} If we plug in
\begin{equation}\Psi = B_1 \varepsilon_0 \otimes \Psi_1 + B_2
\varepsilon_0 \otimes \Psi_2\end{equation} we get \begin{equation}
\frac{i}{2 e^2} \epsilon^{abc} \left( - \cos \psi \bar \Psi_1
\Gamma_{bc} [X^a,\Psi_1] + \sin \psi \bar \Psi_1 \Gamma_{bc}
[X^a,\Psi_2] + \sin \psi \bar \Psi_2 \Gamma_{bc}  [X^a,\Psi_1] +\cos
\psi \bar \Psi_2\Gamma_{bc}  [X^a,\Psi_2] \right)\end{equation} Now
that everything happens in $V_8$ we can reintroduce the
$SO(2,1)\otimes SO(3) \otimes SO(3)$ indices:
\begin{equation}-\frac{1}{e^2} \left( - \cos \psi  \Psi^{A
\alpha}_{1 \dot B} [X^{\dot B \dot C},\Psi_{1A \dot C \alpha}] + 2
\sin \psi \Psi^{A \alpha}_{1 \dot B} [X^{\dot B \dot C},\Psi_{2A
\dot C \alpha}] +\cos \psi  \Psi^{A \alpha}_{2 \dot B} [X^{\dot B
\dot C},\Psi_{2 A \dot C \alpha}]\right) \end{equation} This agrees
with the computation in the text.

\bibliography{notebook}{}
\bibliographystyle{JHEP-2}
\end{document}